\documentclass[aps, pra, twocolumn, amsmath, amssymb, superscriptaddress, nofootinbib]{revtex4-1}

\makeatletter
\def\p@subsection{}
\def\p@subsubsection{}
\makeatother

\usepackage[utf8]{inputenc}
\usepackage{graphicx}
\usepackage{units}
\usepackage{color}
\usepackage[pdftex, colorlinks=true, linkcolor=myblue, citecolor=myblue,urlcolor=myblue]{hyperref}
\usepackage{times}
\usepackage{comment}
\usepackage{graphicx}
\usepackage{amsmath, calc}
\usepackage{mathrsfs}
\usepackage{amsfonts}
\usepackage{amssymb}
\usepackage{bm}
\usepackage{bbm}
\usepackage{color}
\usepackage{array}
\usepackage{units}

\definecolor{myblue}{rgb}{0,0,1}
\definecolor{myred}{rgb}{1,0,0}

\newcommand{\bra}[1]{\langle #1|}
\newcommand{\ket}[1]{|#1\rangle}

\newcommand{\bratwo}[1]{\langle \langle #1||}
\newcommand{\kettwo}[1]{||#1\rangle \rangle}

\begin{document}
\title{Asymmetric coupling between two quantum emitters}

\author{C.~A.~Downing}
\email{downing@unizar.es} 
\affiliation{Departamento de F\'{i}sica de la Materia Condensada, CSIC-Universidad de Zaragoza, Zaragoza E-50009, Spain}

\author{J.~C.~L\'{o}pez~Carre\~{n}o}
\affiliation{Departamento de F\'{i}sica T\'{e}orica de la Materia Condensada and Condensed Matter Physics Center (IFIMAC), Universidad Aut\'{o}noma de Madrid, E-28049 Madrid, Spain}
\affiliation{Faculty of Science and Engineering, University of Wolverhampton, 7 Wulfruna Street, Wolverhampton WV1 1LY, United Kingdom}

\author{A.~I.~Fern\'{a}ndez-Dom\'{i}nguez} 
\affiliation{Departamento de F\'{i}sica T\'{e}orica de la Materia Condensada and Condensed Matter Physics Center (IFIMAC), Universidad Aut\'{o}noma de Madrid, E-28049 Madrid, Spain}

\author{E.~del Valle}
\affiliation{Departamento de F\'{i}sica T\'{e}orica de la Materia Condensada and Condensed Matter Physics Center (IFIMAC), Universidad Aut\'{o}noma de Madrid, E-28049 Madrid, Spain}
\affiliation{Faculty of Science and Engineering, University of Wolverhampton, 7 Wulfruna Street, Wolverhampton WV1 1LY, United Kingdom}

\date{\today}

\begin{abstract}
We study a prototypical model of two coupled two-level systems, where the competition between coherent and dissipative coupling gives rise to a rich phenomenology. In particular, we analyze the case of asymmetric coupling, as well as the limiting case of chiral (or one-way) coupling. We investigate various quantum optical properties of the system, including its steady state populations, power spectrum, and second-order correlation functions, and outline the characteristic features which emerge in each quantity as one sweeps through the nontrivial landscape of effective complex couplings. Most importantly, we reveal instances of population trapping, unexpected spectral features and strong emission correlations.
\end{abstract}


\maketitle



\section{\label{intro}Introduction}

Chirality (or handedness) is an important concept across modern science. Originally used to describe an object which is not identical to its mirror image, chirality now encompasses asymmetries in various guises, including chemical reactions and sub-nuclear processes. The fact that nature is inherently chiral has profound consequences, from the chemistry of primordial biomolecules to the electroweak interaction in the Standard Model~\cite{Wainer2009, Wagniere2007, Guijarro2009}.

The nascent field of chiral quantum optics is concerned with systems where forward and backward propagating photons interact differently with a quantum emitter~\cite{Lodahl2017, Andrews2018, Chang2018}. The most extreme case is chiral (or one-way) coupling~\cite{Gardiner1993, Carmichael1993, Gardiner2004}. Exploiting chiral light-matter interactions is predicted to lead to a host of exciting applications in quantum communication, information and computing, including: non-reciprocal single-photon devices~\cite{Shomroni2014, Yan2018, Tang2019}; optical isolators~\cite{Sayrin2015}; optical circulators~\cite{Sollner2015, Scheucher2016}; integrated quantum optical circuits~\cite{Barik2018, Zhang2018, Fang2019, Barik2019} and quantum networks~\cite{Li2018, Grankin2018, Yan2018b, Zhang2019}. Concurrently, new horizons in more fundamental aspects are expected, such as in quantum entanglement~\cite{Stannigel2012}, unconventional many-body states~\cite{Ramos2014} and emergent quasiparticles~\cite{Chervy2018}.

Novel phenomena stemming from asymmetric coupling have been studied theoretically in a range of systems, including: spin networks~\cite{Pichler2015}; cavity-based photonic devices~\cite{Metelmann2015, Burillo2019}; quantum emitters coupled to plasmonic waveguides~\cite{Ballestero2015, Gonzalez2016}; nanophotonic ring resonators~\cite{Martin2018, Jalali2020}; synthetic phonons~\cite{Peterson2018}; and Janus dipoles~\cite{Picardi2018}. Recently, it was shown that chiral coupling at the nanoscale naturally arises in the system of two circularly-polarized quantum emitters held above a metal surface, where the surface plasmons mediating the emitter-emitter interactions can be controlled in a manifestation of reservoir engineering~\cite{Downing2019}. The earliest works in chiral nanophotonics and chiral plasmonics are reviewed in detail in Refs.~\cite{Bliokh2015, Schaferling2016, Hentschel2017}.

The theoretical frameworks behind a number of the models of chiral coupling are inextricably linked to the formalism of cascaded quantum systems, as independently developed by Gardiner and Carmichael to describe distant source-target quantum systems~\cite{Gardiner1993, Carmichael1993}. In such setups, chiral coupling appears by construction, with the first body (the source) coupling to the second body (the target), while completely forbidding coupling in the reverse direction~\cite{Gardiner2004}. As such, one may use cascaded theory to posit a well-defined criterion for chiral coupling~\cite{Metelmann2015, Metelmann2017, Downing2019}. Meanwhile, a number of papers have appeared recently successfully employing the cascaded formalism to uncover nontrivial photon correlations~\cite{Lopez2015, Lopez2016, Lopez2018, Lopez2019}.

In this work, we introduce a general model of two coupled two-level systems (2LSs)~\cite{AllenBook}. The theory has a wide variety of applications, with similar formalisms being used to describe superconducting qubits~\cite{Pashkin2009}, atomic and molecular spectroscopy~\cite{Wahiddin1998, Munoz2020}, plasmonic dimers~\cite{Bordo2019} and waveguides~\cite{Gonzalez2011, MartinCano2011}. The utility of the theory has allowed for a range of phenomena to be investigated, including entanglement~\cite{Tanas2004, Valle2007, Almutairi2011, Valle2011, Liao2011, Biehs2017}, decoherence~\cite{Grigorenko2005}, quantum processing~\cite{Petrosyan2002} and coherent energy transfer~\cite{Liao2010, McCutcheon2011}.

Our simple model, which importantly includes dissipative coupling~\cite{Wang2020} via an open quantum systems approach, encompasses regimes of coherent, dissipative, chiral and asymmetric coupling. To achieve this rich variety, we allow the coherent and dissipative coupling parameters to be complex quantities. Permitting complex phase degrees of freedom via the coupling parameters is known to greatly increase the depth of physics in quantum systems~\cite{DowningLuis, Burillo2019}, most famously in the Haldane model (where adding complex next-nearest-neighbor hoppings leads to topological non-trivialities~\cite{Haldane1988}). Here we show how modulating the relative strength and the relative phase of the coherent and dissipative couplings allows one to navigate through the landscape of effective couplings, in a manner reminiscent of reservoir engineering~\cite{Poyatos1996}.

\begin{figure*}[tb]
 \includegraphics[width=\linewidth]{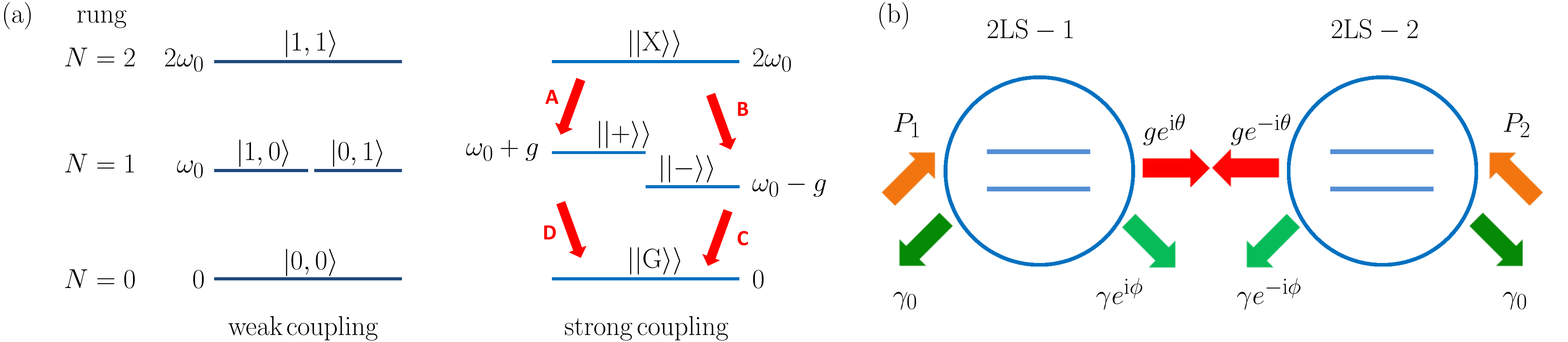}
 \caption{(a) A sketch of the energy ladder of two coupled 2LSs, which is necessarily restricted to three rungs $N=\{0, 1, 2\}$, in the weak (left) and strong (right) coupling regimes [cf. Eq.~\eqref{eq:eigenfrequencies}]. Right: the four red arrows label the two transitions between the $N=2$ and $N=1$ rungs (called A and B), and the two transitions between the $N=1$ and the $N=0$ rungs (called C and D). (b) A cartoon of the system under investigation: a pair of 2LSs (labeled $1$ and $2$) with coherent coupling (red arrows), incoherent pumping (orange arrows), dissipative coupling (light green arrows) and self damping decay (dark green arrows) [cf. Eq.~\eqref{eq:master}].}
 \label{fig:sketch}
\end{figure*}

Our elementary model allows us to calculate analytically the steady state populations of the coupled system, which act as the simplest indicator of chiral coupling and reveals an unconventional population trapping effect (in the limit of strong dissipative coupling). We further calculate the optical spectrum, which showcases different narrow spectral features and significant frequency shifts depending on the coupling regime. We also uncover strong emission correlations via exact expressions for the second-order cross-correlation functions, exposing the quantum nature of the system. Taken together, we provide a systematic analysis of several fundamental quantum optical properties, highlighting how different features are characteristic of each coupling regime.

The rest of this work is organized as follows. We introduce our open quantum system model in Sec.~\ref{model}, and underscore its important limiting case of chiral coupling. We then study the mean populations, power spectrum, and second-order correlation functions of all possible coupling regimes, namely: coherent coupling [Sec.~\ref{SECcoh}], dissipative coupling [Sec.~\ref{SECdiss}], chiral coupling [Sec.~\ref{SECuni}], and asymmetric coupling [Sec.~\ref{SECgen}]. Finally, we draw some conclusions in Sec.~\ref{conc}. We relegate to the Appendices some supporting calculations and technical details.


\section{\label{model}The model}

Our theory is based upon the simple model of two coupled 2LSs originally developed in Ref.~\cite{delValle2010}, and extended to include dissipative coupling and to allow for complex parameters. The model allows us to provide unique physical insight into prototypical chiral and asymmetric quantum systems due to its analyticity. For a comprehensive review of the rich theory of two-atom systems, see Ref.~\cite{Ficek2002}.

In this section, we first introduce the Hamiltonian, and with it the coherent coupling [Sec.~\ref{jambon123}], before unveiling the master equation and the associated dissipative coupling [Sec.~\ref{jambon456}]. We conclude by surveying the coupling landscape of the system [Sec.~\ref{huevos456}], which arises due to the interplay between the coherent and dissipative coupling.


\subsection{\label{jambon123}The Hamiltonian}

We work with the Hamiltonian
\begin{equation}
\label{eq:ham0}
 H = H_0 + H_{\mathrm{c}},
\end{equation}
with the non-interacting part ($\hbar = 1$ throughout the paper)
\begin{equation}
\label{eq:ham1}
 H_0 = \omega_0 \left( \sigma_1^{\dagger} \sigma_1 + \sigma_2^{\dagger} \sigma_2 \right),
\end{equation}
where $\omega_0$ defines the natural resonance frequency of each 2LS. The two 2LSs interact linearly through the dipole-dipole coupling Hamiltonian
\begin{equation}
\label{eq:ham2}
 H_{\mathrm{c}} = g_{12} \sigma_1^{\dagger} \sigma_2 + g_{21} \sigma_2^{\dagger} \sigma_1,
\end{equation}
where the coherent coupling constants, which are in general complex quantities, satisfy $g_{12} = g_{21}^{\ast}$. This property is necessary to ensure Hermiticity, and consequently guarantees reciprocal coupling (which we seek to break later on by introducing dissipation). The lowering (raising) operators of the 2LSs are $\sigma_{i}$ ($\sigma_{i}^{\dagger}$), with $i = \{1, 2\}$, which are subject to the intrinsic condition $\sigma_{i} \sigma_{i} = \sigma_{i}^{\dagger} \sigma_{i}^{\dagger} = 0$. The commutation and anti-commutation relations are
\begin{subequations}
\label{eq:comm}
\begin{alignat}{3}
 [ \sigma_i, \sigma_j^{\dagger} ] = [ \sigma_i, \sigma_j ] &= 0, \quad \text{with}~i \ne j, \\
 \{ \sigma_i, \sigma_i^{\dagger} \} &= 1,
 \end{alignat}
\end{subequations}
which define an algebra of two distinguishable systems. The truncated Hilbert space is four dimensional, encompassing: the ground state $\ket{0, 0}$ with zero excitations; the excited state of each 2LS $\ket{0, 1}$ and $\ket{1, 0}$, which each host a single excitation; and the doubly-excited state $\ket{1, 1}$. Explicitly, it follows from these eigenstates that the lowering operators may be written as
\begin{subequations}
\label{eq:ops}
\begin{alignat}{3}
 \sigma_1 &= \ket{0, 1} \bra{1, 1} + \ket{0, 0} \bra{1, 0}, \\
 \sigma_2 &= \ket{1, 0} \bra{1, 1} + \ket{0, 0} \bra{0, 1},
 \end{alignat}
\end{subequations} 
and the set $\{ \ket{i, j} \}$ is complete, so that $\sum_{i, j} \ket{i, j} \bra{i, j} = 1$.

\begin{figure}[tb]
 \includegraphics[width=\linewidth]{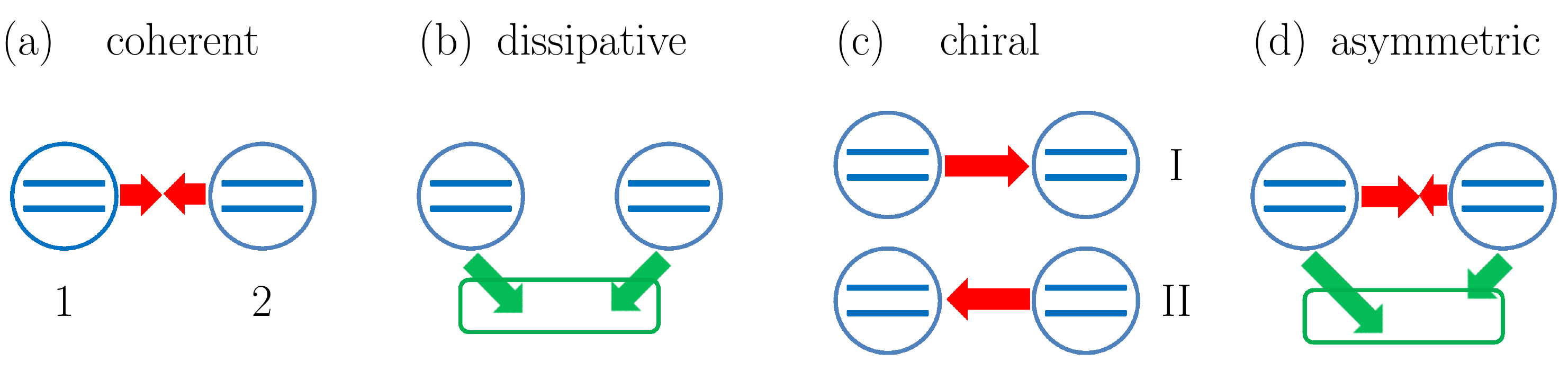}
 \caption{Sketches of the system of two coupled two-level systems, denoted $1$ and $2$, in different coupling regimes [cf. Eq.~\eqref{eq:limitingcases}]. We denote coherent (dissipative) coupling by red (green) arrows, and the common bath by a green rectangle. We consider (a) coherent [cf. Eq.~\eqref{eq:limitingcases1}], (b) dissipative [cf. Eq.~\eqref{eq:limitingcases2}], (c) chiral $\mathrm{I/II}$ (upper/lower) [cf. Eq.~\eqref{eq:limitingcases3} and Eq.~\eqref{eq:limitingcases4}] and (d) asymmetric coupling.}
 \label{fig:landscape}
\end{figure}

The Hamiltonian of Eq.~\eqref{eq:ham0} is straightforwardly diagonalized by Bogoliubov transformation as
\begin{align}
\label{eq:diagham}
 H =&\; \omega_{\mathrm{G}} \kettwo{\mathrm{G}} \bratwo{\mathrm{G}} + \omega_{-} \kettwo{-} \bratwo{-} \nonumber \\
  &+ \omega_{+} \kettwo{+} \bratwo{+} + \omega_{\mathrm{X}} \kettwo{\mathrm{X}} \bratwo{\mathrm{X}},
\end{align}
where we have used double (single) brackets for the eigenstates in the coupled (uncoupled) regime. The ground $\kettwo{\mathrm{G}}$ and doubly-excited $\kettwo{\mathrm{X}}$ eigenstates are given by
\begin{equation}
\label{eq:eigenstates}
 \kettwo{\mathrm{G}} = \ket{0, 0}, \quad \kettwo{\mathrm{X}} = \ket{1, 1},
\end{equation}
and the upper and lower dressed eigenstates are
\begin{equation}
\label{eq:eigenstates2}
 \kettwo{\pm} = \frac{1}{\sqrt{2}} \left( \ket{0, 1} \pm \mathrm{e}^{ \mathrm{i} \theta} \ket{1, 0} \right).
\end{equation}
Here we have introduced the following polar decompositions for the coherent coupling constants [appearing in Eq.~\eqref{eq:ham2}]
\begin{equation}
\label{eq:polardecom}
 g_{12} = g \mathrm{e}^{\mathrm{i} \theta}, \quad g_{21} = g \mathrm{e}^{-\mathrm{i} \theta},
\end{equation}
where $g \ge 0$, thus explicitly accounting for a phase $\theta$ in the coherent interaction. The eigenfrequencies associated with the eigenstates of Eqs.~\eqref{eq:eigenstates}~and~\eqref{eq:eigenstates2} are
\begin{equation}
\label{eq:eigenfrequencies}
 \omega_{\mathrm{G}} = 0, \quad \omega_{\pm} = \omega_0 \pm g, \quad  \omega_{\mathrm{X}} = 2 \omega_0,
\end{equation}
revealing that the dressed levels $\omega_{\pm}$ are separated by the Rabi splitting $2g$, while the ground and doubly-excited levels are unshifted (and $g$ independent). The energy ladder is sketched in the weak and strong coupling regimes in Fig.~\ref{fig:sketch}~(a). Notably, while the dressed state eigenfrequencies $\omega_{\pm}$ are insensitive to the phase $\theta$, the eigenstates of Eq.~\eqref{eq:eigenstates2} (and thus any quantity dependent on them) are influenced by this complex argument. For example, in the simplest case of $\theta = 0~(\pi)$ the interaction of Eq.~\eqref{eq:polardecom} is repulsive~(attractive), giving rise to markedly different behaviors of the dipole moments: the higher frequency state $\omega_{+}$ is associated with in-phase (out-of-phase) dipole moments, and the lower frequency state $\omega_{-}$ corresponds to out-of-phase (in-phase) dipole moments~\cite{Downing2017}.

The effect of dissipation on the system, which as well as introducing finite excitation lifetimes leads to a renormalization of the levels of Eq.~\eqref{eq:eigenfrequencies} [as follows from the fluctuation-dissipation theorem], is discussed next.


\subsection{\label{jambon456}The master equation}

We assume that the couplings of the system to its environment are weak, so that the master equation of the system's density matrix $\rho$ is in the standard Lindblad form~\cite{Gardiner2004, Lembessis2013, Gardiner2014}
\begin{equation}
\label{eq:master}
 \partial_t \rho = \mathrm{i} [ \rho, H ] +  \sum_{i, j = 1, 2} \frac{\gamma_{i j}}{2} \mathscr{L}_{ij} \rho +  \sum_{i = 1, 2} \frac{P_{i}}{2} \left( \mathscr{L}_{ii} \rho \right)^\dagger,
\end{equation}
with the Liouvillian superoperator
\begin{equation}
\label{eq:master2}
\mathscr{L}_{ij} \rho = 2 \sigma_j \rho \sigma_i^{\dagger} -  \sigma_i^{\dagger} \sigma_j \rho - \rho \sigma_i^{\dagger} \sigma_j.
\end{equation}
In Eq.~\eqref{eq:master}, the Hamiltonian $H$ is given by Eq.~\eqref{eq:ham0}, $\gamma_{i j}$ are the self ($i=j$) and collective ($i \ne j$) damping decay rates, and the incoherent pumping rate $P_{i}$ populates 2LS-$i$. Each emitter has its own independent bath, which can correspond to radiation into free space for example, and they also share a common bath [as denoted by the green rectangle in Fig.~\ref{fig:landscape} panels (b) and (d)], which allows for the dissipative coupling.

Assuming the (real-valued) self-damping decay rates to be identical, and utilizing polar decompositions for the (in general, complex-valued) dissipative coupling constants, we write the four damping constants appearing in the second term on the right-hand-side of Eq.~\eqref{eq:master} as
\begin{equation}
\label{eq:damp}
 \gamma_{11} =  \gamma_{22} = \gamma_{0}, \quad \gamma_{12} = \gamma  \mathrm{e}^{\mathrm{i} \phi}, \quad \gamma_{21} = \gamma  \mathrm{e}^{-\mathrm{i} \phi},
\end{equation}
where $\gamma \ge 0$, and we note that in order to have physical dynamics $\gamma \le \gamma_0$~\cite{Gardiner2004}. We thus explicitly account for a phase $\phi$ in the dissipative coupling in the same manner as for the coherent coupling [cf. Eq.~\eqref{eq:polardecom}]. This completes the setup of our combined pumped-dissipative system,  which is sketched in Fig.~\ref{fig:sketch}~(b).

We would like to mention that the form of the parameters of Eqs.~\eqref{eq:polardecom}~and~\eqref{eq:damp} arises naturally in the electromagnetic setup proposed in Ref.~\cite{Downing2019}, whereby two circularly polarized quantum emitters are held above a metal surface hosting plasmons. In the framework of macroscopic quantum electrodynamics, the coherent and dissipative coupling constants $\{ g_{ij}, \gamma_{ij} \}$ describing this system can be directly related to the electromagnetic dyadic Green's function of the system~\cite{Dung2002}. The circular polarization of the quantum emitters gives rise to the complex phase degrees of freedom $\{ \theta, \phi \}$, otherwise the parameters of Eq.~\eqref{eq:damp} are wholly real quantities~\cite{Gonzalez2011, MartinCano2011}. Changing the relative positions of the emitters, or their height above the metallic surface, allows one to tune the emitter-emitter interactions. In doing so, one can traverse the coupling landscape defined by the four fundamental quantities of our model $\{ g, \gamma, \theta, \phi \}$.


\subsection{\label{huevos456}The coupling landscape}

\begin{figure*}[tb]
 \includegraphics[width=\linewidth]{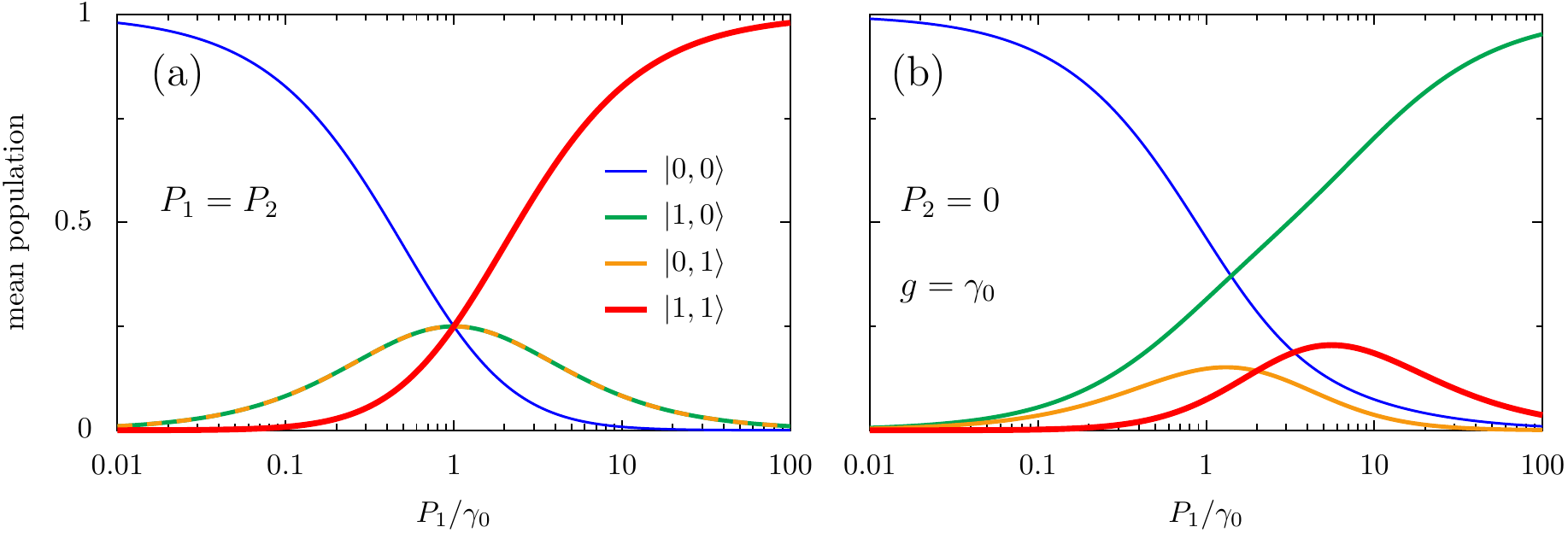}
 \caption{ Mean populations in the coherent coupling regime, as a function of the pumping rate $P_1$, in units of the damping rate $\gamma_0$ [cf. Eq.~\eqref{eq:statespopco}]. Panel (a): the case of symmetric pumping ($P_2 = P_1$), which is independent of the coherent coupling strength $g$. Panel (b): an asymmetric pumping case ($P_2 = 0$), with $g = \gamma_0$. The labeling of the mean population of the state $\ket{i, j}$ is displayed in the legend of panel (a), and states with $N=\{ 0, 1, 2\}$ excitations are shown with increasingly thick lines.}
 \label{fig:popco}
\end{figure*}

In the rest of this work, we will look in detail at the quantum optical properties of the four limiting cases of our model, which reveals the coupling landscape
\begin{subequations}
\label{eq:limitingcases}
\begin{alignat}{3}
g &\ne 0, \quad \gamma = 0, \quad \text{coherent coupling} \label{eq:limitingcases1} \\
 g &= 0, \quad \gamma \ne 0, \quad \text{dissipative coupling} \label{eq:limitingcases2} \\
 \tfrac{g}{\gamma} &= \tfrac{1}{2}, \quad \theta-\phi = \tfrac{\pi}{2}, \quad \text{chiral I coupling} \label{eq:limitingcases3} \\
 \tfrac{g}{\gamma} &= \tfrac{1}{2}, \quad \theta-\phi = \tfrac{3\pi}{2}, \quad \text{chiral II coupling} \label{eq:limitingcases4}
 \end{alignat}
\end{subequations} 
as depicted in Fig.~\ref{fig:landscape}~(a, b, c). We also analyze the asymmetric coupling case shown in panel (d), which has no restrictions on the parameters of the model. Notably, the asymmetric regime was recently shown to include an interesting ``quasichiral'' regime, where the magnitude condition $g/\gamma = 1/2$ of Eqs.~\eqref{eq:limitingcases3}~and~\eqref{eq:limitingcases4} is met but the phase condition on $\theta-\phi$ is not~\cite{Downing2019}. We derive the twin chiral coupling conditions of Eqs.~\eqref{eq:limitingcases3}~and~\eqref{eq:limitingcases4} in Appendix~\ref{nonrec} (see also Refs.~\cite{Metelmann2015, Downing2019} for more details).


\section{\label{SECcoh}Coherent coupling}

In this section, we focus on the simplest nontrivial case, that of purely coherent ($\mathrm{co}$) coupling between the two quantum emitters [as sketched in Fig.~\ref{fig:landscape}~(a)]. We shall consider how the mean populations [Sec.~\ref{SECcohPOP}], correlation functions [Sec.~\ref{SECcohCORR}], and optical spectrum [Sec.~\ref{SECcohSPEC}] behave due to the interplay of the coherent coupling $g$ and the two pumping rates $P_1$ and $P_2$.


\subsection{\label{SECcohPOP}Mean populations}

When considering the mean populations, we make special reference to the steady state ($\mathrm{ss}$) population of a single 2LS in isolation ($\mathrm{iso}$),
\begin{equation}
\label{eq:iso}
n_{\mathrm{iso}} = \langle \sigma^{\dagger} \sigma \rangle_{\mathrm{ss}} =  \frac{P_1}{P_1 + \gamma_0},
\end{equation}
where $P_1$ and $\gamma_0$ are the incoherent pump rate and self-decay decay rate, respectively (see Appendix~\ref{app:asingle2ls} for the background theory). Throughout this work, we shall be interested in how the coupling regime of the pair of coupled 2LSs changes the baseline result of Eq.~\eqref{eq:iso}, which describes a solitary 2LS with lowering (raising) operator $\sigma$ ($\sigma^\dagger$).

In the steady state~\cite{TroianiLaussy2007}, the mean population of the the state $\ket{i, j}$ in the coherent coupling regime is $\rho_{ij}^{\mathrm{co}} = \langle ij | \rho^{\mathrm{co}} | ij \rangle$, which is obtained from the master equation of Eq.~\eqref{eq:master}. The resulting expressions are (see Appendix~\ref{app:Single} for details) 
\begin{subequations}
\label{eq:statespopco}
\begin{align}
 \rho_{0, 0}^{\mathrm{co}} &= \frac{ \gamma_0^2 \left( 1 + \frac{g^2}{\Gamma_+^2} \right) }{\Gamma_1 \Gamma_2 +  4 g^2 },  \\
 \rho_{1, 0}^{\mathrm{co}} &= \frac{ \gamma_0 P_1 + \frac{\gamma_0 P_+ g^2}{2\Gamma_+^2} }{\Gamma_1 \Gamma_2 +  4 g^2 },  \\
 \rho_{0, 1}^{\mathrm{co}} &= \frac{ \gamma_0 P_2  + \frac{\gamma_0 P_+ g^2}{2\Gamma_+^2} }{\Gamma_1 \Gamma_2 +  4 g^2 }, \\
 \rho_{1, 1}^{\mathrm{co}} &= \frac{ P_1 P_2 + \left( \frac{ g P_+ }{ 2\Gamma_+ } \right)^2 }{\Gamma_1 \Gamma_2 +  4 g^2 }. 
 \end{align}
\end{subequations}
Here and in what follows, we make use of the following effective pumping and damping rates
 \begin{subequations}
 \label{eq:m0matrix233}
  \begin{align}
 P_{\pm} &= P_1 \pm P_2, \\
\Gamma_{1, 2} &= \gamma_0 + P_{1, 2}, \\
\Gamma_{\pm} &= \tfrac{1}{4} \left( \Gamma_1 \pm \Gamma_2 \right).
 \end{align}
 \end{subequations}

In the symmetric pump case ($P_1 = P_2$), the populations of Eq.~\eqref{eq:statespopco} take on particularly simple forms, being universal functions of the ratio $P_1/\gamma_0$, and thus independent of the coherent coupling strength $g$. In terms of the isolated 2LS result of Eq.~\eqref{eq:iso}, one finds: the ground state population $\rho_{0, 0}^{\mathrm{co}} = (1- n_{\mathrm{iso}})^2$, the singly-excited populations $\rho_{1, 0}^{\mathrm{co}} = \rho_{0, 1}^{\mathrm{co}} = n_{\mathrm{iso}} (1-n_{\mathrm{iso}})$ and the doubly-excited population $\rho_{1, 1}^{\mathrm{co}} = n_{\mathrm{iso}}^2$. This elementary case is displayed in Fig.~\ref{fig:popco}~(a), where one notices the symmetric (about $P_1 = \gamma_0$) evolution of the populations of all of the states as a function of the pumping rate $P_1$, starting from a wholly occupied ground state $\ket{0, 0}$ (thin blue line) and ending with a wholly occupied doubly-excited state $\ket{1, 1}$ (thick red line), and with a balanced population across all four states at $P_1 = \gamma_0$. These properties for this high symmetry case suggest the found $g$-independence, which is guaranteed by the vanishing coherence between the two 2LSs, $\langle \sigma_1^{\dagger} \sigma_2 \rangle_{\mathrm{ss}} = 0$ (as derived in Appendix~\ref{app:Single}).

For the case of asymmetric pumping ($P_1 \ne P_2$) a $g$-dependence arises in the mean populations of Eq.~\eqref{eq:statespopco}, and there is an asymmetry in the populations of the singly-excited states $\ket{1, 0}$ and $\ket{0, 1}$ (medium green and orange lines respectively), as shown in Fig.~\ref{fig:popco}~(b). In this panel, where $P_2 = 0$ and $g = \gamma_0$, the limiting case with large pumping $P_1 \gg \gamma_0$ in the system is a wholly occupied singly-excited state $\ket{1, 0}$ (medium green line), that which is being incoherently pumped. Clearly, the population imbalance between the two singly-excited states increases with increasingly pumping rate $P_1$, in stark contrast to panel (a). The population imbalance induced by different incoherent pumping rates is crucial in order to obtain nontrivial correlations, as we now discuss.

\begin{figure}[tb]
 \includegraphics[width=\linewidth]{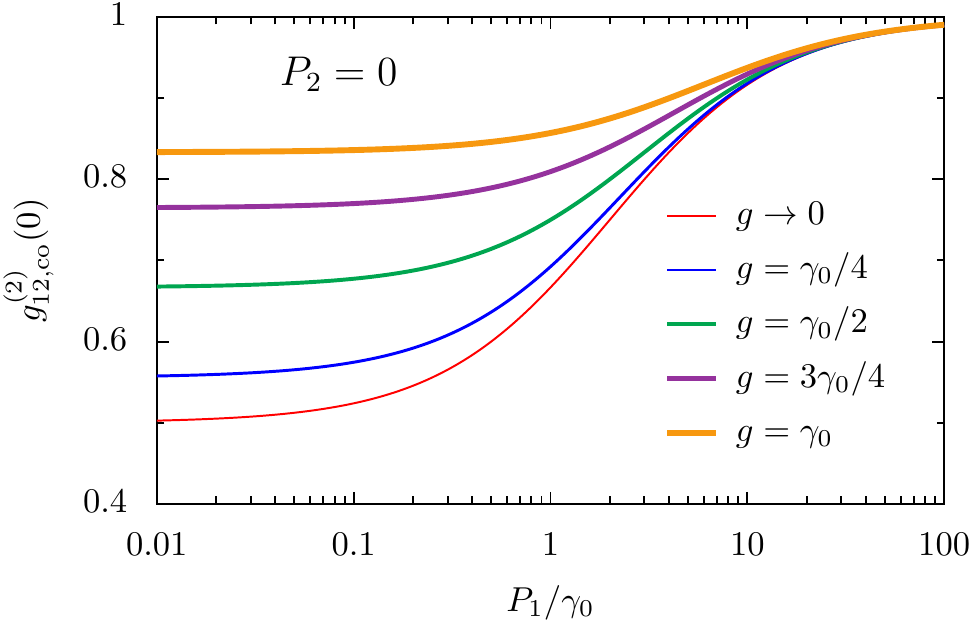}
 \caption{ Second-order cross-correlator in the coherent coupling regime at zero delay $g_{12, \mathrm{co}}^{(2)}(0)$, as a function of the pumping rate $P_1$, in units of the decay rate $\gamma_0$ [cf. Eq.~\eqref{eq:geetwo}]. We show results with asymmetric pumping ($P_2 = 0$), for increasingly strong coherent coupling strengths $g$ (increasingly thick colored lines).  }
 \label{fig:corrco}
\end{figure}


\subsection{\label{SECcohCORR}Correlations}

In order to quantify the correlations in the coupled system, we discuss the normalized second-order cross-correlation function in the steady state ($\mathrm{ss}$), defined by~\cite{Gardiner2014}
\begin{equation}
\label{eq:geetwo_def}
g_{12}^{(2)} (0) = \frac{\langle \sigma_1^{\dagger} \sigma_1 \sigma_2^{\dagger} \sigma_2 \rangle_{\mathrm{ss}}}{  \langle \sigma_1^{\dagger} \sigma_1 \rangle_{\mathrm{ss}}  \langle \sigma_2^{\dagger} \sigma_2 \rangle_{\mathrm{ss}} }.
\end{equation}
This cross-correlator quantifies the probability of simultaneous emissions in the two different systems, 2LS-1 and 2LS-2. When $g_{12}^{(2)} (0) = 1$ the 2LSs emit independently, while $g_{12}^{(2)} (0) = 0$ suggests it is impossible to have two simultaneous emissions in the coupled system. Aside from these extremes, $g_{12, \mathrm{co}}^{(2)} (0) < 1$ describes emission antibunching, reflecting the quantum nature of the system, while $g_{12, \mathrm{co}}^{(2)} (0) > 1$ implies emission bunching. In the coherent coupling regime, we find the cross-correlator (see Appendix~\ref{app:Single} for details) 
\begin{equation}
\label{eq:geetwo}
g_{12, \mathrm{co}}^{(2)} (0) = \frac{ \left( 4 g^2 + \Gamma_1 \Gamma_2 \right) \left( g^2 P_+^2 + 4 P_1 P_2 \Gamma_+^2 \right) }{ 4 \left( g^2 P_+ + P_2 \Gamma_1 \Gamma_+ \right) \left( g^2 P_+ + P_1 \Gamma_2 \Gamma_+ \right) },
\end{equation}
where $P_+, \Gamma_{1, 2}$ and $\Gamma_{+}$ are defined in Eq.~\eqref{eq:m0matrix233}.

With symmetric pumping ($P_2 = P_1$), Eq.~\eqref{eq:geetwo} collapses into $g_{12, \mathrm{co}}^{(2)} (0) = 1$, describing an effectively independent system, due to the vanishing coherence between the two 2LSs, $\langle \sigma_1^{\dagger} \sigma_2 \rangle_{\mathrm{ss}} = 0$ [as is consistent with the highly symmetric mean populations of Fig.~\ref{fig:popco}~(a)].

When the incoherent pumping is asymmetric ($P_2 \ne P_1$) much richer correlations arise due to the inherent population imbalances, as shown in Fig.~\ref{fig:corrco}. In the figure $P_2 = 0$, and $g_{12, \mathrm{co}}^{(2)} (0)$ is shown as a function of $P_1$, where increasingly strong coherent coupling strengths $g$ are denoted by increasingly thick colored lines. Most notably, the quantum nature of the setup leads to the displayed antibunching $g_{12, \mathrm{co}}^{(2)} (0) < 1$, which is increasingly significant for small pumping rates $P_1 \ll \gamma_0$. The minimum value of $g_{12, \mathrm{co}}^{(2)} (0) = 1/2$ is obtained for vanishingly small coherent coupling $g \to 0$, since it corresponds to the case of maximal population imbalance (here the order of limits is important, $g \to 0$ is the final limit to be taken). Larger pumping rates $P_1 \gg \gamma_0$ wash out any correlations, since the system simplifies into supporting the singly-excited state $\ket{1, 0}$ only, with the other states are unoccupied, as follows from the mean populations of Fig.~\ref{fig:popco}~(b).


\subsection{\label{SECcohSPEC}Spectrum}

A fundamental quantity enabling one to characterize the coupling regime is the spectrum of the system. The normalized optical spectrum of 2LS-1 reads~\cite{delValle2010}
\begin{equation}
\label{eq:specdddddddddddddtrum}
S_1 (\omega) = \frac{\langle \sigma_1^{\dagger}(\omega) \sigma_1(\omega) \rangle}{ \langle \sigma_1^{\dagger} \sigma_1 \rangle_{\mathrm{ss}} },
\end{equation}
and is formally derived in Refs.~\cite{ValleLaussy2009, ValleBook2010, ValleLaussy2011, Kavokin2007, Gardiner2014}. The optical spectrum of Eq.~\eqref{eq:specdddddddddddddtrum} may be written as
\begin{equation}
\label{eq:spectrum}
S_1 (\omega) = \sum_{p=\mathrm{A}, \mathrm{B}, \mathrm{C}, \mathrm{D}} S_{1}^{p} (\omega), 
\end{equation}
which has been decomposed into four lineshapes, labelled by the index $p$, given by
\begin{equation}
\label{eq:spectrumao}
S_{1}^{p} (\omega) = \frac{1}{\pi} \frac{\tfrac{\gamma_p}{2}~L_p - \left( \omega - \omega_p \right)~K_p}{ \left( \tfrac{\gamma_p}{2} \right)^2 + \left( \omega - \omega_p \right)^2  },
\end{equation}
due to the four possible transitions in the system, as follows from the four dimensional Hilbert space of Eq.~\eqref{eq:diagham}. These transitions $\{ \mathrm{A}, \mathrm{B}, \mathrm{C}, \mathrm{D} \}$ are denoted by red arrows in Fig.~\ref{fig:sketch}~(a, right), which is sketched for the coherently coupled regime~\cite{delValle2010, ValleBook2010, Downing2019}. In Eq.~\eqref{eq:spectrumao}, the Lorentzian ($L_p$) and dispersive ($K_p$) weighting coefficients are real numbers, while $\omega_p$ and $\gamma_p$ define the effective frequency shifts and broadenings of the system (for further details see Appendix~\ref{app:Double}).

The spectrum $S_2 (\omega)$ of the second 2LS, as well as generalizations of Eq.~\eqref{eq:specdddddddddddddtrum} for operators such as $\left( \sigma_1 + \sigma_2 \right) / \sqrt{2}$,  may be calculated in the same manner~\cite{Downing2019}, but are not presented here due to their bulky nature and our focus in this work of the fundamental analytical theory.

In what follows, we consider Eq.~\eqref{eq:spectrum} in the vanishing pump limit ($P_{1}, P_{2} \ll \gamma_0$), and we make special reference to the spectrum of a single 2LS in isolation (see Appendix~\ref{app:asingle2ls} for the derivation)
\begin{equation}
\label{eq:speciso}
  S_{\mathrm{iso}} (\omega) = \frac{1}{\pi} \frac{ \gamma_0/2 }{ \left( \gamma_0 / 2 \right)^2 + ( \omega - \omega_0 )^2}.
\end{equation}
This Lorentzian expression for the optical spectrum of course displays no frequency shifts from $\omega_0$, or renormalization effects from the bare broadening $\gamma_0$. Deviations from Eq.~\eqref{eq:speciso} as one traverses the coupling landscape allows for the characterization of various coupling regimes of interest.

\begin{figure}[tb]
 \includegraphics[width=\linewidth]{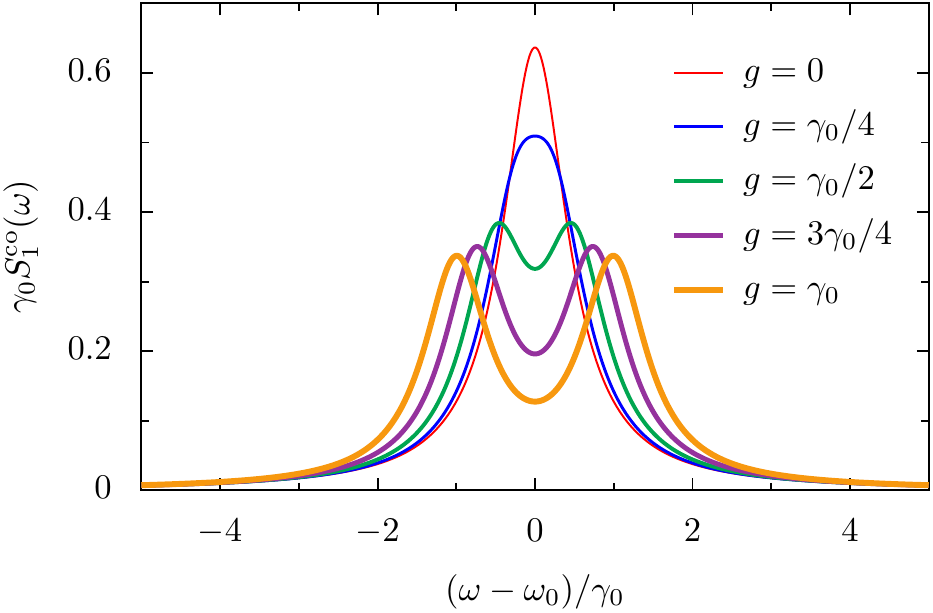}
 \caption{ Spectrum of 2LS-1 in the coherent coupling regime $S_1^{\mathrm{co}}(\omega)$, in units of the inverse damping rate $\gamma_0^{-1}$, for increasingly strong coherent coupling strengths $g$ (increasingly thick colored lines) [cf. Eq.~\eqref{eq:spectrumererh}].  }
 \label{fig:specco}
\end{figure}

For the coherent coupling parameters of Eq.~\eqref{eq:limitingcases1}, we obtain the following simple expressions for the frequencies $\omega_p$ and damping rates $\gamma_p$ appearing in Eq.~\eqref{eq:spectrumao} [the calculation is performed in Appendix~\ref{app:Double}]
\begin{subequations}
\label{eq:nreigen2sdwewewewewwwwsd222}
\begin{alignat}{2}
 \tfrac{1}{2} \gamma_{\mathrm{A}}^{\mathrm{co}} + \mathrm{i} \omega_{\mathrm{A}}^{\mathrm{co}} &= \tfrac{3}{2} \gamma_{0} + \mathrm{i} \left( \omega_0 + g \right),  \\
  \tfrac{1}{2} \gamma_{\mathrm{B}}^{\mathrm{co}} + \mathrm{i} \omega_{\mathrm{B}}^{\mathrm{co}} &= \tfrac{3}{2} \gamma_{0} + \mathrm{i} \left( \omega_0 - g \right),  \\
   \tfrac{1}{2} \gamma_{\mathrm{C}}^{\mathrm{co}} + \mathrm{i} \omega_{\mathrm{C}}^{\mathrm{co}} &= \tfrac{1}{2} \gamma_{0} + \mathrm{i} \left( \omega_0 + g \right),  \\
    \tfrac{1}{2} \gamma_{\mathrm{D}}^{\mathrm{co}} + \mathrm{i} \omega_{\mathrm{D}}^{\mathrm{co}} &=  \tfrac{1}{2} \gamma_{0} + \mathrm{i} \left( \omega_0 - g \right).
  \end{alignat}
\end{subequations}
Equation~\eqref{eq:nreigen2sdwewewewewwwwsd222}, which are essentially the eigenvalues of the Liouvillian, describes a pair of peaks with broadenings $\gamma_{\mathrm{A}}^{\mathrm{co}} = \gamma_{\mathrm{B}}^{\mathrm{co}} = 3 \gamma_0$ and a pair of peaks with broadenings $\gamma_{\mathrm{C}}^{\mathrm{co}} = \gamma_{\mathrm{D}}^{\mathrm{co}} = \gamma_0$. Each pair of peaks are split by the Rabi frequency $2 g$, as is consistent with the Hamiltonian dynamics of Eq.~\eqref{eq:eigenfrequencies}. We also obtain the equal Lorentzian weighting coefficients (corresponding to transitions to the ground state)
\begin{subequations}
\label{eq:elle33}
\begin{alignat}{2}
   L_{\mathrm{C}}^{\mathrm{co}}  &= \tfrac{1}{2},  \\
    L_{\mathrm{D}}^{\mathrm{co}} &=  \tfrac{1}{2},
  \end{alignat}
\end{subequations}
since we are in a highly symmetrical, reciprocal case. The coefficients associated with the labels $\mathrm{A}$ and $\mathrm{B}$ are zero. This is because they correspond to the two transitions from the upper energy level $\ket{1, 1}$, which is unpopulated in the vanishing pump limit in which we work, to the hybridized states involving $\ket{0, 1}$ and $\ket{1, 0}$. This fact is rigorously proved in Ref.~\cite{delValle2010}, and is implied by the red arrows indicating the transitions in system in Fig.~\ref{fig:sketch}~(a, right). The coefficients of Eq.~\eqref{eq:elle33} also reveal that the spectrum is purely Lorentzian with no dispersive components ($K_{p}^{\mathrm{co}}=0$).

Substitution of Eqs.~\eqref{eq:nreigen2sdwewewewewwwwsd222}~and~\eqref{eq:elle33} into Eq.~\eqref{eq:spectrum} yields the optical spectrum in the coherent coupling regime
\begin{equation}
\label{eq:spectrumererh}
 S_1^{\mathrm{co}} (\omega) = \frac{1}{2 \pi} \sum_{\tau = \pm 1} \frac{\gamma_0 /2}{ \left( \tfrac{\gamma_0}{2} \right)^2 + \left( \omega - \omega_0 + \tau g \right)^2  }.
\end{equation}
Of course, Eq.~\eqref{eq:spectrumererh} recovers the uncoupled result of Eq.~\eqref{eq:speciso} for vanishing coherent coupling ($g \to 0$).

We plot the spectrum of Eq.~\eqref{eq:spectrumererh} in Fig.~\ref{fig:specco} for increasingly strong coherent coupling strengths $g$ (increasingly thick colored lines). One notices that the twin Lorentzian contributions give rise to a Rabi doublet shape distinctive of strong coupling, which only becomes hidden when the peaks (symmetric about $\omega_0$) merge for smaller ratios of $g/\gamma_0 < 1/2$. The effects of non-negligible pumping rates on the spectrum, which may be either symmetric or asymmetric, are described in detail in Ref.~\cite{delValle2010}.


\section{\label{SECdiss}Dissipative coupling}

\begin{figure*}[tb]
 \includegraphics[width=\linewidth]{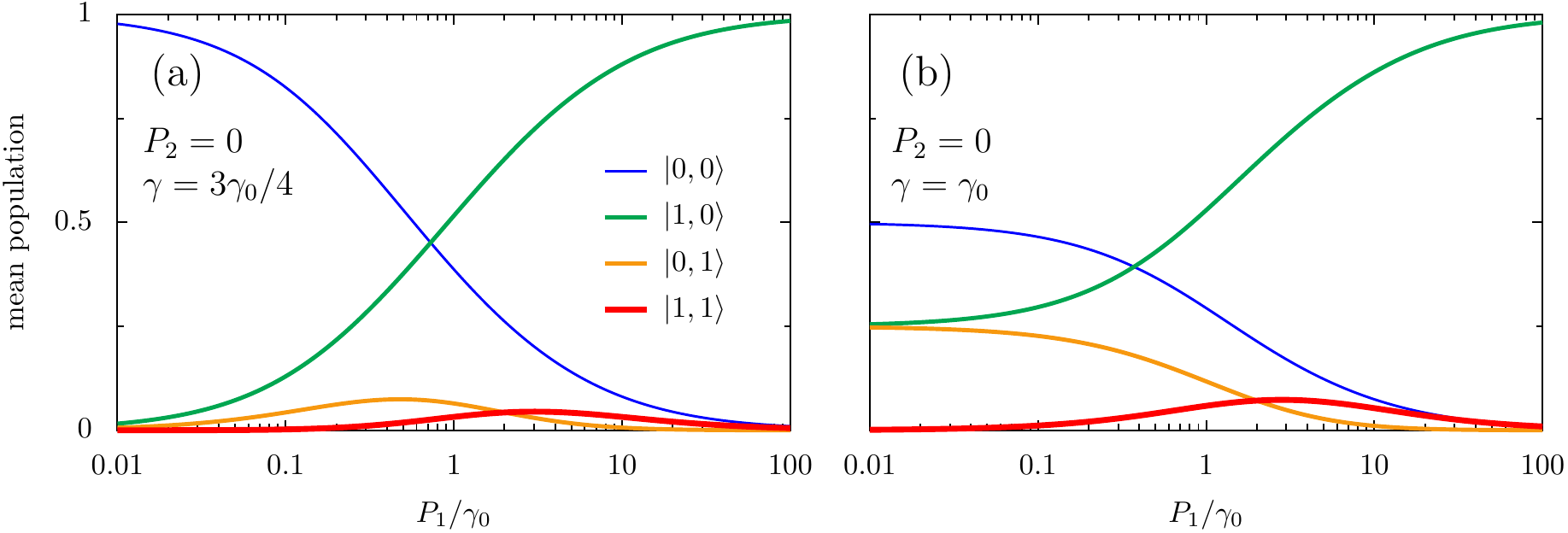}
 \caption{ Mean populations in the dissipative coupling regime, as a function of the pumping rate $P_1$, in units of the damping rate $\gamma_0$ [cf. Eq.~\eqref{eq:statespopdiss}]. Panel (a): the dissipative coupling strength $\gamma = 3\gamma_0/4$. Panel (b): maximal coupling, $\gamma = \gamma_0$. The labeling of the mean population of the state $\ket{i, j}$ is displayed in the legend of panel (a), and states with $N=\{ 0, 1, 2\}$ excitations are shown with increasingly thick lines. In the figure, we consider asymmetric pumping, with $P_2 = 0$.}
 \label{fig:popdiss}
\end{figure*}

In this section, we contemplate the simplest case with nontrivial dissipation, namely where there is dissipative ($\mathrm{ds}$) coupling between the two quantum emitters but no coherent coupling [as drawn in Fig.~\ref{fig:landscape}~(b)]. We will investigate how the populations [Sec.~\ref{SECdissPOP}], correlation functions [Sec.~\ref{SECdissCORR}] and optical spectrum [Sec.~\ref{SECdissSPEC}] change due to the competition between the dissipative coupling strength $\gamma$ and the pumping rates $P_1$ and $P_2$.


\subsection{\label{SECdissPOP}Mean populations}

The mean population $\rho_{ij}^{\mathrm{ds}}$ of the state $\ket{i, j}$ may be found from the master equation of Eq.~\eqref{eq:master} (see Appendix~\ref{app:Single}). The results are given by 
\begin{subequations}
\label{eq:statespopdiss}
\begin{align}
 \rho_{00}^{\mathrm{ds}} &= \frac{ 4 \Gamma_+^2 \left( \gamma_0^2 - \gamma^2 \right) + \gamma^2 \left( \gamma_0 P_+ + \frac{P_1^2 + 6 P_1 P_2 + P_2^2}{4} \right) }{\gamma^2 \left( P_1 \Gamma_1 + P_2 \Gamma_2 \right) + 4 \Gamma_+^2 \left( \Gamma_1 \Gamma_2 - \gamma^2 \right)  },  \\
 \rho_{10}^{\mathrm{ds}} &= \frac{ P_1 \left( 4 \gamma_0 \Gamma_+^2 + P_- \gamma^2 \right) - P_- \gamma^2 \left( \Gamma_+ + \frac{P_-}{4} \right) }{\gamma^2 \left( P_1 \Gamma_1 + P_2 \Gamma_2 \right) + 4 \Gamma_+^2 \left( \Gamma_1 \Gamma_2 - \gamma^2 \right)  },  \\
 \rho_{01}^{\mathrm{ds}} &= \frac{ P_2 \left( 4 \gamma_0 \Gamma_+^2 - P_- \gamma^2 \right) + P_- \gamma^2 \left( \Gamma_+ - \frac{P_-}{4} \right) }{\gamma^2 \left( P_1 \Gamma_1 + P_2 \Gamma_2 \right) + 4 \Gamma_+^2 \left( \Gamma_1 \Gamma_2 - \gamma^2 \right)  }, \\
 \rho_{11}^{\mathrm{ds}} &= \frac{ 4 P_1 P_2 \Gamma_+^2 + \left( \frac{\gamma P_-}{2} \right)^2 }{\gamma^2 \left( P_1 \Gamma_1 + P_2 \Gamma_2 \right) + 4 \Gamma_+^2 \left( \Gamma_1 \Gamma_2 - \gamma^2 \right)  }, 
 \end{align}
\end{subequations}
where the effective rates $P_{\pm}, \Gamma_{1}, \Gamma_{2}$ and $\Gamma_+$ are defined in Eq.~\eqref{eq:m0matrix233}.

We plot the mean populations of Eq.~\eqref{eq:statespopdiss} in Fig.~\ref{fig:popdiss} as a function of the incoherent pumping $P_1$ into 2LS-1, for asymmetric pumping ($P_2 = 0$). In panel (a), where the dissipative coupling strength $\gamma = 3\gamma_0/4$, the population evolutions are reminiscent of the asymmetric coherent coupling case of Fig.~\ref{fig:popco}~(b). The principle difference is the lower populations of the states $\ket{0, 1}$ and $\ket{1, 1}$ (medium orange line and thick red line respectively) at intermediate pumping rates $P_1 \simeq \gamma_0$. However in Fig.~\ref{fig:popdiss}~(b), with maximal dissipative coupling $\gamma = \gamma_0$, there is a striking population trapping effect in the limit of weak pumping $P_1 \ll \gamma_0$, which has no analogue in the coherent coupling regime. Remarkably, here the mean population of the ground state $\rho_{00}^{\mathrm{ds}} \simeq 1/2$ when $P_1 \ll \gamma_0$ (thin blue line), due to the nonzero populations of the two singly-excited states with $\rho_{10}^{\mathrm{ds}} = \rho_{01}^{\mathrm{ds}} \simeq 1/4$ (medium green and orange lines). This trapping phenomena has arisen due to the quenching, with large dissipative coupling, of transitions from the intermediate states on the $N=1$ rung of the energy ladder [cf. Fig.~\ref{fig:sketch}~(a)] to the ground state on the $N=0$ rung (the weights of such processes are proportional to $\gamma_0 - \gamma$).

Population trapping has been noticed before in other contexts, such as in driven three-level systems~\cite{Radmore1982, Dalton1982, Swain1982}, and for two atoms with different resonance frequencies~\cite{Akram2000}. In our driven-dissipative setup, its effects have already been shown to be important for quantum entanglement~\cite{Gonzalez2011, MartinCano2011}. We shall see shortly in Sec.~\ref{SECdissSPEC} that this effect also impacts greatly on the optical spectrum of the system.

Finally, we note that the symmetric pumping configuration ($P_2 = P_1$) in the dissipative coupling regime only presents obvious changes to the results presented in Fig.~\ref{fig:popdiss}. Namely, the results of Fig.~\ref{fig:popdiss} become symmetrized in the same manner as in going from panel (a) to panel (b) in Fig.~\ref{fig:popco}. Therefore, we relegate this supplementary plot to App.~\ref{app:additional_plot}.


\subsection{\label{SECdissCORR}Correlations}

\begin{figure*}[tb]
 \includegraphics[width=\linewidth]{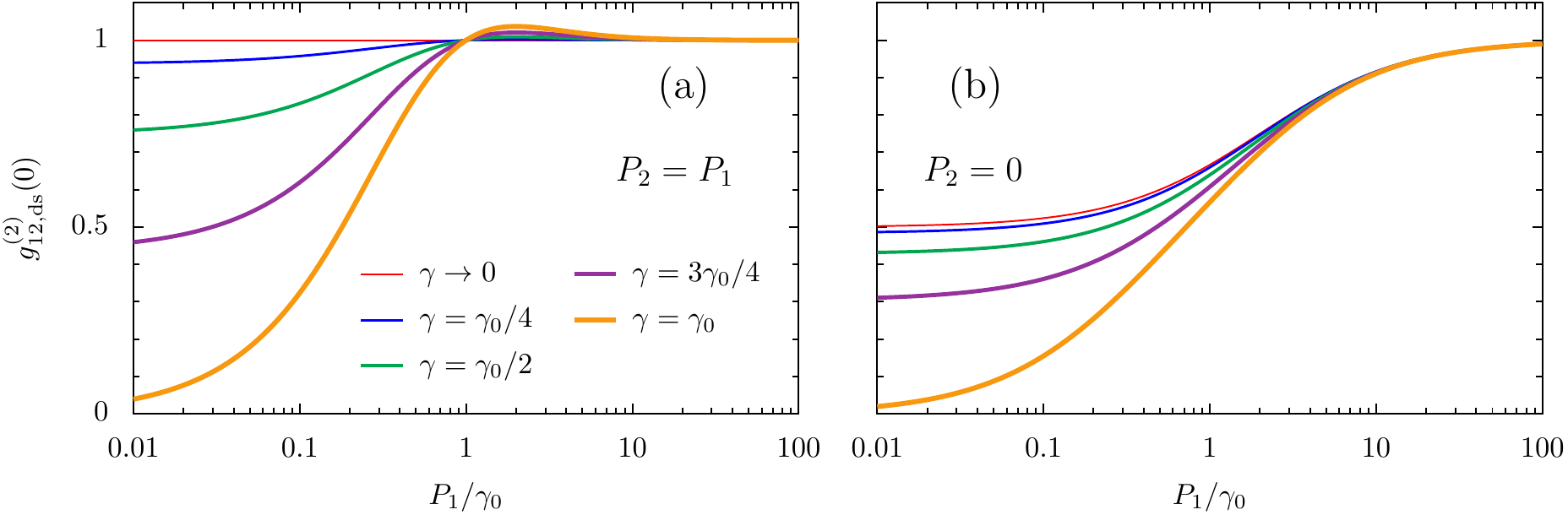}
 \caption{ Second-order cross-correlator in the dissipative coupling regime at zero delay $g_{12, \mathrm{ds}}^{(2)}(0)$, as a function of the pumping rate $P_1$, in units of the decay rate $\gamma_0$ [cf. Eq.~\eqref{eq:geetwodiss}]. Increasingly strong dissipative coupling strengths $\gamma$ are denoted by increasingly thick colored lines. We show results with symmetric pumping ($P_2 = P_1$) in panel (a) and asymmetric pumping ($P_2 = 0$) in panel (b).  }
 \label{fig:corrdiss}
\end{figure*}

The second-order coherence allows one to adjudicate on the probability of simultaneous emissions from 2LS-1 and 2LS-2 [cf. Eq.~\eqref{eq:geetwo_def}]. We find, in the dissipative coupling regime, the following expression for the cross-correlator at zero delay (see Appendix~\ref{app:Single} for the calculations) 
\begin{widetext}
\begin{equation}
\label{eq:geetwodiss}
g_{12, \mathrm{ds}}^{(2)} (0) = \frac{ \left( 4 P_1 P_2 \Gamma_+^2 + \left( \frac{\gamma P_-}{2} \right)^2 \right) \bigg( \gamma^2 \left( P_1 \Gamma_1 + P_2 \Gamma_2 \right) -4 \Gamma_+^2 \left( \gamma^2 - \Gamma_1 \Gamma_2 \right) \bigg) }{ \bigg( P_- \gamma^2 \left( P_1 - \Gamma_+ \right) + 4 P_1 \Gamma_2 \Gamma_+^2 \bigg) \bigg( P_- \gamma^2 \left( \Gamma_+ - P_2 \right) + 4 P_2 \Gamma_1 \Gamma_+^2 \bigg) },
\end{equation}
\end{widetext}
where the quantities $P_{\pm}, \Gamma_{1}, \Gamma_{2}$ and $\Gamma_+$ are given by Eq.~\eqref{eq:m0matrix233}.

We plot Eq.~\eqref{eq:geetwodiss} in Fig.~\ref{fig:corrdiss}, with symmetric pumping ($P_2 = P_1$) in panel (a) and with asymmetric pumping ($P_2 = 0$) in panel (b). In the figure, increasingly strong dissipative coupling strengths $\gamma$ are denoted by increasingly thick colored lines. In the symmetric regime of panel (a), Eq.~\eqref{eq:geetwodiss} collapses into $g_{12, \mathrm{ds}}^{(2)} (0) = 1 + \gamma^2 (P_1 - \gamma_0) / (P_1 + \gamma_0)^3$. Therefore, in the absence of any dissipative coupling $\gamma \to 0$, the system behaves effectively independently and $g_{12, \mathrm{ds}}^{(2)} (0) = 1$, as shown by the thin red line in panel (a). Once there is some nonzero dissipative coupling $\gamma \ne 0$ (thicker colored lines) antibunching is displayed with weak coupling $P_1 \ll \gamma_0$, a manifestation of the quantum nature of the system. Surprisingly, bunching $g_{12, \mathrm{ds}}^{(2)} (0) > 1$ is also possible for stronger dissipative coupling, and reaches its maximum value when $P_1 = 2 \gamma_0$. See for example the case of maximal dissipative coupling $\gamma = \gamma_0$ (thick orange line), and the bunching displayed when $P_1 \gtrsim \gamma_0$.

In Fig.~\ref{fig:corrdiss}~(b), with asymmetric pumping ($P_2 = 0$), the situation is quite different. Now antibunching $g_{12, \mathrm{ds}}^{(2)} (0) < 1$ is always exhibited, even in the limit of vanishing dissipative coupling $\gamma \to 0$ (thin red line) when $g_{12, \mathrm{ds}}^{(2)} (0) = (P_1 + \gamma_0) / (P_1 + 2\gamma_0)$. Of course, in the strong pumping limit $P_1 \gg \gamma_0$ the correlations are washed out and $g_{12, \mathrm{ds}}^{(2)} (0) \simeq 1$, since only the state $\ket{1, 0}$ is supported [medium green line in Fig.~\ref{fig:popdiss}~(b)]. These results, and their stark contrast to those of the coherent coupling regime in Fig.~\ref{fig:corrco}, suggest emission correlations and quantum spectroscopy as an important instrument to discriminate the coupling landscape~\cite{Dorfman2016}.


\subsection{\label{SECdissSPEC}Spectrum}

We now consider the optical spectra achievable with dissipative coupling only. With the dissipative coupling parameters of Eq.~\eqref{eq:limitingcases2}, we obtain the following simple expressions for the frequencies $\omega_p$ and damping rates $\gamma_p$ appearing in Eq.~\eqref{eq:spectrumao} [the calculation is performed in Appendix~\ref{app:Double}]
\begin{subequations}
\label{eq:nreigen2ssasssssssdsdsdsd222}
\begin{alignat}{2}
 \tfrac{1}{2} \gamma_{\mathrm{A}}^{\mathrm{ds}} + \mathrm{i} \omega_{\mathrm{A}}^{\mathrm{ds}}  &= \tfrac{1}{2} \left( 3 \gamma_0 - \gamma \right) + \mathrm{i} \omega_0,  \\
  \tfrac{1}{2} \gamma_{\mathrm{B}}^{\mathrm{ds}} + \mathrm{i} \omega_{\mathrm{B}}^{\mathrm{ds}}  &= \tfrac{1}{2} \left( 3 \gamma_0 + \gamma \right) + \mathrm{i} \omega_0,  \\
   \tfrac{1}{2} \gamma_{\mathrm{C}}^{\mathrm{ds}} + \mathrm{i} \omega_{\mathrm{C}}^{\mathrm{ds}}  &= \tfrac{1}{2} \left( \gamma_0 - \gamma \right) + \mathrm{i} \omega_0, \label{eq:thickasabrick2} \\
    \tfrac{1}{2} \gamma_{\mathrm{D}}^{\mathrm{ds}} + \mathrm{i} \omega_{\mathrm{D}}^{\mathrm{ds}}  &= \tfrac{1}{2} \left( \gamma_0 + \gamma \right) + \mathrm{i} \omega_0.
  \end{alignat}
\end{subequations}
Equation~\eqref{eq:nreigen2ssasssssssdsdsdsd222} describes energy levels completely unshifted in frequency ($\omega_{p}^{\mathrm{ds}} = \omega_0$), due to the lack of any coherent coupling. Instead, the dissipative coupling acts to induce super-radiant ($\gamma_{\mathrm{D}}^{\mathrm{ds}} = \gamma_0 + \gamma$) and sub-radiant ($\gamma_{\mathrm{C}}^{\mathrm{ds}} = \gamma_0 - \gamma$) broadening contributions to the spectrum.

The spectral decomposition of Eq.~\eqref{eq:spectrumao} is defined by the following unequal weighting coefficients
\begin{subequations}
\label{eq:elle334343}
\begin{alignat}{2}
    L_{\mathrm{C}}^{\mathrm{ds}}  &= \tfrac{\gamma_0 + \gamma}{2 \gamma_0},  \\
    L_{\mathrm{D}}^{\mathrm{ds}}  &= \tfrac{\gamma_0 - \gamma}{2 \gamma_0}.
  \end{alignat}
\end{subequations}
The coefficients associated with $\mathrm{A, B}$ labeling are zero, as is in the coherently coupled case of Eq.~\eqref{eq:elle33}, since we are working in the vanishing pumping rate limit where transitions from the unpopulated doubly-excited state $\ket{1, 1}$ do not contribute (as was shown in Ref.~\cite{delValle2010}). Most notably, in the maximally dissipatively coupled limit of $\gamma \to \gamma_0$, there is only one nonzero contribution to the spectrum since $L_{\mathrm{C}}^{\mathrm{ds}} \to 1$ and $L_{\mathrm{D}}^{\mathrm{ds}} \to 0$ from Eq.~\eqref{eq:elle334343}. This is a manifestation of the population trapping effect found in Fig.~\ref{fig:popdiss}~(b), where a transition from a singly-populated level to the ground state has been suppressed. Population trapping has a long history, see for example Refs.~\cite{Aspect1988, Santori2006, Agarwal2006}, and its appearance via this dissipative mechanism was alluded to before in Refs.~\cite{Gonzalez2011, MartinCano2011}.

\begin{figure}[tb]
 \includegraphics[width=\linewidth]{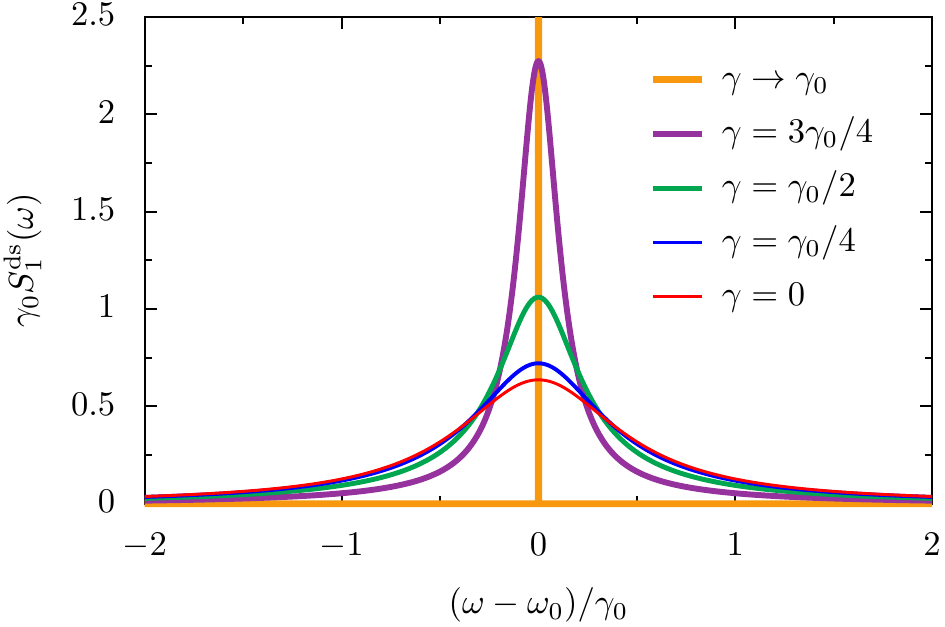}
 \caption{ Spectrum of 2LS-1 in the dissipative coupling regime $S_1^{\mathrm{ds}}(\omega)$, in units of the inverse damping rate $\gamma_0^{-1}$, for increasingly strong dissipative coupling strengths $\gamma$ (increasingly thick colored lines) [cf. Eq.~\eqref{eq:spectrumersdsdserh}].   }
 \label{fig:specdiss}
\end{figure}

The coefficients of Eq.~\eqref{eq:elle334343} and expressions of Eq.~\eqref{eq:nreigen2ssasssssssdsdsdsd222} lead to the optical spectrum
\begin{equation}
\label{eq:spectrumersdsdserh}
  S_1^{\mathrm{ds}} (\omega) = \frac{1}{4 \pi \gamma_0} \sum_{\tau = \pm 1} \frac{ \gamma_0^2-\gamma^2 }{ \left( \tfrac{\gamma_0 + \tau \gamma}{2} \right)^2 + \left( \omega - \omega_0 \right)^2}.
\end{equation}
Of course, in the uncoupled limit ($\gamma \to 0$) one recovers the single 2LS result of Eq.~\eqref{eq:speciso}. In the opposite, maximally dissipative limit ($\gamma \to \gamma_0$) there is a just a single contribution ($L_{\mathrm{C}}^{\mathrm{ds}} \to 1$, $L_{\mathrm{D}}^{\mathrm{ds}} \to 0$) and Eq.~\eqref{eq:spectrumersdsdserh} tends towards becoming a delta spectral peak
\begin{equation}
\label{eq:spectrumersdsdserh545}
  S_1^{\mathrm{ds}} (\omega) = \delta (\omega - \omega_0), \quad \gamma \to \gamma_0,
\end{equation}
where $\delta (x)$ is the Dirac delta function.

\begin{figure*}[tb]
 \includegraphics[width=\linewidth]{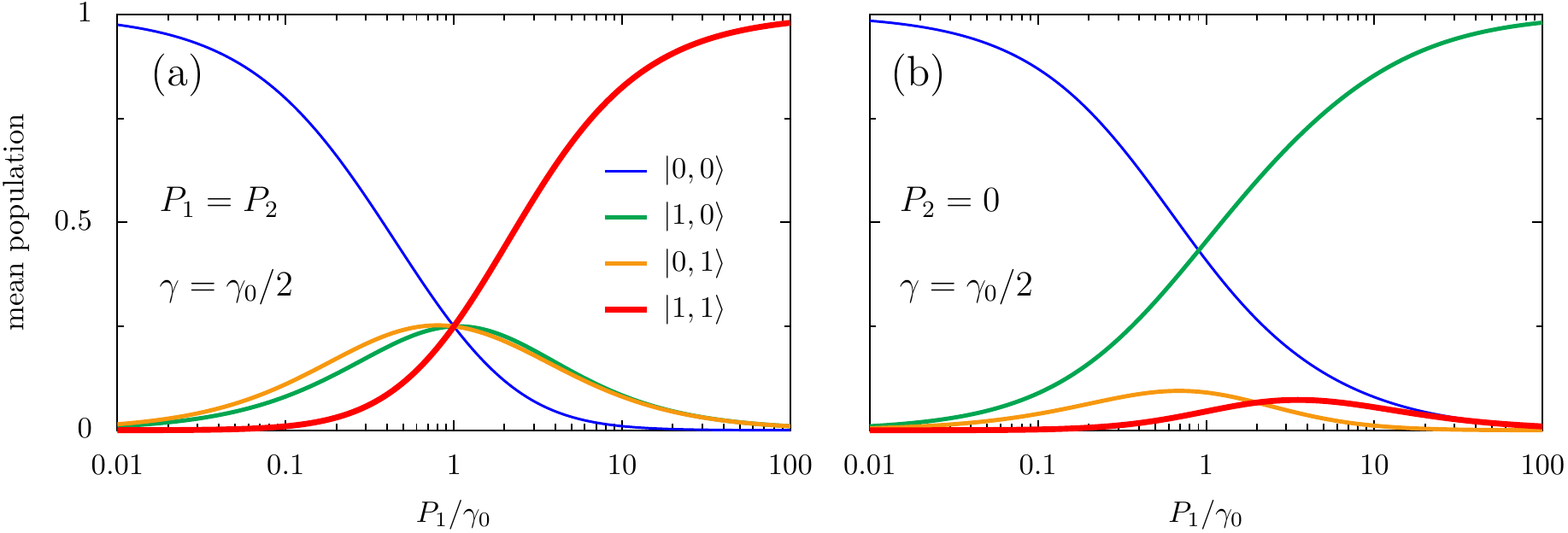}
 \caption{ Mean populations in the chiral-1 coupling regime, as a function of the pumping rate $P_1$, in units of the damping rate $\gamma_0$ [cf. Eq.~\eqref{eq:statespopuni}]. We show results with symmetric pumping ($P_2 = P_1$) in panel (a) and asymmetric pumping ($P_2 = 0$) in panel (b), for the dissipative coupling strength $\gamma = \gamma_0/2$. The labeling of the mean population of the state $\ket{i, j}$ is displayed in the legend of panel (a), and states with $N=\{ 0, 1, 2\}$ excitations are shown with increasingly thick lines.}
 \label{fig:popuni}
\end{figure*}

We plot the spectrum of Eq.~\eqref{eq:spectrumersdsdserh} in Fig.~\ref{fig:specdiss}, where increasingly strong dissipative coupling strengths $\gamma$ are denoted by increasingly thick colored lines. The plot shows the tendency towards a delta peak (thick orange line) with increasing dissipative coupling, and the characteristic singlet structure pinned at the unshifted resonance frequency $\omega_0$. 


\section{\label{SECuni}Chiral coupling}

Here we ruminate on the special limiting case of chiral coupling, that is when the coupling between the two quantum emitters goes in one direction only [as pictured in Fig.~\ref{fig:landscape}~(c)]. This nonreciprocal situation occurs due to the exact compensation of the back action from one of the quantum emitters, which arises due to a careful balance between both the relative magnitudes and relative phases of the coherent and dissipative coupling, as discussed in detail in Appendix~\ref{nonrec} and Refs.~\cite{Metelmann2015, Downing2019}. Controlling the directionality of coupling is important for the realization of non-reciprocal nanophotonic devices, such as unidirectional waveguides and circulators~\cite{Amico2019, Huang2020}, and our dimer model represents the simplest system which can exhibit such chirality.

In what follows, we ponder how the mean populations (Sec.~\ref{SECuniPOP}), correlation functions (Sec.~\ref{SECuniCORR}) and optical spectrum (Sec.~\ref{SECuniSPEC}) change due to the contest between the dissipative coupling strength $\gamma$ and the pumping rates $P_1$ and $P_2$, when the system is chirally coupled according to the relations of Eq.~\eqref{eq:limitingcases3}.


\subsection{\label{SECuniPOP}Mean populations}

The mean population $\rho_{ij}^{\mathrm{ch, I}}$ of the state $\ket{i, j}$, in the steady state, are obtained from the master equation of Eq.~\eqref{eq:master} (see Appendix~\ref{app:Single}). The resultant expressions, when Eq.~\eqref{eq:limitingcases3} are fulfilled, read
\begin{subequations}
\label{eq:statespopuni}
\begin{align}
 \rho_{00}^{\mathrm{ch, I}} &= \frac{ P_1 P_2 \gamma^2 + P_+ \gamma_0^3 + \gamma_0^4 + \left( \frac{\gamma_0 P_+}{2} \right)^2 }{ 2 \Gamma_1 \left( P_1 \gamma^2 + 2 \Gamma_2 \Gamma_+^2 \right) },  \\
 \rho_{10}^{\mathrm{ch, I}} &= \frac{P_1}{2 \Gamma_1} \frac{ P_1 \gamma^2 + 4 \gamma_0 \Gamma_+^2 }{P_1 \gamma^2 + 2 \Gamma_2 \Gamma_+^2},  \\
 \rho_{01}^{\mathrm{ch, I}} &= \frac{ 4 P_1 \gamma^2 \Gamma_+ + 4 P_2 \gamma_0 \Gamma_+^2 - \gamma^2 P_1 \left( P_1 + 2 P_2 \right) }{ 2 \Gamma_1 \left( P_1 \gamma^2 + 2 \Gamma_2 \Gamma_+^2 \right) }, \\
 \rho_{11}^{\mathrm{ch, I}} &= \frac{P_1}{2 \Gamma_1} \frac{ P_1 \gamma^2 + 4 P_2 \Gamma_+^2 }{P_1 \gamma^2 + 2 \Gamma_2 \Gamma_+^2}, 
 \end{align}
\end{subequations}
where the effective rates $P_{\pm}, \Gamma_{1},  \Gamma_{2}$ and $\Gamma_+$ are defined in Eq.~\eqref{eq:m0matrix233}.

Most importantly, the sum $\rho_{1, 0}^{\mathrm{ch, I}} + \rho_{1, 1}^{\mathrm{ch, I}} = n_1^{\mathrm{ch, I}} = P_1/\Gamma_1$, meaning that the probability for the 2LS-1 to be excited (that is, the system is either in the state $\ket{1, 0}$ or $\ket{1, 1}$) is exactly the same as in an isolated system, as given by Eq.~\eqref{eq:iso}. This is a hallmark of the chiral coupling regime at the fundamental level of the populations of the system.

We plot the mean populations of Eq.~\eqref{eq:statespopuni} in Fig.~\ref{fig:popuni} as a function of the incoherent pumping $P_1$ into 2LS-1, for symmetric pumping ($P_2 = P_1$) in panel (a) and asymmetric pumping ($P_2 = 0$) in panel (b).  While $\theta-\phi = \pi/2$ and $g = \gamma/2$ necessarily in this chiral case, the dissipative coupling strength is chosen as $\gamma = \gamma_0/2$ in both panels. The effect of introducing the asymmetry in the pumping is similar to in the coherent coupling case of Fig.~\ref{fig:popco}, and no population trapping may occur (even in the limit of  maximal dissipative coupling, $\gamma \to \gamma_0$) in contrast to the purely dissipative case of Fig.~\ref{fig:popdiss}. Thus the mean populations look superficially similar to other coupling regimes, and one is forced to look at other quantities to find distinguishing features for chiral coupling.


\subsection{\label{SECuniCORR}Correlations}

We now investigate the second-order coherence in the chiral coupling-I regime [cf. Eq.~\eqref{eq:geetwo_def}]. The cross-correlator reads (see Appendix~\ref{app:Single} for details) 
\begin{equation}
\label{eq:geetwouni}
g_{12, \mathrm{ch, I}}^{(2)} (0) = \frac{ \Gamma_1 \left( P_1 \gamma^2 + 4 P_2 \Gamma_+^2 \right) }{ 4 P_2 \Gamma_1 \Gamma_+^2 - 2 P_1 \gamma^2 \left( P_2 - 2 \Gamma_+ \right) },
\end{equation}
where $P_{\pm}, \Gamma_{1}, \Gamma_{2}$ and $\Gamma_+$ are effective rates, as introduced in Eq.~\eqref{eq:m0matrix233}.

\begin{figure}[tb]
 \includegraphics[width=\linewidth]{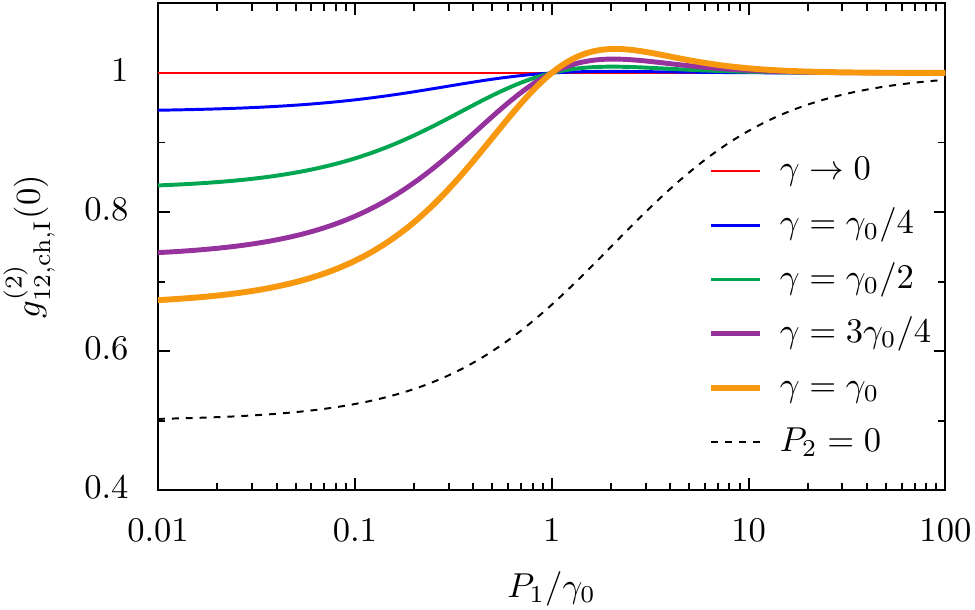}
 \caption{  Second-order cross-correlator in the chiral-1 coupling regime $g_{12, \mathrm{ch, I}}^{(2)}(0)$ at zero delay, as a function of the pumping rate $P_1$, in units of the decay rate $\gamma_0$ [cf. Eq.~\eqref{eq:geetwouni}]. Solid lines: symmetric pumping ($P_2 = P_1$), for increasingly strong dissipative coupling strengths $\gamma$ (increasingly thick colored lines). Dashed line: an asymmetric pumping case ($P_2 = 0$), which is independent of $\gamma$.  }
 \label{fig:corruni}
\end{figure}

We plot Eq.~\eqref{eq:geetwouni} in Fig.~\ref{fig:corruni}, with symmetric pumping ($P_2 = P_1$) demonstrated by the solid lines and with asymmetric pumping ($P_2 = 0$) described by the dashed line. For symmetric pumping, increasingly strong dissipative coupling strengths $\gamma$ are denoted by increasingly thick colored lines. The symmetric case is similar to the results in the dissipatively coupled regime, as shown in Fig.~\ref{fig:corrdiss}, where antibunching is dominant when $P_1 \ll \gamma_0$ and the correlations are washed out such that $g_{12, \mathrm{ch, I}}^{(2)} (0) \to 1$ when $P_1 \gg \gamma_0$. However, the asymmetric case (dashed line) is independent of the dissipative couping strength $\gamma$. It universally presents antibunching behavior bounded by  $g_{12, \mathrm{ch, I}}^{(2)} (0) \in [1/2, 1]$ and is governed by the expression $g_{12, \mathrm{ch, I}}^{(2)} (0) = (P_1 + \gamma_0)/(P_1 + 2\gamma_0)$. The chirality of the setup here is not immediately obvious in these correlations, due to their similarity to the purely dissipatively coupled case of Fig.~\ref{fig:corrdiss}, but chirality is most apparent when considering the optical spectrum as we now do.


\subsection{\label{SECuniSPEC}Spectrum}

The optical spectrum is perhaps the easiest way to identify chiral coupling. For the chiral (ch) coupling parameters of Eq.~\eqref{eq:limitingcases3}~and~\eqref{eq:limitingcases4}, of cases $\mathrm{I}$ and $\mathrm{II}$ respectively, we find the effective frequencies and broadenings appearing in Eq.~\eqref{eq:spectrumao} (see Appendix~\ref{app:Double} for details)
\begin{subequations}
\label{eq:nreigedsdsdsdn2222}
\begin{alignat}{2}
 \tfrac{1}{2} \gamma_{\mathrm{A}}^{\mathrm{ch}} + \mathrm{i} \omega_{\mathrm{A}}^{\mathrm{ch}} &= \tfrac{3}{2} \gamma_{0} + \mathrm{i} \omega_0,  \\
  \tfrac{1}{2} \gamma_{\mathrm{B}}^{\mathrm{ch}} + \mathrm{i} \omega_{\mathrm{B}}^{\mathrm{ch}} &= \tfrac{3}{2} \gamma_{0} + \mathrm{i} \omega_0,  \\
   \tfrac{1}{2} \gamma_{\mathrm{C}}^{\mathrm{ch}} + \mathrm{i} \omega_{\mathrm{C}}^{\mathrm{ch}} &= \tfrac{1}{2} \gamma_{0} + \mathrm{i} \omega_0,  \\
    \tfrac{1}{2} \gamma_{\mathrm{D}}^{\mathrm{ch}} + \mathrm{i} \omega_{\mathrm{D}}^{\mathrm{ch}} &= \tfrac{1}{2} \gamma_{0} + \mathrm{i} \omega_0.
  \end{alignat}
\end{subequations}
Equation~\eqref{eq:nreigedsdsdsdn2222} describes four frequency unshifted ($\omega_{p}^{\mathrm{ch}} = \omega_0$) transitions, which are either super-radiant ($\gamma_{\mathrm{A}}^{\mathrm{ch}} = \gamma_{\mathrm{B}}^{\mathrm{ch}} = 3 \gamma_0$) or radiant ($\gamma_{\mathrm{C}}^{\mathrm{ch}} = \gamma_{\mathrm{D}}^{\mathrm{ch}} = \gamma_0$).

\begin{figure}[tb]
 \includegraphics[width=\linewidth]{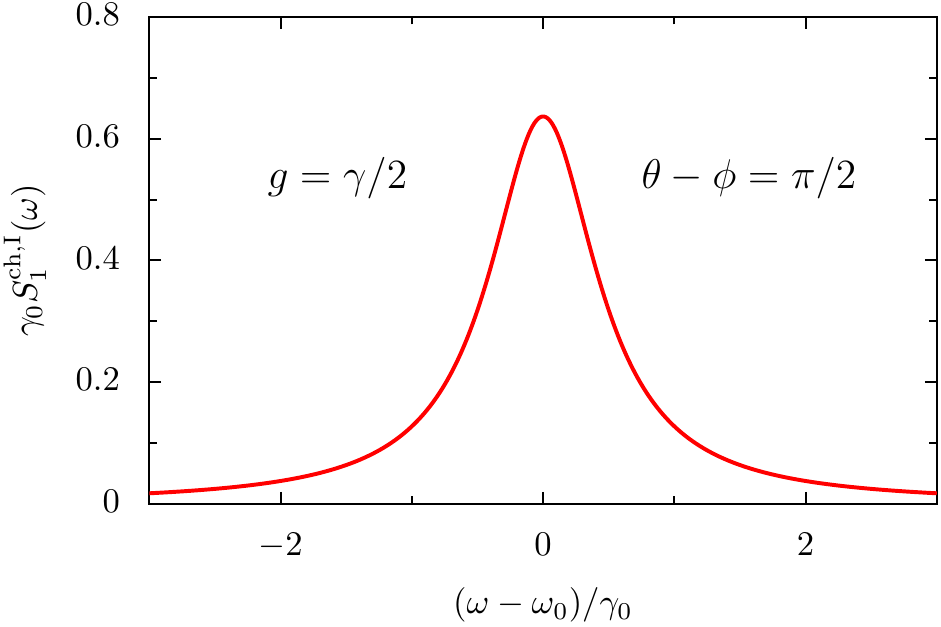}
 \caption{ Spectrum of 2LS-1 in the chiral-1 coupling regime $S_1^{\mathrm{ch, I}}(\omega)$, in units of the inverse damping rate $\gamma_0^{-1}$ [cf. Eq.~\eqref{eq:spectrumersdswwwewedserh}].  }
 \label{fig:specuni}
\end{figure}

We now consider chiral case $\mathrm{I}$, corresponding to 2LS-1 coupling to 2LS-2 in a one-way manner [see the upper sketch in Fig.\ref{fig:landscape}~(c)]. We find the simplest possible result for the Lorentzian weighting coefficient
\begin{equation}
\label{eq:elle11}
 L_{\mathrm{D}}^{\mathrm{ch}, \mathrm{I}} = 1, 
\end{equation}
while all of the other coefficients are zero. It immediately follows that the spectrum of Eq.~\eqref{eq:spectrum} is identical to that of a single 2LS in isolation, explicitly [cf. Eq.~\eqref{eq:speciso}]
\begin{equation}
\label{eq:spectrumersdswwwewedserh}
  S_1^{\mathrm{ch, I}} (\omega) =  S_{\mathrm{iso}} (\omega),
\end{equation}
which is of course independent of both the coherent $g$ and dissipative coupling $\gamma$ strengths, due to their exact compensation by design. The remarkably simple result of Eq.~\eqref{eq:spectrumersdswwwewedserh} is a signature of chiral coupling and is shown in Fig.~\ref{fig:specuni} for completeness.

\section{\label{SECgen}Asymmetric coupling}

In this section, we examine the coupling between the two quantum emitters in the most general manner, as sketched in Fig.~\ref{fig:sketch}~(b).  We present how the mean populations (Sec.~\ref{SECgenPOP}), correlation functions (Sec.~\ref{SECgenCORR}) and optical spectrum (Sec.~\ref{SECspecSPEC}) evolve across the coupling landscape, which is formed by the coherent coupling strength $g$, the dissipative coupling strength $\gamma$, and crucially the phase difference $\theta-\phi$ between these two quantities. Tuning these parameters allows one to induce asymmetry into the coupling, and thus traverse the entire coupling landscape [cf. Fig.~\ref{fig:landscape}].


\subsection{\label{SECgenPOP}Mean populations}

\begin{figure}[tb]
 \includegraphics[width=\linewidth]{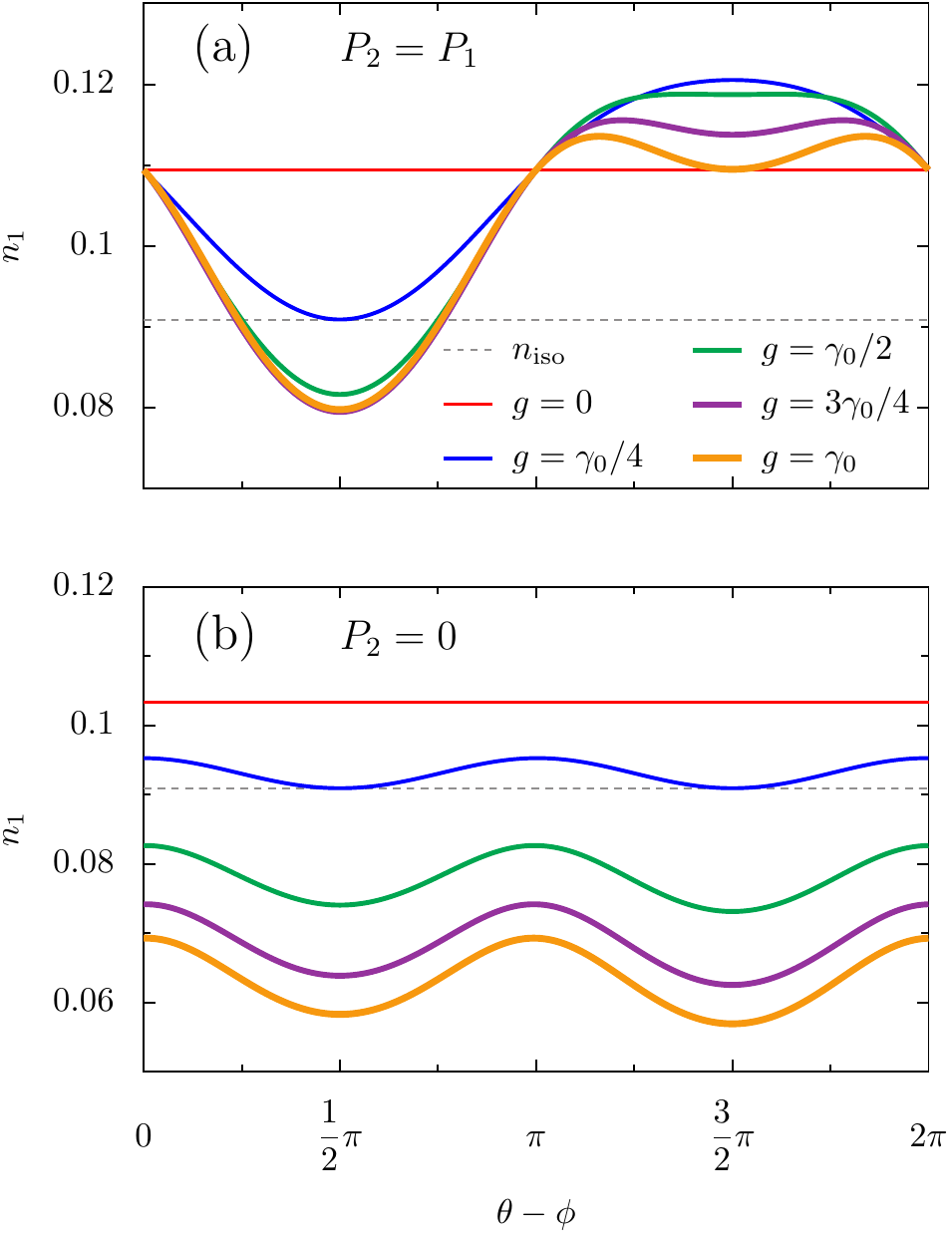}
 \caption{ Mean population $n_1$ of the first 2LS, as a function of the relative phase $\theta-\phi$, for increasingly strong coherent coupling strengths $g$ (increasingly thick solid lines) [cf. Eq.~\eqref{eq:mike1}]. Panel (a): the case of symmetric pumping rates ($P_2 = P_1$). Panel (b): an asymmetric pumping case ($P_2 = 0$). Dashed gray line: the population of an isolated 2LS, $n_{\mathrm{iso}}$ [cf. Eq.~\eqref{eq:iso}]. In the figure, $P_1 = \gamma_0/10$ and the dissipative coupling strength $\gamma = \gamma_0/2$, so that the blue line fulfills the chiral magnitude condition of $g/\gamma = 1/2$. }
 \label{fig:popgen}
\end{figure}

The most general expressions for the steady state correlators, those without any restrictions on the system parameters, read (see Appendix~\ref{app:Single} for the theory)
\begin{widetext}
\begin{subequations}
\label{eq:statespopgen}
\begin{align}
 n_1 &= \frac{ 4 P_1 \Gamma_2 \Gamma_+^2 + \gamma^2 P_1 P_- + \Gamma_+ \Big( 4 g^2 P_+ - \gamma^2 P_- \Big) + 2 g \gamma \Big( P_1 P_+ - 2 P_2 \Gamma_+ \Big) \sin \left( \theta - \phi \right) }{ g^2 \gamma^2 P_+ / \Gamma_+ + \gamma^2 \Big( P_1 \Gamma_1 + P_2 \Gamma_2 - 2 g^2 \Big) + 4 \Gamma_+^2 \Big( \Gamma_1 \Gamma_2 + 4 g^2 - \gamma^2  \Big) + g \gamma Q }, \label{eq:mike1} \\
 n_2 &= \frac{ 4 P_2 \Gamma_1 \Gamma_+^2 - \gamma^2 P_2 P_- + \Gamma_+ \Big( 4 g^2 P_+ + \gamma^2 P_- \Big) - 2 g \gamma \Big( P_2 P_+ - 2 P_2 \Gamma_+ \Big) \sin \left( \theta - \phi \right) }{ g^2 \gamma^2 P_+ / \Gamma_+ + \gamma^2 \Big( P_1 \Gamma_1 + P_2 \Gamma_2 - 2 g^2 \Big) + 4 \Gamma_+^2 \Big( \Gamma_1 \Gamma_2 + 4 g^2 - \gamma^2  \Big) + g \gamma Q },  \label{eq:mike2} \\
 n_{12} &= \frac{ \mathrm{e}^{\mathrm{i} \phi - 2 \theta } g^2 P_+ \gamma \left( P_+ /  \Gamma_+ - 2 \right) / 2 - \mathrm{i} \mathrm{e}^{\mathrm{i} \theta - 2 \phi } g \gamma^2 P_- (P_+/\Gamma_+ -2)/4 + F + G }{ g^2 \gamma^2 P_+ / \Gamma_+ + \gamma^2 \Big( P_1 \Gamma_1 + P_2 \Gamma_2 - 2 g^2 \Big) + 4 \Gamma_+^2 \Big( \Gamma_1 \Gamma_2 + 4 g^2 - \gamma^2  \Big) + g \gamma Q }, \label{eq:mike3} \\
  n_{\mathrm{X}} &= \frac{ g^2 P_+^2 + \left( \frac{\gamma P_-}{2} \right)^2 + 4 P_1 P_2 \Gamma_+^2 + g \gamma P_+ P_- \sin \left( \theta - \phi \right) }{ g^2 \gamma^2 P_+ / \Gamma_+ + \gamma^2 \Big( P_1 \Gamma_1 + P_2 \Gamma_2 - 2 g^2 \Big) + 4 \Gamma_+^2 \Big( \Gamma_1 \Gamma_2 + 4 g^2 - \gamma^2 \Big) + g \gamma Q }, \label{eq:mike4}
\end{align} 
\end{subequations}
where we have introduced the auxiliary functions
\begin{subequations}
\label{eq:sfsdfsdf}
\begin{align}
 Q &= g \gamma \Big( P_+ / \Gamma_+ -2 \Big) \cos \left( 2 \theta - 2 \phi \right) + 2 \Big( P_1 \Gamma_1 - P_2 \Gamma_2 \Big) \sin \left( \theta - \phi \right), \label{eq:sdssdsdsq} \\
   F &= -\mathrm{i} \mathrm{e}^{- \mathrm{i} \theta} \frac{g}{4} \Big( \gamma^2 P_- P_+ /\Gamma_+ - 2 \gamma^2 P_-  + 8 \Gamma_+ \left( P_1 \Gamma_2 - P_2 \Gamma_1 \right) \Big), \\
   G &= \mathrm{e}^{- \mathrm{i} \phi} \frac{\gamma}{2} \Big( g^2 P_+ \left( P_+ / \Gamma_+ - 2  \right) + 2 \Gamma_+ \left( 4 P_1 P_2 - P_2 \Gamma_1 - P_1 \Gamma_2 \right) \Big),
\end{align}  
\end{subequations}
\end{widetext}
and the effective rates $P_{\pm}, \Gamma_{1}, \Gamma_{2}$ and $\Gamma_+$ are given by Eq.~\eqref{eq:m0matrix233}.

We note that $n_1 = \langle \sigma_{1}^{\dagger} \sigma_{1} \rangle_{\mathrm{ss}}$ and $n_2 = \langle \sigma_{2}^{\dagger} \sigma_{2} \rangle_{\mathrm{ss}}$ refer to the steady state populations of 2LS-1 and 2LS-2 respectively, rather than referring to the population $\rho_{ij}$ of a certain state $\ket{i, j}$. The coherence reads $n_{12} = \langle \sigma_{1}^{\dagger} \sigma_{2} \rangle_{\mathrm{ss}}$, and the joint probability that both 2LSs are excited is $n_{\mathrm{X}} = \langle \sigma_{1}^{\dagger} \sigma_{1} \sigma_{2}^{\dagger} \sigma_{2} \rangle_{\mathrm{ss}} = \rho_{11}$. The probabilities of having only 2LS-1 or 2LS-2 excited are found via the relations $\rho_{1, 0} = n_{1} - n_{\mathrm{X}}$ and $\rho_{0, 1} = n_{2} - n_{\mathrm{X}}$ respectively, while the population of the ground state with zero excitations is given by $\rho_{0, 0}  = 1 + n_{\mathrm{X}} - n_1 - n_2$. 

We plot the 2LS-1 population $n_1$ of Eq.~\eqref{eq:mike1} as a function of $\theta-\phi$ in Fig.~\ref{fig:popgen}, with symmetric pumping ($P_2 = P_1$) in panel (a) and asymmetric pumping ($P_2 = 0$) in panel (b). We show results for increasingly strong coherent coupling strengths $g$ with increasingly thick lines, while the dissipative coupling strength $\gamma = \gamma_0/2$ and pumping rate $P_1 = \gamma_0/10$ are both held constant. We plot as a guide to the eye $n_{\mathrm{iso}}$, the population of an isolated 2LS from Eq.~\eqref{eq:iso}, as the gray dashed line. Most apparent in both panels is the equivalence of $n_1$ and $n_{\mathrm{iso}}$ precisely at the chiral coupling conditions of $\theta-\phi = \pi/2$ and $g = \gamma/2$ (medium blue line at $\theta-\phi = \pi/2$). The intersection of $n_1$ and $n_{\mathrm{iso}}$ at other points in parameter space (non-blue lines) in panel (a) is not associated with a mapping to a cascaded master equation describing a source and a target.

In the symmetric pumping configuration of Fig.~\ref{fig:popgen}~(a), with vanishing coherent coupling $g=0$ (thin red line) the population $n_1$ is of course independent of the relative phase $\theta-\phi$. Increasing the coherent coupling up to the chiral magnitude condition of $g/\gamma = 1/2$ (medium blue line) sees $n_1 \ge n_{\mathrm{iso}}$ for all phases. However, for coupling ratios $g/\gamma > 1/2$ (the green, purple and orange thicker lines) one notices that while $n_1$ is usually greater than $n_{\mathrm{iso}}$, there is a region of relative phases near to the chiral phase condition of $\theta-\phi = \pi/2$ at which $n_1 \le n_{\mathrm{iso}}$.

The asymmetric pumping setup of Fig.~\ref{fig:popgen}~(b) exhibits different behavior for larger coupling ratios $g/\gamma > 1/2$ (the thicker green, purple and orange lines), which always satisfy $n_1 < n_{\mathrm{iso}}$ because there is not enough pumping into the system to reach the isolated 2LS result of $n_{\mathrm{iso}}$. Panel (b) is symmetric about $\theta-\phi = \pi$ because there is no pump being passed from 2LS-2 from 2LS-1 to be adjusted by the relative phase, whereas panel (a) presents an asymmetry in $\theta-\phi$ because the coupling directionality is important when there is nonzero pump $P_2 \ne 0$ into 2LS-2, which can then be redistributed in the system.


\subsection{\label{SECgenCORR}Correlations}

The most general second-order coherence function of the system is found by simply dividing Eq.~\eqref{eq:mike4} by Eq.~\eqref{eq:mike1} and Eq.~\eqref{eq:mike2}, leading to the normalized cross-correlator at zero delay 
\begin{widetext}
 \begin{align}
\label{eq:geetwogenn}
g_{12}^{(2)} (0) =&~~~~\bigg( g^2 P_+^2 + \gamma^2 P_-^2/4 + 4 P_1 P_2 \Gamma_+^2 + g \gamma P_+ P_- \sin \left( \theta - \phi \right) \bigg) \nonumber  \\ 
 &\times \bigg( g^2 \gamma^2 P_+ / \Gamma_+ + \gamma^2 \left( P_1 \Gamma_1 + P_2 \Gamma_2 - 2 g^2 \right) + 4 \Gamma_+^2 \left( 4 g^2 - \gamma^2 + \Gamma_1 \Gamma_2 \right) + g \gamma Q \bigg) \nonumber  \\ 
  &\times \bigg( 4 g^2 P_+ \Gamma_+ + P_1 P_- \gamma^2 - \gamma^2 P_- \Gamma_+ + 4 P_1 \Gamma_2 \Gamma_+^2 + 2 g \gamma \left( P_1 P_+ - 2 P_2 \Gamma_+ \right) \sin \left( \theta - \phi \right)  \bigg)^{-1} \nonumber  \\ 
&\times \bigg( 4 g^2 P_+ \Gamma_+ - P_2 P_- \gamma^2 + \gamma^2 P_- \Gamma_+ + 4 P_2 \Gamma_1 \Gamma_+^2 - 2 g \gamma \left( P_2 P_+ - 2 P_1 \Gamma_+ \right) \sin \left( \theta - \phi \right)  \bigg)^{-1},
 \end{align} 
\end{widetext}
where the auxiliary function $Q$ is defined in Eq.~\eqref{eq:sdssdsdsq}.

\begin{figure}[tb]
 \includegraphics[width=\linewidth]{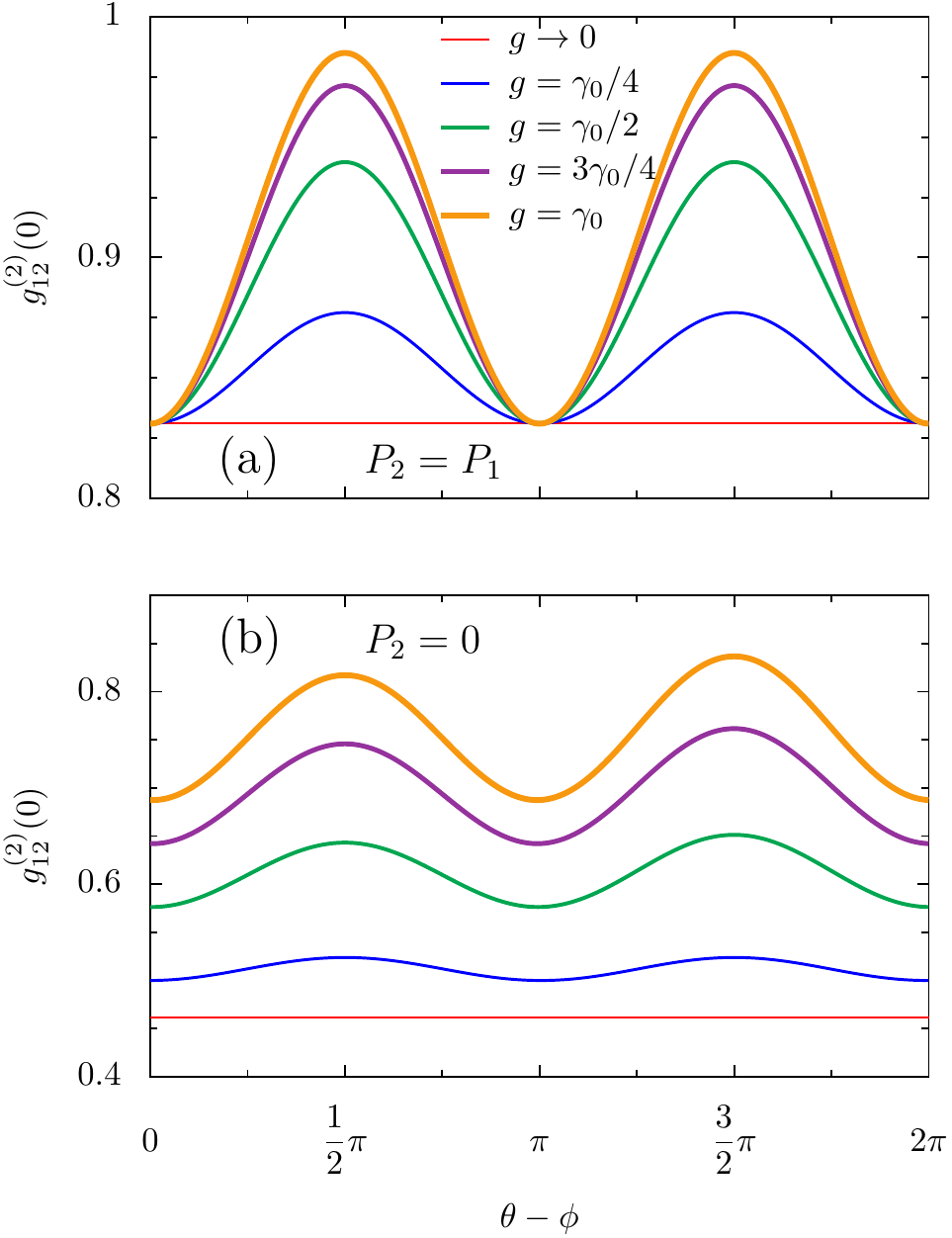}
 \caption{ Second-order cross-correlator in the asymmetric coupling regime at zero delay $g_{12}^{(2)}(0)$, as a function of the relative phase $\theta-\phi$. We show results with symmetric pumping rates ($P_2 = P_1$) in panel (a) and asymmetric pumping rates ($P_2 = 0$) in panel (b), for increasingly strong coherent coupling strengths $g$ (increasingly thick colored lines) [cf. Eq.~\eqref{eq:geetwogenn}]. In the figure, $P_1 = \gamma_0/10$ and the dissipative coupling strength $\gamma = \gamma_0/2$, so that the blue line fulfills the chiral magnitude condition of $g/\gamma = 1/2$.}
 \label{fig:corrgen}
\end{figure}

We plot in Fig.~\ref{fig:corrgen} the second-order cross-correlator of Eq.~\eqref{eq:geetwogenn}, as a function of the relative phase $\theta-\phi$. We show results with symmetric pumping rates ($P_2 = P_1$) in panel (a) and asymmetric pumping rates ($P_2 = 0$) in panel (b), for increasingly strong coherent coupling strengths $g$ (increasingly thick colored lines). The incoherent pumping rate $P_1 = \gamma_0/10$ and the dissipative coupling strength $\gamma = \gamma_0/2$ are both held constant.

In the regime of Fig.~\ref{fig:corrgen}~(a), and with vanishing coherent coupling $g \to 0$ (thin red line), Eq.~\eqref{eq:geetwogenn} reduces to $g_{12}^{(2)} (0) = 1 - \gamma^2 (\gamma_0 - P_1)/ (\gamma_0 + P_1)^3$ and defines the minimum of $g_{12}^{(2)} (0)$ for nonzero $g$. With non-vanishing coherent coupling (non-red lines), the degree of antibunching may be tuned as a function of the relative phase, with local maxima at the nonreciprocal relative phases of $\theta-\phi = \{ \pi/2, 3\pi/2 \}$ and minima at the reciprocal phases $\theta-\phi = \{ 0, \pi, 2\pi \}$. This behavior arises because at the nonreciprocal phases the system is close to chiral coupling, such that the system behaves most similarly to an independent system with $g_{12}^{(2)} (0) = 1$, while the opposite is true for phases far from the chiral phase condition.

The asymmetric pumping case of Fig.~\ref{fig:corrgen}~(b) displays a similar behavior, but there is no longer equivalence in $g_{12}^{(2)} (0)$ of all coupling strengths (all lines) at the reciprocal relative phases of $\theta-\phi = \{ 0, \pi, 2 \pi \}$. As in panel (a), stronger coherent coupling strengths lead to a greater variety in the magnitude of the coherence as one sweeps across the relative phases $\theta-\phi$.


\subsection{\label{SECspecSPEC}Spectrum}

\begin{figure*}[tb]
 \includegraphics[width=\linewidth]{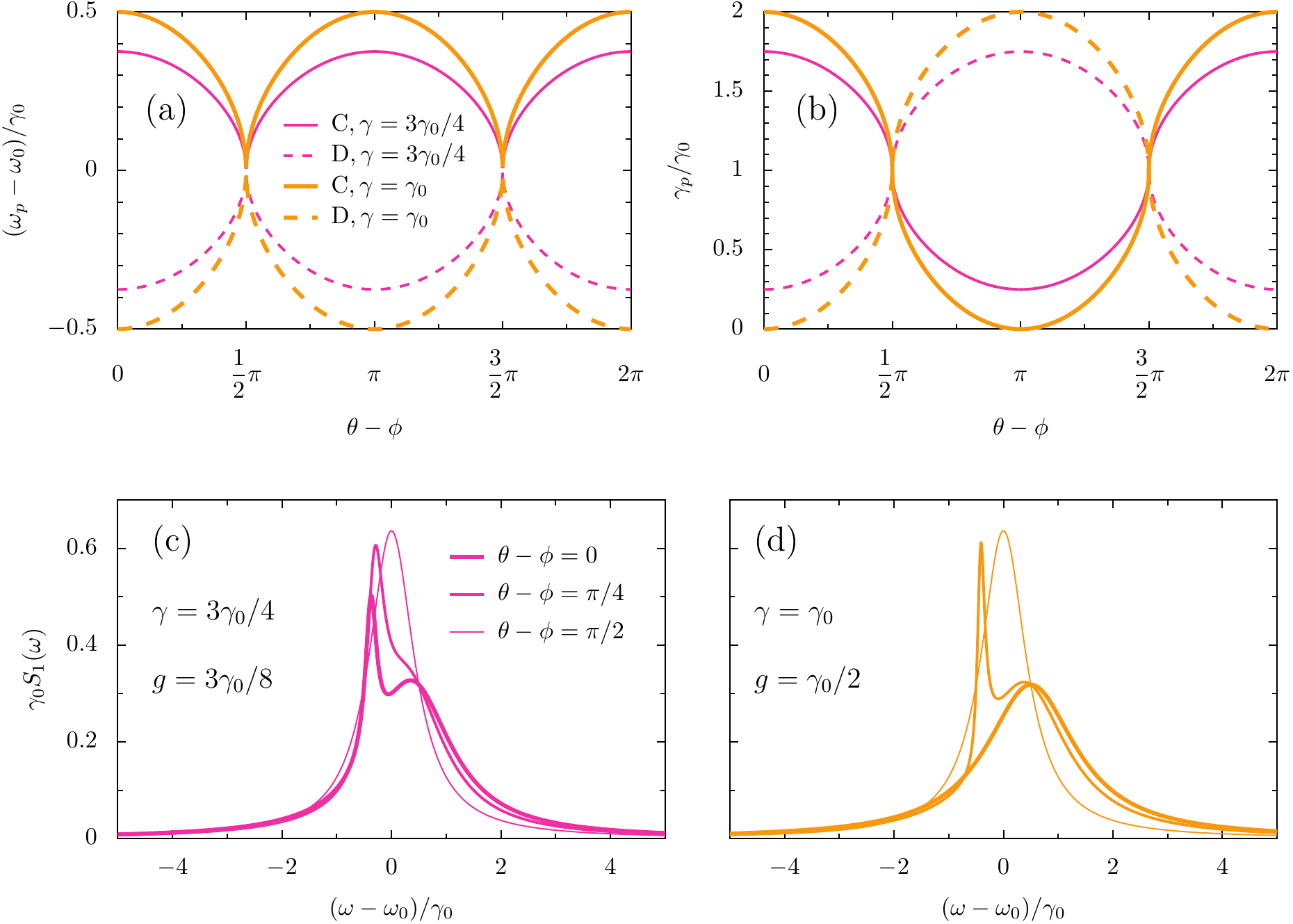}
 \caption{ Upper panels: Relative frequencies $\omega_{p}-\omega_0$  [panel (a)] and broadenings $\gamma_{p}$ [panel (b)], both in units of the damping rate $\gamma_0$, as a function of the relative phases $\theta-\phi$, with $g/\gamma = 1/2$ [cf. Eq.~\eqref{eq:nraaaaaaaaaeigen222sds2}]. Thin pink lines: the dissipative (coherent) coupling strength $\gamma = \gamma_0/2$ ($g = 3\gamma_0/8$). Thick orange lines: $\gamma = \gamma_0$ ($g = \gamma_0/2$). Solid (dashed) lines: the index $p = \mathrm{C} (\mathrm{D})$ [cf. Eq.~\eqref{eq:spectrumao}]. Lower panels: spectrum of 2LS-1 in the asymmetric coupling regime $S_1(\omega)$, in units of the inverse damping rate $\gamma_0^{-1}$ [cf. Eq.~\eqref{eq:spectrum}]. We show results for $\theta-\phi = \{ 0, \pi/4, \pi/2 \}$ (increasingly thin lines). Panel (c): $\gamma = 3\gamma_0/4$ and $g = 3\gamma_0/8$, corresponding to the pink lines in the upper panels. Panel (d): $\gamma = \gamma_0$ and $g = \gamma_0/2$, corresponding to orange lines in the upper panels.  }
 \label{fig:specgen}
\end{figure*}

We now consider the optical spectrum in the most general coupling case, which allows for a full consideration of asymmetric coupling effects. The frequencies $\omega_p$ and broadenings $\gamma_p$ defining the spectrum of 2LS-1, as decomposed like in Eq.~\eqref{eq:spectrumao}, read (see Appendix~\ref{app:Double} for details)
\begin{subequations}
\label{eq:nraaaaaaaaaeigen222sds2}
\begin{alignat}{2}
 \tfrac{1}{2} \gamma_{\mathrm{A}} + \mathrm{i} \omega_{\mathrm{A}} &= \tfrac{3}{2} \gamma_{0} + \mathrm{i} \left( \omega_0 + \Omega^{\ast} \right), \\
  \tfrac{1}{2} \gamma_{\mathrm{B}} + \mathrm{i} \omega_{\mathrm{B}} &= \tfrac{3}{2} \gamma_{0} + \mathrm{i} \left( \omega_0 - \Omega^{\ast} \right),  \\
   \tfrac{1}{2} \gamma_{\mathrm{C}} + \mathrm{i} \omega_{\mathrm{C}} &= \tfrac{1}{2} \gamma_{0} + \mathrm{i} \left( \omega_0 + \Omega \right) ,  \\
    \tfrac{1}{2} \gamma_{\mathrm{D}} + \mathrm{i} \omega_{\mathrm{D}} &= \tfrac{1}{2} \gamma_{0} + \mathrm{i} \left( \omega_0 - \Omega \right).
  \end{alignat}
\end{subequations}
where we have introduced the complex frequency 
\begin{equation}
\label{eq:complexfrequency}
 \Omega = \sqrt{g^2 - \left( \tfrac{\gamma}{2} \right)^2 - \mathrm{i} g \gamma \cos \left( \theta - \phi \right) }.
\end{equation}
Of course, the most general parameters of Eq.~\eqref{eq:nraaaaaaaaaeigen222sds2} recover the priorly addressed special cases of Eqs.~\eqref{eq:nreigen2sdwewewewewwwwsd222},~\eqref{eq:nreigen2ssasssssssdsdsdsd222},~and~\eqref{eq:nreigedsdsdsdn2222} under the appropriate conditions. Equation~\eqref{eq:nraaaaaaaaaeigen222sds2} also naturally arises from a non-Hermitian Hamiltonian approach, as is detailed in App.~\ref{app:first_mom}. Furthermore, we note that while throughout this work we have neglected cross-Kerr type interactions, they do not lead to meaningful changes to the results presented in Eq.~\eqref{eq:nraaaaaaaaaeigen222sds2}, a fact which is justified in App.~\ref{app:interactions}.

In general, the complex frequency $\Omega$ as given by Eq.~\eqref{eq:complexfrequency} may contribute to both the frequency shifts $\omega_{p}$ and broadenings $\gamma_{p}$ making up the spectral curves comprising the full spectrum of Eq.~\eqref{eq:spectrumao}. One notices that dissipative coupling $\gamma$ only leads to frequency shifts $\omega_{p}$ if the coherent couping $g$ is nonzero, while the coherent coupling $g$ only modifies the collective damping rates $\gamma_{p}$ if the dissipative couping $\gamma$ is nonzero, due to the form of Eq.~\eqref{eq:complexfrequency}. Furthermore, the complex frequency of Eq.~\eqref{eq:complexfrequency} vanishes for the specific cases where the chiral conditions of Eq.~\eqref{eq:limitingcases3}~or~\eqref{eq:limitingcases4} are satisfied. 

We plot the key quantities $\{ \omega_{p} , \gamma_{p} \}$ of Eq.~\eqref{eq:nraaaaaaaaaeigen222sds2} as a function of the relative phase $\theta - \phi$ in the upper panels of Fig.~\ref{fig:specgen}, where the chiral magnitude condition of $g = \gamma/2$ is satisfied but the phase condition is not necessarily fulfilled, this is the so-called ``quasichiral'' regime of Ref.~\cite{Downing2019}. The thin pink lines denote when the dissipative (coherent) coupling strength $\gamma = 3\gamma_0/4$ ($g = 3\gamma_0/8$), while the thick orange lines mark the case of maximal dissipative (coherent) coupling $\gamma = \gamma_0$ ($g = \gamma_0/2$). Solid (dashed) lines are associated with the index $p$ being equal to $\mathrm{C}$ ($\mathrm{D}$), which correspond to the two parts of the spectral decomposition entering the spectrum [cf. Eq.~\eqref{eq:spectrum}]. The quantities linked to $p = \{ \mathrm{A},\mathrm{B} \}$ are not shown since they have zero spectral weight in the vanishing pump limit in which we work~\cite{delValle2010}.

Fig.~\ref{fig:specgen}~(a) shows how the frequencies $\omega_p$ entering the spectrum decomposition of Eq.~\eqref{eq:spectrumao} change as a function of the relative phase. Most notably, $\omega_{\mathrm{C}} = \omega_{\mathrm{D}} = \omega_0$ at the chiral phase conditions of $\theta-\phi = \{ \pi/2, 3\pi/2\}$, since the spectrum of an isolated 2LS must be recovered. Otherwise, $\omega_p$ presents both red ($\omega_p < \omega_0$) and blue ($\omega_p > \omega_0$) shifts from the resonance frequency of a single 2LS $\omega_0$, which are stronger with increasing $\gamma$ (orange lines as compared to pink), giving remarkable freedom for the positions $\omega_p$ of the resultant spectral peaks.

Fig.~\ref{fig:specgen}~(b) displays the associated broadenings $\gamma_p$ which enter Eq.~\eqref{eq:spectrumao}. Panel~(b) shows how both super-radiant ($\gamma_p > \gamma_0$) and sub-radiant ($\gamma_p < \gamma_0$) transitions are possible. Of course, at the chiral phase conditions $\gamma_{\mathrm{C}} = \gamma_{\mathrm{D}} = \gamma_0$, ensuring the spectrum of an isolated 2LS arises. Most interestingly, vanishingly small broadenings $\gamma_p \ll \gamma_0$ arise at the nonreciprocal phases $\theta-\phi = \{ 0, \pi, 2\pi \}$ for the maximal dissipative coupling case (orange lines), suggesting the creation of extremely sharp spectral features.

Taken together, panels (a) and (b) of Fig.~\ref{fig:specgen} imply that a thin spectral peak, defined by $\gamma_p \ll \gamma_0$, can be associated with either a red-shifted ($\omega_p < \omega_0$) or a blue-shifted ($\omega_p > \omega_0$) frequency change from $\omega_0$. This feature should be highly apparent in the full spectrum $S_1 (\omega)$ for strong dissipative coupling $\gamma \simeq \gamma_0$ as the relative phase approaches a reciprocal phase difference $\theta-\phi = \{ 0, \pi, 2\pi \}$.

We plot the spectrum $S_1 (\omega)$ [cf. Eq.~\eqref{eq:spectrum}] in the lower panels of Fig.~\ref{fig:specgen}, for three relative phases $\theta - \phi = \{ 0, \pi/4, \pi/2 \}$, which are marked by increasingly thin lines. In panel (c), we choose the parameters $\gamma = 3\gamma_0/4$ and $g = 3\gamma_0/8$, which correspond to the thin pink lines in the upper panels of Fig.~\ref{fig:specgen}. Most notably, while the spectrum in panel (c) is comprised of two peaks associated with the indices $p = \{ \mathrm{C}, \mathrm{D} \}$, this fact is most pronounced for the case $\theta - \phi = 0$ (thick pink lines). This is because here the red-shifted ($\omega_{\mathrm{D}} < \omega_0$) narrow spectral peak is highly sub-radiant ($\gamma_{\mathrm{D}} \ll \gamma_0$), as follows from the dashed pink lines in Fig.~\ref{fig:specgen}~(a, b) at $\theta - \phi = 0$. Meanwhile, the $p = \mathrm{C}$ broad contribution to the spectrum is super-radiant ($\gamma_{\mathrm{C}} > \gamma_0$) and presents a blue-shift ($\omega_{\mathrm{C}} > \omega_0$), as is captured by the solid pink lines in Fig.~\ref{fig:specgen}~(a, b) at $\theta - \phi = 0$.

With the increased phase difference of $\theta - \phi = \pi/4$ (medium pink line) in panel (c), this extremely sharp spectral feature begins to be lost. This behavior can be traced back from the dashed pink lines in the upper panels of Fig.~\ref{fig:specgen} at $\theta - \phi = \pi/4$, which show an increased broadening $\gamma_{\mathrm{D}}$ and decreased redshift $\omega_{\mathrm{D}} - \omega_0$. Finally, in the chiral coupling limit of $\theta - \phi = \pi/2$ in panel (c) (thin pink line) a standard Lorentzian spectrum is recovered, since the coupling is purely unidirectional. This is congruent with the constituent parameters $\omega_{\mathrm{C}} = \omega_{\mathrm{D}} = \omega_0$ and $\gamma_{\mathrm{C}} = \gamma_{\mathrm{D}} = \gamma_0$, as follows from the solid and dashed pink lines in the upper panels in Fig.~\ref{fig:specgen} at $\theta - \phi = \pi/2$.

In Fig.~\ref{fig:specgen}~(d) the same result is of course produced for the chirally coupled case (thin orange line), where $\theta - \phi = \pi/2$. However, in panel~(d) the narrow peak is much more noticeable at the intermediate phase $\theta - \phi = \pi/4$ (medium orange line), due to the maximal coupling constants considered ($\gamma = \gamma_0$ and $g = \gamma_0/2$), which correspond to the thick orange lines in the upper panels. When $\theta - \phi = 0$ (thick orange line), the narrow peak is lost with maximal dissipative coupling since the $p = \mathrm{D}$ contribution to the spectrum has zero spectral weight ($L_{\mathrm{D}} = K_{\mathrm{D}} = 0$) in this limit. The narrow peak appears with any nonzero relative phase (as illustrated by the medium orange line). 

The modulation of the optical spectrum as a function of the relative phase, and the emergence of a narrow peak as illustrated by the lower panels of Fig.~\eqref{fig:specgen}, is a remarkable manifestation of asymmetric coupling between the pair of 2LSs, and offers the opportunity for the experimental detection of the diverse coupling landscape in the coupled system.


\section{\label{conc}Conclusions}

We have introduced an analytic model of two coupled two-level systems, where the phases of both the coherent and dissipative couplings are of paramount importance. Depending on both the relative strength and phase difference between the coherent and dissipative couplings, we have shown that the model evolves through a rich coupling landscape, including: coherent, dissipative, chiral (or one-way) and asymmetric coupling.

The required phases attached to the complex coupling parameters between two bodies may be realized in range of systems, such as with ultracold atoms~\cite{Aidelsburger2013, Celi2014} or photonic resonators~\cite{Fang2012b, Fang2013, Tzuang2014}, where tunable, synthetic magnetic fields have been implemented by, for example, harmonically modulating in time the coupling, or by using laser-assisted tunneling in optical potentials.

We have analyzed several fundamental quantum optical properties of the model as a function of the type of coupling, namely the steady state populations, optical spectrum and second-order correlation functions. We have found some remarkable properties including unexpected spectral features, population trapping, and strong emission correlations, all of which may act as signatures of chiral and asymmetric coupling in future experiments.

Our work on the simplest possible coupled system, that of a dimer, helps to provides insight into more complicated systems, such as chirally coupled chains~\cite{Mirza2017, Jen2019, Mirza2018, Buonaiuto2019, Lin2020, Jenny2020, Jen2019b}. Our results also pave the way for future work on chirally coupled metasurfaces, as the young field of chiral quantum optics continues to evolve.


\section*{Acknowledgments}
This work has been funded by the Spanish MINECO, via the Juan de la Cierva program [CAD], the CLAQUE project (FIS2015-64951-R) [AIFD and EdV], and the `Mar\'{i}a de Maeztu' Programme for Units of Excellence in R \& D (CEX2018-000805-M) [AIFD and EdV]. This work was also supported by a 2019 Leonardo Grant for researchers and cultural creators, BBVA foundation [AIFD]. We thank L. Mart\'{i}n-Moreno and D. Zueco for fruitful discussions, and S.~Bilan for technical support.


\appendix


\section{\label{nonrec}Chiral coupling conditions}

In this appendix, we derive the conditions for chiral coupling, following the prescription of Metelmann and Clerk as described in Ref.~\cite{Metelmann2015}.

The master equation of a cascaded quantum system, where system $2 (1)$ is being driven from the output from system $1 (2)$ [cf. Fig.~\ref{fig:sketch}~(b)], is given by~\cite{Gardiner1993, Carmichael1993, Gardiner2004}
\begin{align}
\label{eq:cascade}
 \partial_t \rho =&~\mathrm{i} [\rho , H_0] +  \frac{\gamma_{0}}{2} \mathscr{L}_{11} \rho +  \frac{\gamma_{0}}{2} \mathscr{L}_{22} \rho \nonumber  \\
&+ \beta \gamma_{0} \left( \mathrm{e}^{\mathrm{i} \eta} [ \sigma_{1 (2)} \rho, \sigma_{2 (1)}^{\dagger} ] + \mathrm{e}^{-\mathrm{i} \eta} [ \sigma_{2 (1)} , \rho \sigma_{1(2)}^{\dagger}] \right),
\end{align}
where $H_0$, the non-interacting part of the Hamiltonian of the coupled 2LSs, is given by Eq.~\eqref{eq:ham1} and the Liouvillian superoperator $\mathscr{L}_{ij}$ by Eq.~\eqref{eq:master2}. In Eq.~\eqref{eq:cascade}, the subscripts $i(j)$ refer to the two directions of driving, which we label case $\mathrm{I}$ ($\mathrm{II}$), $\beta$ is a nonnegative real number and $\eta$ is an arbitrary phase. Notably, in the original derivations of Refs.~\cite{Gardiner1993, Carmichael1993} these quantities were chosen as $\beta = 1$ (to describe the maximum possible chiral coupling strength) and $\eta = 0$.

Let us now consider the joint decay operator~\cite{Gardiner2004, Lopez2016, Lopez2018}
\begin{equation}
\label{eq:jointdecay}
\xi = \sqrt{ \nu_1 \gamma_{0}} \mathrm{e}^{\mathrm{i}\eta} \sigma_1 + \sqrt{ \nu_2 \gamma_{0}} \sigma_2,
\end{equation}
where $\nu_{1, 2}$ parameterize the strength of the collective damping decay rate $\gamma$, such that $0 \le \nu_{1, 2} \le 1$. Upon employing the operator of Eq.~\eqref{eq:jointdecay} in the master equation of Eq.~\eqref{eq:cascade}, one obtains the Lindblad form
\begin{align}
\label{eq:cascaded2}
 \partial_t \rho =&~\mathrm{i} [\rho , H_0]  + \frac{1}{2} \mathscr{L}_{\xi} \rho \nonumber \\
&+   \frac{(1-\nu_1) \gamma_{0}}{2} \mathscr{L}_{11} \rho + \frac{(1-\nu_2) \gamma_{0}}{2} \mathscr{L}_{22} \rho \nonumber \\
&+ \frac{\beta \gamma_{0} }{2} \left( \mathrm{e}^{\mathrm{i} \eta} [ \rho, \sigma_{2 (1)}^{\dagger} \sigma_{1 (2)}] - \mathrm{e}^{-\mathrm{i} \eta} [ \rho, \sigma_{1(2)}^{\dagger} \sigma_{2 (1)} ] \right).
\end{align}
Here $\beta = \sqrt{\nu_1 \nu_2}$, which implies that $0 \le \beta \le 1$. Upon expanding out the general and specific master equations of Eq.~\eqref{eq:master} and Eq.~\eqref{eq:cascade} respectively, and assigning all of the like terms (see also Refs.~\cite{Metelmann2015, Downing2019}) one readily finds the conditions on the system parameters to be in the chiral coupling regime 
\begin{subequations}
\label{eq:conditions}
\begin{align}
\frac{g}{\gamma} &= \frac{1}{2}, \label{eq:conditions2}  \\
 \theta - \phi &=
\begin{cases} \tfrac{\pi}{2}, \quad \text{case}~\mathrm{I}:~\text{2LS-1~drives~2LS-2}, \\ 
\tfrac{3 \pi}{2}, \quad \text{case}~\mathrm{II}:~\text{2LS2-2~drives~2LS-1},
\end{cases} \label{eq:conditions1}
 \end{align}
\end{subequations} 
where we considered the vanishing pump limit ($P_1, P_2 \to 0$) and used $\eta = \mp \phi$ in Eq.~\eqref{eq:cascade}. We also have the physical condition $\gamma = \beta \gamma_{0}$, or equivalently [due to Eq.~\eqref{eq:cascaded2}] the inequality
\begin{equation}
\label{eq:cascade3443}
0 \le \gamma \le \gamma_0,
\end{equation}
which ensures the magnitude of the dissipative coupling is never greater than the self decay rate. Notably, the limiting case of maximal dissipative coupling ($\beta = 1$, or $\gamma = \gamma_0$) leads to peculiar effects such as population trapping, and is therefore of special interest throughout this work.


\section{A single 2LS with incoherent pumping}
\label{app:asingle2ls}

\begin{figure}[htb]
 \includegraphics[width=\linewidth]{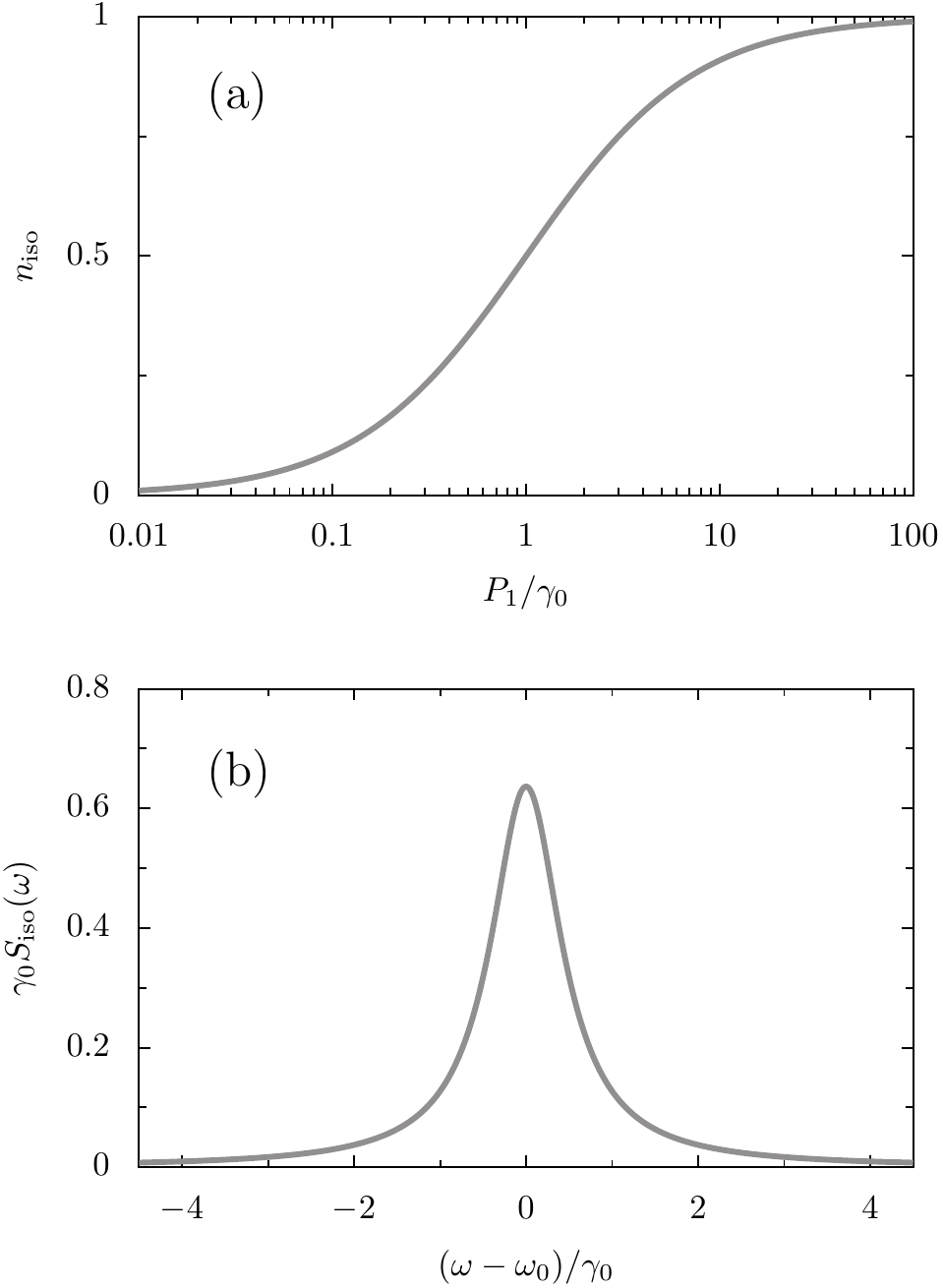}
 \caption{ Panel (a): mean population $n_{\mathrm{iso}}$ of an isolated 2LS as a function of the pumping rate $P_1$, in units of the damping rate $\gamma_0$ [cf. Eq.~\eqref{eq:iso}]. Panel (b): spectrum $S_{\mathrm{iso}}(\omega)$ of an isolated 2LS, in units of the inverse damping rate $\gamma_0^{-1}$, as a function of the frequency $\omega$ (which is measured from the 2LS resonance frequency $\omega_0$) [cf. Eq.~\eqref{eq:speciso}]. }
 \label{fig:popiso}
\end{figure}

In this appendix, we briefly detail some results for a single 2LS subject to incoherent pumping. The derived expressions are used as comparisons to the behavior of the system of two coupled 2LSs studied throughout the main text.

The Hamiltonian and master equation of an isolated ($\mathrm{iso}$) 2LS read~\cite{Lopez2016}
 \begin{subequations}
\label{eqqvn1}
\begin{alignat}{3}
 H_{\mathrm{iso}} &= \omega_0 \sigma^{\dagger} \sigma, \label{eqqvn1a} \\
  \partial_t \rho_{\mathrm{iso}} &= \mathrm{i} [ \rho, H ] +  \frac{\gamma_0}{2} \left( 2 \sigma \rho \sigma^{\dagger} -  \sigma^{\dagger} \sigma \rho - \rho \sigma^{\dagger} \sigma \right) \nonumber \\ 
&~~~~+  \frac{P_1}{2} \left( 2 \sigma^{\dagger} \rho \sigma -  \sigma \sigma^{\dagger}  \rho - \rho \sigma \sigma^{\dagger} \right),  \label{eqqvnb}
\end{alignat}
 \end{subequations}
with the 2LS resonance frequency $\omega_0$, the self-damping decay rate $\gamma_0$, and the incoherent pumping rate $P_1$.

The master equation of Eq.~\eqref{eqqvnb} directly leads to the following equation of motion for the population correlator
 \begin{equation}
\label{eqqvn2}
  \partial_t \langle \sigma^{\dagger} \sigma \rangle = P_1 - \left( P_1 + \gamma_0 \right) \langle \sigma^{\dagger} \sigma  \rangle.
   \end{equation}
When $t \to \infty$ the system reaches its steady state ($\mathrm{ss}$), and setting $\partial_t \langle \sigma^{\dagger} \sigma \rangle = 0$ in Eq.~\eqref{eqqvn2} intermediately yields the expression for the steady state population of a single 2LS. This result is given as Eq.~\eqref{eq:iso} in the main text, where we rename $\langle \sigma^{\dagger} \sigma \rangle_{\mathrm{ss}} = n_{\mathrm{iso}}$. We plot $n_{\mathrm{iso}}$ in Fig.~\ref{fig:popiso}~(a), showing the evolution of the population as a function of the pumping rate $P_1$.

Application of the quantum regression formula~\cite{Gardiner2004, Gardiner2014}, along with the master equation of Eq.~\eqref{eqqvnb}, yields the two-time equation of motion for a single 2LS
 \begin{equation}
\label{eqqvn3}
  \partial_\tau \langle \sigma^{\dagger} (\tau + t) \sigma (t) \rangle = - \left( \mathrm{i} \omega_0 + \tfrac{1}{2} \left[ P_1 + \gamma_0 \right] \right) \langle \sigma^{\dagger} (\tau + t) \sigma (t) \rangle.
 \end{equation}
Upon integrating Eq.~\eqref{eqqvn3}, and using the Wiener-Khinchin theorem~\cite{Gardiner2014, Kavokin2007}, one finds the optical spectrum of a single 2LS. The result is $S_{\mathrm{iso}} (\omega)$, the normalized spectrum given as Eq.~\eqref{eq:speciso} in the main text (in the limit of vanishing pumping, $P_1 \to 0$). We plot $S_{\mathrm{iso}} (\omega)$ in Fig.~\ref{fig:popiso}~(b), displaying the standard Lorentzian lineshape centered at $\omega_0$, and with broadening $\gamma_0$.


\section{Single-time dynamics}
\label{app:Single}

In this appendix, we derive the single-time dynamics of the correlators associated with the mean populations of the two coupled 2LSs, in the spirit of Refs.~\cite{Laussy2008, ValleLaussy2009, Laussy2009, Valle2009, Villas2011, ValleLaussy2011, Kavokin2007, ValleSuper2011}. In particular, we exploit the quantum regression theorem, which allows one to find the dynamics of a desired correlator from the mean values of some observable in time~\cite{Gardiner2004, Gardiner2014}. Specifically, we calculate the steady state populations of the coupled system in the manner of Ref.~\cite{delValle2010}, and investigate how the various coupling regimes (see Fig.~\ref{fig:landscape}) affect the populations of the system. 


\subsection{\label{myeqofmo123}Equation of motion}

Using the master equation~\eqref{eq:master} and the relation $\partial_t \langle O \rangle = \mathrm{Tr} \left( O \partial_t \rho \right)$ for any operator $O$, we arrive at the equation of motion
\begin{equation}
\label{eq:of_motion}
\frac{\mathrm{d}}{\mathrm{d} t} \mathbf{u}  = \mathbf{P} - \mathbf{M} \mathbf{u},
\end{equation}
for the 5-vector of correlators $\mathbf{u}$ and drive term $\mathbf{P}$, with
\begin{equation}
\label{eq:umatrix}
\mathbf{u} =
\begin{pmatrix}
 \langle \sigma_1^{\dagger} \sigma_1 \rangle \\
 \langle \sigma_2^{\dagger} \sigma_2 \rangle \\
  \langle \sigma_1^{\dagger} \sigma_2 \rangle  \\
  \langle \sigma_2^{\dagger} \sigma_1 \rangle  \\
  \langle \sigma_1^{\dagger} \sigma_1 \sigma_2^{\dagger} \sigma_2 \rangle
 \end{pmatrix}, 
 \quad  \mathbf{P} =
 \begin{pmatrix}
 P_1 \\
 P_2 \\
 0 \\
 0 \\
 0
 \end{pmatrix}. 
 \end{equation}
In Eq.~\eqref{eq:of_motion}, the one-time regression matrix is given by
\begin{equation}
\label{eq:m0matrix}
\mathbf{M} =
\begin{pmatrix}
  \Gamma_{1} & 0 & \tilde{g}_{+}  & \tilde{g}_{+}^{\ast} & 0 \\
  0 & \Gamma_{2} & \tilde{g}_{-} & \tilde{g}_{-}^{\ast} & 0 \\
  \tilde{g}_{-}^{\ast} & \tilde{g}_{+}^{\ast} & 2 \Gamma_{+} & 0 & - 2 \gamma \mathrm{e}^{ -\mathrm{i} \phi} \\
  \tilde{g}_{-} & \tilde{g}_{+} & 0 & 2 \Gamma_{+} & - 2 \gamma \mathrm{e}^{ \mathrm{i} \phi} \\
  -P_2 & -P_1 & 0 & 0 & 4 \Gamma_{+}
 \end{pmatrix},
 \end{equation}
where $\Gamma_{1, 2}$ and $\Gamma_{+}$ are given by Eq.~\eqref{eq:m0matrix233}, and the generalized coupling constants are defined via
 \begin{equation}
\label{eq:m0matrsdsdix}
\tilde{g}_{\pm} = \pm \mathrm{i} g \mathrm{e}^{\mathrm{i} \theta} + \tfrac{1}{2} \gamma \mathrm{e}^{\mathrm{i} \phi}.
 \end{equation}
In the steady state ($\mathrm{ss}$), we directly obtain five quantities from the derived equation of motion Eq.~\eqref{eq:of_motion} via the formal solution
\begin{equation}
 \label{eq:timer}
 \mathbf{u}_{\mathrm{ss}} = \mathbf{M}^{-1} \mathbf{P}.
\end{equation}
Namely, we find: the probabilities of having the first and second 2LSs excited, $n_{1} = \langle \sigma_{1}^{\dagger} \sigma_{1} \rangle_{\mathrm{ss}}$ and $n_{2} = \langle \sigma_{2}^{\dagger} \sigma_{2} \rangle_{\mathrm{ss}}$ respectively; the coherence between the two 2LSs, $n_{12} = \langle \sigma_1^{\dagger} \sigma_2 \rangle_{\mathrm{ss}}$ and $n_{21} = n_{12}^{\ast}$; and the joint probability that both 2LSs are excited, $n_{\mathrm{X}} = \langle \sigma_1^{\dagger} \sigma_1 \sigma_2^{\dagger} \sigma_2 \rangle_{\mathrm{ss}}$. Indirectly, we also have access to the probabilities of having only 2LS-1 or 2LS-2 excited, $\rho_{1, 0} = n_{1} - n_{\mathrm{X}}$ and $\rho_{0, 1} = n_{2} - n_{\mathrm{X}}$; and the population of the ground state with zero excitations, $\rho_{0, 0}  = 1 + n_{\mathrm{X}} - n_1 - n_2$. Of course, unitarity is always observed since $\rho_{0, 0} + \rho_{1, 0} + \rho_{0, 1} + \rho_{1, 1} = 1$, where $\rho_{1, 1} = n_{\mathrm{X}}$.

The most general solutions from Eq.~\eqref{eq:timer} are given by Eq.~\eqref{eq:statespopgen} in the main text. We now go on to investigate the aforementioned steady state populations for several limiting cases, namely for: coherent coupling [Sec.~\ref{direct123}], dissipative coupling [Sec.~\ref{diss123}], and chiral coupling [Sec.~\ref{reciprocal2}].


\subsection{\label{direct123}Coherent coupling}

With the coherent ($\mathrm{co}$) coupling parameters of Eq.~\eqref{eq:limitingcases1}, we obtain from Eq.~\eqref{eq:timer} the following simple expressions
\begin{subequations}
\label{eq:populations2a}
\begin{align}
 n_1^{\mathrm{co}} &= \frac{P_1 \Gamma_2 + g^2 \frac{P_+}{\Gamma_+} }{\Gamma_1 \Gamma_2 +  4 g^2 }, \label{eq:del1} \\
 n_2^{\mathrm{co}} &= \frac{P_2 \Gamma_1 + g^2 \frac{P_+}{\Gamma_+} }{\Gamma_1 \Gamma_2 +  4 g^2 }, \label{eq:del2} \\
 n_{12}^{\mathrm{co}} &= \frac{ \mathrm{i} g \mathrm{e}^{-\mathrm{i} \theta}}{2 \Gamma_+} \frac{ \Gamma_1 P_2 - \Gamma_2 P_1 }{ \Gamma_1 \Gamma_2 + 4 g^2},  \label{eq:del3} \\
 n_{\mathrm{X}}^{\mathrm{co}} &= \frac{P_1 P_2 + \left( \frac{g P_+}{2\Gamma_+} \right)^2 }{\Gamma_1 \Gamma_2 + 4 g^2 }. \label{eq:del4} 
 \end{align}
\end{subequations}
When one substitutes $\theta = 0$ into the complex-valued coherence of Eq.~\eqref{eq:del3}, one recovers Eqs.~(14, 15) of Ref.~\cite{delValle2010}, where real-valued coupling parameters were considered. Furthermore, dividing Eq.~\eqref{eq:del4} by Eq.~\eqref{eq:del1} and Eq.~\eqref{eq:del2} yields the cross-correlator given as Eq.~\eqref{eq:geetwo} in the main text.


\subsection{\label{diss123}Dissipative coupling}

When the dissipative ($\mathrm{ds}$) coupling parameters of Eq.~\eqref{eq:limitingcases2} are fulfilled, we obtain from Eq.~\eqref{eq:timer} the compact expressions
\begin{subequations}
\label{eq:populationsd343434issi}
\begin{align}
 n_1^{\mathrm{ds}} &= \frac{ P_- \gamma^2 \left( P_1 - \Gamma_+ \right) + 4 P_1 \Gamma_2 \Gamma_+^2 }{\gamma^2 \left( P_1 \Gamma_1 + P_2 \Gamma_2 \right) + 4 \Gamma_+^2 \left( \Gamma_1 \Gamma_2 - \gamma^2 \right)  }, \label{eq:kjk1}  \\
 n_2^{\mathrm{ds}} &= \frac{ P_- \gamma^2 \left(  \Gamma_+ - P_2 \right) + 4 P_2 \Gamma_1 \Gamma_+^2 }{\gamma^2 \left( P_1 \Gamma_1 + P_2 \Gamma_2 \right) + 4 \Gamma_+^2 \left( \Gamma_1 \Gamma_2 - \gamma^2 \right)  }, \label{eq:kjk2} \\
 n_{12}^{\mathrm{ds}} &= \frac{ \Gamma_+ \gamma \mathrm{e}^{-\mathrm{i} \phi}  \left( 4 P_1 P_2 - P_1 \Gamma_2 - P_2 \Gamma_1 \right) }{\gamma^2 \left( P_1 \Gamma_1 + P_2 \Gamma_2 \right) + 4 \Gamma_+^2 \left( \Gamma_1 \Gamma_2 - \gamma^2 \right)  }, \label{eq:kjk3} \\
 n_{\mathrm{X}}^{\mathrm{ds}} &= \frac{ 4 P_1 P_2 \Gamma_+^2 + \left( \frac{\gamma P_-}{2} \right)^2 }{\gamma^2 \left( P_1 \Gamma_1 + P_2 \Gamma_2 \right) + 4 \Gamma_+^2 \left( \Gamma_1 \Gamma_2 - \gamma^2 \right)  }. \label{eq:kjk4}
 \end{align}
\end{subequations}
The cross-correlator Eq.~\eqref{eq:geetwodiss} in the main text is found by dividing Eq.~\eqref{eq:kjk4} by Eq.~\eqref{eq:kjk1} and Eq.~\eqref{eq:kjk2}.


\subsection{\label{reciprocal2}Chiral coupling}

At the chiral (ch) parameters, we obtain for case $\mathrm{I}$ [cf. Eq.~\eqref{eq:limitingcases3}] the following expressions from Eq.~\eqref{eq:timer}
\begin{subequations}
\label{eq:populations3bb}
\begin{align}
 n_1^{\mathrm{ch, I}} &= \frac{P_1}{\Gamma_1}, \label{eq:cdr1} \\
 n_2^{\mathrm{ch, I}} &= \frac{2 P_2 \Gamma_+^2 - P_1 \gamma^2 \left( P_2 - 2 \Gamma_+ \right) / \Gamma_1}{P_1 \gamma^2 + 2 \Gamma_2 \Gamma_+^2}, \label{eq:cdr2} \\
 n_{12}^{\mathrm{ch, I}} &= \frac{ \mathrm{e}^{-\mathrm{i} \phi} \gamma P_1 \Gamma_+ }{\Gamma_1} \frac{ 2 P_2 - \Gamma_2 }{P_1 \gamma^2 + 2 \Gamma_2 \Gamma_+^2}, \label{eq:cdr3}  \\
 n_{\mathrm{X}}^{\mathrm{ch, I}} &= \frac{P_1}{2 \Gamma_1} \frac{ P_1 \gamma^2 + 4 P_2 \Gamma_+^2 }{P_1 \gamma^2 + 2 \Gamma_2 \Gamma_+^2}. \label{eq:cdr4} 
 \end{align}
\end{subequations}
Most importantly, Eq.~\eqref{eq:cdr1} showcases that the population of the first 2LS is identical to that of a single 2LS in isolation [cf. Eq.~\eqref{eq:iso}], a hallmark of chiral coupling. The second 2LS population in Eq.~\eqref{eq:cdr2} is in general enhanced due to the one-way nature of the coupling in favor of 2LS-2 [see Fig.~\ref{fig:landscape}~(c)]. Dividing Eq.~\eqref{eq:cdr4} by Eq.~\eqref{eq:cdr1} and Eq.~\eqref{eq:cdr2} yields the normalized cross-correlator Eq.~\eqref{eq:geetwouni} of the main text. 

The expressions for case $\mathrm{II}$ [cf. Eq.~\eqref{eq:limitingcases4}] are found by interchanging the indices ($1 \rightleftharpoons 2$) everywhere, such that it is of course also possible to have one-way coupling in the opposite direction, characterized by $n_2^{\mathrm{ch, II}} = P_2/\Gamma_2$.


\section{Two-time dynamics}
\label{app:Double}

In this appendix, we calculate various two-time correlators of interest, which gives us access to both the power spectrum of the coupled system and its underlying structure (via its spectral decomposition). We use the same theoretical framework as in Refs.~\cite{Valle208, Degenfeld2012, Valle211}. Our modus operandi is underpinned by the quantum regression theorem~\cite{Gardiner2004, Gardiner2014}, in the same manner as Appendix~\ref{app:Single}.

Similar to Ref.~\cite{delValle2010}, we focus on the spectrum $S_1(\omega)$ of the first emitter 2LS-1, since all the expressions for 2LS-2 may be found by natural interchanges of $1$ and $2$. Furthermore, the theory may be generalized to analyze other modes of interest~\cite{Downing2019}. The equation of motion for the pertinent correlators reads
\begin{equation}
\label{eq:populations2322}
\frac{\partial}{\partial \tau} \mathbf{v} (t, t + \tau) + \mathbf{Q} \mathbf{v} (t, t + \tau) = 0,
\end{equation}
where the correlators are contained in the 4-vector
\begin{equation}
\label{eq:vmatrix}
\mathbf{v}  =
\begin{pmatrix}
  \langle \sigma_1^{\dagger} (t)~\sigma_1 (t + \tau ) \rangle \\
 \langle \sigma_1^{\dagger} (t)~\sigma_2 (t + \tau ) \rangle  \\
 \langle \sigma_1^{\dagger} (t)~\sigma_1^{\dagger} \sigma_1 \sigma_2 (t + \tau ) \rangle \\
 \langle \sigma_1^{\dagger} (t)~\sigma_1 \sigma_2^{\dagger} \sigma_2 (t + \tau ) \rangle
 \end{pmatrix}. 
 \end{equation}
The two-time regression matrix in Eq.~\eqref{eq:populations2322} reads
\begin{widetext}
\begin{equation}
\label{eq:m1matrix}
\mathbf{Q} =
\begin{pmatrix}
  \mathrm{i} \omega_0 + \tfrac{1}{2} \Gamma_{1} & \tilde{g}_{+} & - 2 \tilde{g}_{+}  & 0  \\
  \tilde{g}_{-}^{\ast} & \mathrm{i} \omega_0 + \tfrac{1}{2} \Gamma_{2} & 0  & - 2 \tilde{g}_{-}^{\ast} \\
  0 & -P_1 & \mathrm{i} \omega_0 + \Gamma_{1} + \tfrac{1}{2} \Gamma_{2}  & \tilde{g}_{+}^{\ast} \\
 -P_2 & 0 & \tilde{g}_{-}  & \mathrm{i} \omega_0 + \Gamma_{2}  + \tfrac{1}{2} \Gamma_{1}
 \end{pmatrix},
 \end{equation}
 \end{widetext}
where the effective broadenings $\Gamma_{1}$ and $\Gamma_{2}$ and coupling constants $\tilde{g}_{\pm}$ are given by Eq.~\eqref{eq:m0matrix233} and Eq.~\eqref{eq:m0matrsdsdix} respectively. The exact solution of Eq.~\eqref{eq:populations2322} reads
\begin{equation}
\label{eq:populatwewewewions2322}
\mathbf{v} (t, t + \tau) = \sum_{p=\mathrm{A}, \mathrm{B}, \mathrm{C}, \mathrm{D}} c_p \mathbf{v}_p^E \mathrm{e}^{-\left(\mathrm{i} \omega_p + \gamma_p /2 \right) \tau} ,
\end{equation}
where the $p$-th complex eigenvalue of $-\mathbf{Q}$ is $\lambda_p$, with associated eigenvector $\mathbf{v}_{p}^E$. The complex eigenfrequencies $\lambda_p$ may be decomposed as the damping rates $\gamma_p = - 2 \mathrm{Re} \left( \lambda_{p} \right)$ and the frequency shifts $\omega_p = - \mathrm{Im} \left( \lambda_{p} \right)$, producing the exponent in Eq.~\eqref{eq:populatwewewewions2322}. The four constants $c_p$ are obtained from the boundary conditions $\sum_p \mathbf{v}_p^E c_p = \left(n_1, n_{12}, 0, n_{\mathrm{X}} \right)^{\mathrm{T}}$, where the required steady state expressions $n_i$ are given in Eq.~\eqref{eq:statespopgen}.

With regard to the optical spectrum decomposition of Eq.~\eqref{eq:spectrumao}, the coefficients $L_p$ and $K_p$ may be found via the relation $L_p + \mathrm{i} K_p = c_p \mathbf{v}_p^E [1]/ n_1$, where $\mathbf{x}[1]$ refers to the first element of the column vector $\mathbf{x}$, and $n_1$ is given in its most general form by Eq.~\eqref{eq:mike1}. In the various limiting cases we have focused on throughout, the eigenvalues of Eq.~\eqref{eq:m1matrix} are given by Eqs.~\eqref{eq:nreigen2sdwewewewewwwwsd222},~\eqref{eq:nreigen2ssasssssssdsdsdsd222},~\eqref{eq:nreigedsdsdsdn2222},~and~\eqref{eq:nraaaaaaaaaeigen222sds2} in the main text.


\section{Supplementary results for the mean populations in the dissipative coupling regime}
\label{app:additional_plot}

\begin{figure*}[tb]
 \includegraphics[width=\linewidth]{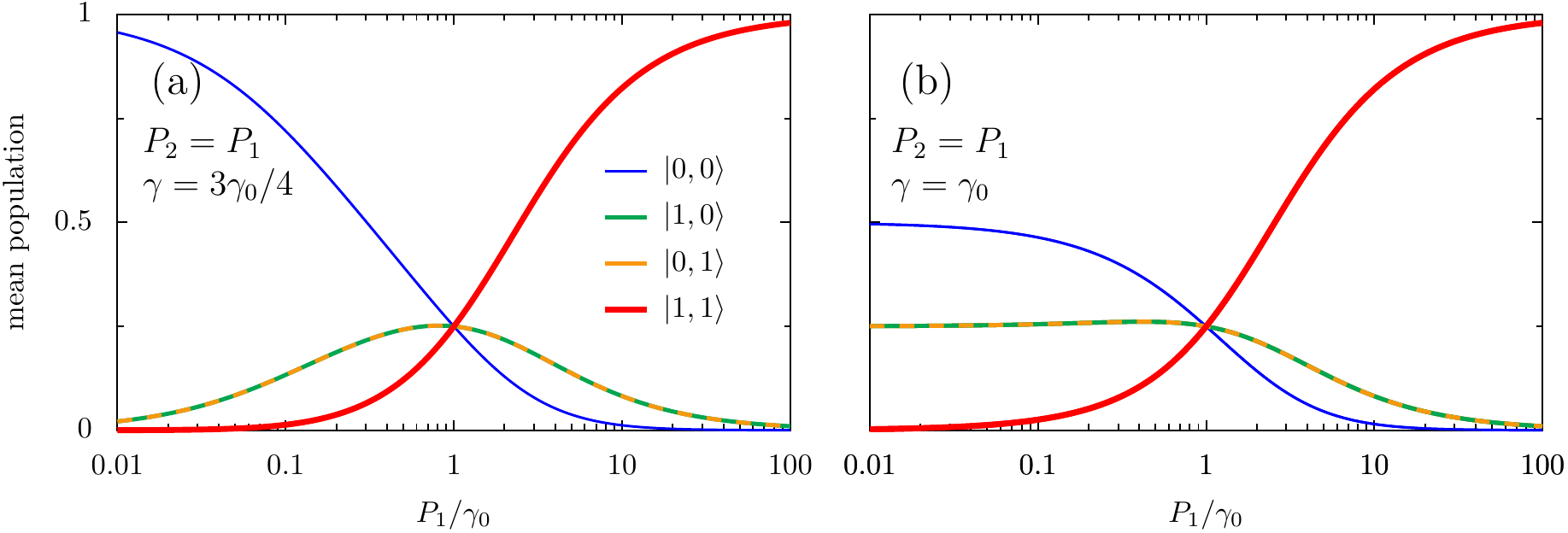}
 \caption{ Mean populations in the dissipative coupling regime, as a function of the pumping rate $P_1$, in units of the damping rate $\gamma_0$ [cf. Eq.~\eqref{eq:statespopdiss}]. Panel (a): the dissipative coupling strength $\gamma = 3\gamma_0/4$. Panel (b): maximal coupling, $\gamma = \gamma_0$. The labeling of the mean population of the state $\ket{i, j}$ is displayed in the legend of panel (a), and states with $N=\{ 0, 1, 2\}$ excitations are shown with increasingly thick lines. In the figure, we consider symmetric pumping ($P_2 = P_1$).}
 \label{fig:popdissapp}
\end{figure*}

In Sec.~\ref{SECdissPOP} of the main text, we noted that the results for the mean populations in the dissipative coupling regime with symmetric pumping ($P_2 = P_1$) are not surprising, at least once the asymmetrically pumped results are known [cf. Fig.~\ref{fig:popdiss}].

We show explicitly the symmetric pumping configuration results in Fig.~\ref{fig:popdissapp}, with high [maximal] dissipative coupling $\gamma = 3\gamma_0/4$ [$\gamma = \gamma_0$] displayed in panel (a) [(b)]. Most importantly, there is a population trapping effect in panel (b), in the limit of vanishing pumping, in exactly the same manner as in Fig.~\ref{fig:popdiss}~(b) of the main text. Broadly, the key features in Fig.~\ref{fig:popdissapp} are simply symmetrized analogues of the behavior shown in Fig.~\ref{fig:popdiss}.

\section{The first moments of the dimer}
\label{app:first_mom}

The calculation of Appendix~\ref{app:Double} may be used to immediately write down the equation of motion for the first moments of the dimer, which are encapsulated by the equation
\begin{equation}
\label{eqapp:of_motion_io}
\mathrm{i} \partial_t \hat{\psi} = \mathcal{H} \hat{\psi}, \quad
\end{equation}
where the first moments are contained within the four-dimensional object [cf. Eq.~\eqref{eq:vmatrix}]
\begin{equation}
\label{eqapp:of_motion_iwewo}
\hat{\psi} =
\begin{pmatrix}
  \langle \sigma_1 \rangle  \\
  \langle \sigma_2 \rangle  \\
  \langle \sigma_1^\dagger \sigma_1 \sigma_2 \rangle \\
  \langle \sigma_1 \sigma_2^\dagger \sigma_2  \rangle
 \end{pmatrix}.
\end{equation}
The first moments matrix $\mathcal{H}$ in Eq.~\eqref{eqapp:of_motion_io} reads [cf. Eq.~\eqref{eq:m1matrix}]
\begin{widetext}
\begin{equation}
\label{eqapp:of_motion_io_2}
\mathcal{H} = \begin{pmatrix}
   \omega_0 - \mathrm{i} \frac{\Gamma_1}{2} && -\mathrm{i} \tilde{g}_{+} && 2 \mathrm{i} \tilde{g}_{+} && 0  \\
  - \mathrm{i} \tilde{g}_{-}^\ast && \omega_0 -\mathrm{i} \frac{\Gamma_2}{2} && 0 && 2 \mathrm{i} \tilde{g}_{-}^\ast  \\
  0 && \mathrm{i} P_1 && \omega_0 - \mathrm{i} \left( \Gamma_1 + \frac{\Gamma_2}{2} \right) && -\mathrm{i} \tilde{g}_{+}^\ast \\
   \mathrm{i} P_2 && 0 && -\mathrm{i} \tilde{g}_{-} && \omega_0 - \mathrm{i} \left( \Gamma_2 + \frac{\Gamma_1}{2} \right)
 \end{pmatrix},
\end{equation}
\end{widetext}
where the generalized coupling constants $\tilde{g}_{\pm}$ are defined in Eq.~\eqref{eq:m0matrsdsdix}, and the generalized damping rates $\Gamma_1$ and $\Gamma_2$ by Eq.~\eqref{eq:m0matrix233}. The Hamiltonian-like matrix $\mathcal{H}$ is in general non-Hermitian, and therefore has four complex eigenvalues. In the limit of vanishing pumping ($P_1, P_2 \to 0$), the four eigenvalues $\epsilon_i$ of Eq.~\eqref{eqapp:of_motion_io_2} are given by 
\begin{widetext}
\begin{align}
\label{eqapp:of_motion_io_2sdsd}
 \epsilon_1 &=  \omega_0 + \sqrt{g^2 - \left( \tfrac{\gamma}{2} \right)^2 + \mathrm{i} g \gamma \cos \left( \theta - \phi \right) } + \mathrm{i} \frac{3\gamma_{0}}{2}, \\
  \epsilon_2 &=  \omega_0 - \sqrt{g^2 - \left( \tfrac{\gamma}{2} \right)^2 + \mathrm{i} g \gamma \cos \left( \theta - \phi \right) } + \mathrm{i} \frac{3\gamma_{0}}{2}, \\
   \epsilon_3 &=  \omega_0 + \sqrt{g^2 - \left( \tfrac{\gamma}{2} \right)^2 - \mathrm{i} g \gamma \cos \left( \theta - \phi \right) } + \mathrm{i} \frac{\gamma_{0}}{2} , \\
    \epsilon_4 &=  \omega_0 - \sqrt{g^2 - \left( \tfrac{\gamma}{2} \right)^2 - \mathrm{i} g \gamma \cos \left( \theta - \phi \right) } + \mathrm{i} \frac{\gamma_{0}}{2}. \label{eqapp:of_motion_io_2sdsdP}
\end{align}
\end{widetext}
The eigenenergies $\epsilon_i$ of Eq.~\eqref{eqapp:of_motion_io_2sdsd}-\eqref{eqapp:of_motion_io_2sdsdP} may be readily split into real and imaginary parts, which allows for their interpretation as energy levels and inverse lifetimes, as was done in the spectral calculation of Appendix~\ref{app:Double}, and its associated result of Eq.~\eqref{eq:nraaaaaaaaaeigen222sds2}.

Equations~\eqref{eqapp:of_motion_io_2sdsd}-\eqref{eqapp:of_motion_io_2sdsdP} highlight how the interplay of the coherent coupling $g$ and dissipative coupling $\gamma$ crucially determines fundamental properties of the system, and how their relative phase difference $\theta-\phi$ is of the utmost importance. 


\section{Cross-Kerr interactions}
\label{app:interactions}

In what follows, we briefly account for the introduction of an interaction term between the two quantum emitters via cross-Kerr coupling.

We supplement the Hamiltonian of Eq.~\eqref{eq:ham0} with the interaction term $H_{\mathrm{I}}$, so that it reads $H = H_0 + H_{\mathrm{c}}+H_{\mathrm{I}}$, where
\begin{equation}
\label{eq:ham_int}
 H_{\mathrm{I}} = -\chi \sigma_1^{\dagger} \sigma_1 \sigma_2^{\dagger} \sigma_2,
\end{equation}
where $\chi$ is the non-linear cross-Kerr frequency. This interaction leads to a renormalization of the doubly-occupied energy level from $\omega_{\mathrm{X}} = 2 \omega_0$ [cf. Eq.~\eqref{eq:eigenfrequencies}] to 
\begin{equation}
\label{eq:ham_int2}
 \omega_{\mathrm{X}} = 2 \omega_0 - \chi.
\end{equation}
Equation~\eqref{eq:ham_int2} breaks the symmetry of the energy ladder about $\omega_0$, the most symmetric case with $\chi = 0$ is sketched in Fig.~\ref{fig:sketch}~(a).

The mean populations and the cross-correlator are unaffected by the extra term of Eq.~\eqref{eq:ham_int}, since the cross-Kerr coupling $\chi$ does not enter the matrix $\mathbf{M}$ in the one-time equation of motion given by Eq.~\eqref{eq:of_motion}. However, the optical spectrum is influenced by $\chi$. Generalized to account for interactions, the two-time equation of motion of Eq.~\eqref{eq:populations2322} sees four additions to the matrix $\mathbf{Q}$. Explicitly, four matrix elements $\mathbf{Q}_{i,j}$ in Eq.~\eqref{eq:m1matrix} need to be updated: $\mathbf{Q}_{1,4} \to \mathbf{Q}_{1,4} - \mathrm{i} \chi$, $\mathbf{Q}_{2,3} \to \mathbf{Q}_{2,3} - \mathrm{i} \chi$, $\mathbf{Q}_{3,3} \to \mathbf{Q}_{3,3} - \mathrm{i} \chi$ and $\mathbf{Q}_{4,4} \to \mathbf{Q}_{4,4} - \mathrm{i} \chi$. The resulting eigenvalues of $\mathbf{Q}$ yield the following frequencies $\omega_p$ and broadenings $\gamma_p$ which shape the spectrum [cf. Eq.~\eqref{eq:nraaaaaaaaaeigen222sds2}] 
\begin{subequations}
\label{eq:222gvhvh2}
\begin{alignat}{2}
 \tfrac{1}{2} \gamma_{\mathrm{A}} + \mathrm{i} \omega_{\mathrm{A}} &= \tfrac{3}{2} \gamma_{0} + \mathrm{i} \left( \omega_0 - \chi + \Omega^{\ast} \right), \\
  \tfrac{1}{2} \gamma_{\mathrm{B}} + \mathrm{i} \omega_{\mathrm{B}} &= \tfrac{3}{2} \gamma_{0} + \mathrm{i} \left( \omega_0 - \chi - \Omega^{\ast} \right),  \\
   \tfrac{1}{2} \gamma_{\mathrm{C}} + \mathrm{i} \omega_{\mathrm{C}} &= \tfrac{1}{2} \gamma_{0} + \mathrm{i} \left( \omega_0 + \Omega \right) ,  \\
    \tfrac{1}{2} \gamma_{\mathrm{D}} + \mathrm{i} \omega_{\mathrm{D}} &= \tfrac{1}{2} \gamma_{0} + \mathrm{i} \left( \omega_0 - \Omega \right),
  \end{alignat}
\end{subequations}
where $\Omega$ is defined in Eq.~\eqref{eq:complexfrequency}. Notably, the effect of interactions is only felt through the replacement $\omega_0 \to \omega_0 - \chi$ in the frequency shifts associated with the labels $\mathrm{A}$ and $\mathrm{B}$ in Eq.~\eqref{eq:222gvhvh2}, as follows from Eq.~\eqref{eq:ham_int2}. Since these contributions to the spectrum describe optical transitions from the doubly-excited level $\omega_{\mathrm{X}}$, those which are unpopulated in the vanishing pumping limit we consider, we can safely neglect interactions of the form of Eq.~\eqref{eq:ham_int} in the main text without loss of generality.



\begin{thebibliography}{100}


\bibitem{Wainer2009}
W.~J.~Wainer and I.~Lough, 
\textit{Chirality in Natural and Applied Science} (Blackwell, Oxford, 2002).

\bibitem{Wagniere2007}
G.~H.~Wagniere, 
\textit{On Chirality and the Universal Asymmetry} (Wiley, Weinheim, 2007).

\bibitem{Guijarro2009}
A.~Guijarro and M.~Yus, 
\textit{The Origin of Chirality in the Molecules of Life} (Royal Society of Chemistry, Cambridge, 2009).

\bibitem{Lodahl2017}
P.~Lodahl, S.~Mahmoodian, S.~Stobbe, A.~Rauschenbeutel, P.~Schneeweiss, J.~Volz, H.~Pichler and P.~Zoller,
Chiral quantum optics,
\href{https://doi.org/10.1038/nature21037}
{Nature \textbf{541}, 473 (2017)}.

\bibitem{Andrews2018}
D.~L.~Andrews,
Quantum formulation for nanoscale optical and material chirality: Symmetry issues, space and time parity, and observables,
\href{https://doi.org/10.1088/2040-8986/aaaa56}
{J. Opt. \textbf{20}, 033003 (2018)}.

\bibitem{Chang2018}
D.~E.~Chang, J.~S.~Douglas, A.~Gonzalez-Tudela, C.-L.~Hung, and H.~J.~Kimble,
Colloquium: Quantum matter built from nanoscopic lattices of atoms and photons,
\href{https://doi.org/10.1103/RevModPhys.90.031002}
{Rev. Mod. Phys. \textbf{90}, 031002 (2018)}.



\bibitem{Gardiner1993}
C.~W.~Gardiner,
Driving a quantum system with the output field from another driven quantum system,
\href{https://doi.org/10.1103/PhysRevLett.70.2269}
{Phys. Rev. Lett. \textbf{70}, 2269 (1993)}.

\bibitem{Carmichael1993}
H.~J.~Carmichael,
Quantum trajectory theory for cascaded open systems,
\href{https://doi.org/10.1103/PhysRevLett.70.2273}
{Phys. Rev. Lett. \textbf{70}, 2273 (1993)}.

\bibitem{Gardiner2004}
C.~Gardiner and P.~Zoller,
\textit{Quantum Noise} (Springer, Berlin, 2004).

\bibitem{Shomroni2014}
I.~Shomroni, S.~Rosenblum, Y.~Lovsky, O.~Bechler, G.~Guendelman, and B.~Dayan,
All---optical routing of single photons by a one-atom switch controlled by a single photon,
\href{https://doi.org/10.1126/science.1254699}
{Science \textbf{345}, 903 (2014)}.


\bibitem{Yan2018}
W.-B.~Yan, W.-Y.~Ni, J.~Zhang, F.-Y.~Zhang, and H.~Fan,
Tunable single-photon diode by chiral quantum physics,
\href{https://doi.org/10.1103/PhysRevA.98.043852}
{Phys. Rev. A \textbf{98}, 043852 (2018)}.

\bibitem{Tang2019}
L.~Tang, J.~Tang, W.~Zhang, G.~Lu, H.~Zhang, Y.~Zhang, K.~Xia, and M.~Xiao,
On-chip chiral single-photon interface: Isolation and unidirectional emission,
\href{https://doi.org/10.1103/PhysRevA.99.043833}
{Phys. Rev. A \textbf{99}, 043833 (2019)}.


\bibitem{Sayrin2015}
C.~Sayrin, C.~Junge, R.~Mitsch, B.~Albrecht, D.~O'Shea, P.~Schneeweiss, J.~Volz, and A.~Rauschenbeutel,
Nanophotonic optical isolator controlled by the internal state of cold atoms,
\href{https://doi.org/10.1103/PhysRevX.5.041036}
{Phys. Rev. X \textbf{5}, 041036 (2015)}.


\bibitem{Sollner2015}
I.~Sollner, S.~Mahmoodian, S.~L.~Hansen, L.~Midolo, A.~Javadi, G.~Kirsanske, T.~Pregnolato, H.~El-Ella, E.~H.~Lee, J.~D.~Song, S.~Stobbe and 
P.~Lodahl,
Deterministic photon---emitter coupling in chiral photonic circuits,
\href{https://doi.org/10.1038/nnano.2015.159}
{Nature Nanotechnology \textbf{10}, 775 (2015)}.


\bibitem{Scheucher2016}
M.~Scheucher, A.~Hilico, E.~Will, J.~Volz, and A.~Rauschenbeutel,
Quantum optical circulator controlled by a single chirally coupled atom,
\href{https://doi.org/10.1126/science.aaj2118}
{Science \textbf{354}, 1577 (2016)}.


\bibitem{Barik2018}
S.~Barik, A.~Karasahin, C.~Flower, T.~Cai, H.~Miyake, W.~DeGottardi, M.~Hafezi, and E.~Waks,
A topological quantum optics interface,
\href{https://doi.org/10.1126/science.aaq0327}
{Science \textbf{359}, 666 (2018)}.


\bibitem{Zhang2018}
S.~Zhang, Y.~Hu, G.~Lin, Y.~Niu, K.~Xia, J.~Gong and S.~Gong, 
Thermal-motion-induced non-reciprocal quantum optical system,
\href{https://doi.org/10.1038/s41566-018-0269-2}
{Nature Photonics \textbf{12}, 744 (2018)}.

\bibitem{Fang2019}
L.~Fang, H.-Z.~Luo, X.-P.~Cao, S.~Zheng, X.-L.~Cai, and J.~Wang, 
Ultra-directional high-efficiency chiral silicon photonic circuits,
\href{https://doi.org/10.1364/OPTICA.6.000061}
{Optica \textbf{6}, 61 (2019)}.


\bibitem{Barik2019}
S.~Barik, A.~Karasahin, S.~Mittal, E.~Waks, and M.~Hafezi,
Chiral quantum optics using a topological resonator,
\href{https://doi.org/10.1103/PhysRevB.101.205303}
{Phys. Rev. B \textbf{101}, 205303 (2020)}.


\bibitem{Li2018}
T.~Li, A.~Miranowicz, X.~Hu, K.~Xia, and F.~Nori,
Quantum memory and gates using a $\Lambda$-type quantum emitter coupled to a chiral waveguide,
\href{https://doi.org/10.1103/PhysRevA.97.062318}
{Phys. Rev. A \textbf{97}, 062318 (2018)}.


\bibitem{Grankin2018}
A.~Grankin, P.~O.~Guimond, D.~V.~Vasilyev, B.~Vermersch, and P.~Zoller,
Free-space photonic quantum link and chiral quantum optics,
\href{https://doi.org/10.1103/PhysRevA.98.043825}
{Phys. Rev. A \textbf{98}, 043825 (2018)}.

\bibitem{Yan2018b}
C.-H.~Yan, Y.~Li, H.~Yuan, and L.~F.~Wei,
Targeted photonic routers with chiral photon-atom interactions,
\href{https://doi.org/10.1103/PhysRevA.98.043825}
{Phys. Rev. A \textbf{97}, 023821 (2018)}.

\bibitem{Zhang2019}
F.~Zhang, J.~Ren, L.~Shan, X.~Duan, Y.~Li, T.~Zhang, Q.~Gong, and Y.~Gu,
Chiral cavity quantum electrodynamics with coupled nanophotonic structures,
\href{https://doi.org/10.1103/PhysRevA.100.053841}
{Phys. Rev. A \textbf{100}, 053841 (2019)}.


\bibitem{Stannigel2012}
K.~Stannigel, P.~Rabl and P.~Zoller,
Driven-dissipative preparation of entangled states in cascaded quantum-optical networks,
\href{https://doi.org/10.1088/1367-2630/14/6/063014}
{New J. Phys. \textbf{14}, 063014 (2012)}.

\bibitem{Ramos2014}
T.~Ramos, H.~Pichler, A.~J.~Daley, and P.~Zoller,
Quantum spin dimers from chiral dissipation in cold-atom chains,
\href{https://doi.org/10.1103/PhysRevLett.113.237203}
{Phys. Rev. Lett. \textbf{113}, 237203 (2014)}.

\bibitem{Chervy2018}
T.~Chervy, S.~Azzini, E.~Lorchat, S.~Wang, Y.~Gorodetski, J.~A.~Hutchison, S.~Berciaud, T.~W.~Ebbesen, and C.~Genet,
Room temperature chiral coupling of valley excitons with spin-momentum locked surface plasmons,
\href{https://doi.org/10.1021/acsphotonics.7b01032}
{ACS Photonics \textbf{5},1281 (2018)}.

\bibitem{Pichler2015}
H.~Pichler, T.~Ramos, A.~J.~Daley, and P.~Zoller,
Quantum optics of chiral spin networks,
\href{https://doi.org/10.1103/PhysRevA.91.042116}
{Phys. Rev. A \textbf{91}, 042116 (2015)}.


\bibitem{Metelmann2015}
A.~Metelmann and A.~A.~Clerk,
Nonreciprocal photon transmission and amplification via reservoir engineering,
\href{https://doi.org/10.1103/PhysRevX.5.021025}
{Phys. Rev. X \textbf{5}, 021025 (2015)}.

\bibitem{Burillo2019}
E.~Sanchez-Burillo, C.~Wan, D.~Zueco, and A.~Gonzalez-Tudela,
Chiral quantum optics in photonic sawtooth lattices,
\href{https://doi.org/10.1103/PhysRevResearch.2.023003}
{Phys. Rev. Research \textbf{2}, 023003 (2020)}.


\bibitem{Ballestero2015}
C.~Gonzalez-Ballestero, A.~Gonzalez-Tudela, F.~J.~Garcia-Vidal, and E.~Moreno,
Chiral route to spontaneous entanglement generation,
\href{https://doi.org/10.1103/PhysRevB.92.155304}
{Phys. Rev. B \textbf{92}, 155304 (2015)}.

\bibitem{Gonzalez2016}
C.~Gonzalez-Ballestero, E.~Moreno, F.~J.~Garcia-Vidal, and A.~Gonzalez-Tudela,
Nonreciprocal few-photon routing schemes based on chiral waveguide-emitter couplings,
\href{https://doi.org/10.1103/PhysRevA.97.062318}
{Phys. Rev. A \textbf{94}, 063817 (2016)}.

\bibitem{Martin2018}
D.~Martin-Cano, H.~R.~Haakh, and N.~Rotenberg,
Chiral emission into nanophotonic resonators,
\href{https://doi.org/10.1021/acsphotonics.8b01555}
{ACS Photonics \textbf{6}, 961, (2019)}.

\bibitem{Jalali2020}
M.~Jalali~Mehrabad, A.~P.~Foster, R.~Dost, A.~M.~Fox, M.~S.~Skolnick, and L.~R.~Wilson,
Chiral topological photonics with an embedded quantum emitter,
\href{https://arxiv.org/abs/1912.09943}
{arXiv:1912.09943}.


\bibitem{Peterson2018}
C.~W.~Peterson, S.~Kim, J.~T.~Bernhard and G.~Bahl,
Synthetic phonons enable nonreciprocal coupling to arbitrary resonator networks,
\href{https://doi.org/10.1126/sciadv.aat0232}
{Science Advances \textbf{4}, 6, (2018)}.


\bibitem{Picardi2018}
M.~F. Picardi, A.~V.~Zayats, and F.~J.~Rodriguez-Fortuno,
Janus and Huygens dipoles: near-field directionality beyond spin-momentum locking,
\href{https://doi.org/10.1103/PhysRevLett.120.117402}
{Phys. Rev. Lett. \textbf{120}, 117402 (2018)}.

\bibitem{Downing2019}
C.~A.~Downing, J.~C.~L\'{o}pez~Carre\~{n}o, F.~P.~Laussy, E.~del Valle, and A.~I.~Fern\'{a}ndez-Dom\'{i}nguez,
Quasichiral interactions between quantum emitters at the nanoscale,
\href{https://doi.org/10.1103/PhysRevLett.122.057401}
{Phys. Rev. Lett. \textbf{122}, 057401 (2019)}.

\bibitem{Bliokh2015}
K.~Y.~Bliokh, F.~J.~Rodr\'{i}guez-Fortu\~{n}o, F.~Nori, and A.~V.~Zayats,
Spin-orbit interactions of light,
\href{https://doi.org/10.1038/NPHOTON.2015.201}
{Nat. Photonics \textbf{9}, 796 (2015)}.

\bibitem{Schaferling2016}
M.~Schaferling, 
\textit{Chiral Nanophotonics} (Springer, Cham, 2016).

\bibitem{Hentschel2017}
M.~Hentschel, M.~Schaferling, X.~Duan, H,~Giessen and N.~Liu,
Chiral plasmonics,
\href{https://doi.org/10.1126/sciadv.1602735}
{Sci Adv. \textbf{3}, e1602735 (2017)}.

\bibitem{Metelmann2017}
A.~Metelmann and A.~A.~Clerk,
Nonreciprocal quantum interactions and devices via autonomous feedforward,
\href{https://doi.org/10.1103/PhysRevA.95.013837}
{Phys. Rev. A \textbf{95}, 013837 (2017)}.


\bibitem{Lopez2015}
J.~C.~L\'{o}pez~Carre\~{n}o  and F.~P.~Laussy,
Excitation with quantum light. I. Exciting a harmonic oscillator,
\href{https://doi.org/10.1103/PhysRevA.94.063825}
{Phys. Rev. A \textbf{94}, 063825 (2016)}.


\bibitem{Lopez2016}
J.~C.~L\'{o}pez~Carre\~{n}o, C.~S\'{a}nchez Mu\~{n}oz, E.~del~Valle and F.~P.~Laussy,
Excitation with quantum light. II. Exciting a two-level system,
\href{https://doi.org/10.1103/PhysRevA.94.063826}
{Phys. Rev. A \textbf{94}, 063826 (2016)}.



\bibitem{Lopez2018}
J.~C.~L\'{o}pez~Carre\~{n}o, E.~del Valle, and F.~P~ Laussy,
Frequency-resolved Monte Carlo,
\href{https://doi.org/10.1038/s41598-018-24975-y}
{Scientific Reports \textbf{8}, 6975 (2018)}.

\bibitem{Lopez2019}
J.~C.~L\'{o}pez~Carre\~{n}o, E.~Z.~Casalengua, E.~del Valle, and F.~P~ Laussy,
Impact of detuning and dephasing on a laser-corrected subnatural-linewidth single-photon source,
\href{https://doi.org/10.1088/1361-6455/aaf68d}
{J. Phys. B: At. Mol. Opt. Phys. \textbf{52}, 035504 (2019)}.


\bibitem{AllenBook}
L.~Allen, and J.~H.~Eberly,
\textit{Optical Resonance and Two-Level Atoms} (Wiley, New York 1975).



\bibitem{Pashkin2009}
Y.~A.~Pashkin, O.~Astafiev, T.~Yamamoto, Y.~Nakamura, and J.~S.~Tsai,
Josephson charge qubits: a brief review,
\href{https://doi.org/10.1007/s11128-009-0101-5}
{Quant. Info. Proc. \textbf{8}, 55 (2009)}.

\bibitem{Wahiddin1998}
M.~R.~B.~Wahiddin, Z.~Ficek, U.~Akram, K.~T.~Lim, and M.~R.~Muhamad,
Squeezing-induced complete transparency in two-level systems,
\href{https://doi.org/10.1103/PhysRevA.57.2072}
{Phys. Rev. A \textbf{57}, 2072 (1998)}.

\bibitem{Munoz2020}
C.~S\'{a}nchez~Mu\~{n}oz, and F.~Schlawin,
Photon correlation spectroscopy as a witness for quantum coherence,
\href{https://doi.org/10.1103/PhysRevLett.124.203601}
{Phys. Rev. Lett. \textbf{124}, 203601 (2020)}.


\bibitem{Bordo2019}
V.~G.~Bordo,
Quantum plasmonics of metal nanoparticles,
\href{https://doi.org/10.1364/JOSAB.36.000323}
{J. Opt. Soc. Am. B \textbf{36}, 323 (2019)}.



\bibitem{Gonzalez2011}
A.~Gonzalez-Tudela, D.~Martin-Cano, E.~Moreno, L.~Martin-Moreno, C.~Tejedor, and F.~J.~Garcia-Vidal,
Entanglement of two qubits mediated by one-dimensional plasmonic waveguides,
\href{https://doi.org/10.1103/PhysRevLett.106.020501}
{Phys. Rev. Lett. \textbf{106}, 020501 (2011)}.

\bibitem{MartinCano2011}
D.~Martin-Cano, A.~Gonzalez-Tudela, L.~Martin-Moreno, F.~J.~Garcia-Vidal, C.~Tejedor, and E.~Moreno,
Dissipation-driven generation of two-qubit entanglement mediated by plasmonic waveguides,
\href{https://doi.org/10.1103/PhysRevB.84.235306}
{Phys. Rev. B \textbf{84}, 235306 (2011)}.

\bibitem{Tanas2004}
R.~Tanas, and Z.~Ficek,
Stationary two-atom entanglement induced by nonclassical two-photon correlations,
\href{https://doi.org/10.1088/1464-4266/6/6/022}
{J. Opt. B: Quantum Semiclass. Opt. \textbf{6}, S610 (2004)}.

\bibitem{Valle2007}
E.~del Valle, F.~P.~Laussy, F.~Troiani, and C.~Tejedor,
Entanglement and lasing with two quantum dots in a microcavity,
\href{https://doi.org/10.1103/PhysRevB.76.235317}
{Phys. Rev. B \textbf{76}, 235317 (2007)}.

\bibitem{Almutairi2011}
K.~Almutairi, R.~Tanas, and Z.~Ficek,
Generating two-photon entangled states in a driven two-atom system,
\href{https://doi.org/10.1103/PhysRevA.84.013831}
{Phys. Rev. A \textbf{84}, 013831 (2011)}.


\bibitem{Valle2011}
E.~del Valle,
Steady-state entanglement of two coupled qubits,
\href{https://doi.org/10.1364/JOSAB.28.000228}
{Journal of the Optical Society of America B \textbf{28}, 228 (2011)}.

\bibitem{Liao2011}
J.-Q.~Liao, J.-F.~Huang, and L.-M.~Kuang,
Quantum thermalization of two coupled two-level systems in eigenstate and bare-state representations,
\href{https://doi.org/10.1103/PhysRevA.83.052110}
{Phys. Rev. A \textbf{83}, 052110 (2011)}.


\bibitem{Biehs2017}
S.-A.~Biehs and G.~S.~Agarwal,
Qubit entanglement across epsilon-near-zero media,
\href{https://doi.org/10.1103/PhysRevA.83.052110}
{Phys. Rev. A \textbf{96}, 022308 (2017)}.



\bibitem{Grigorenko2005}
I.~A.~Grigorenko and D.~V.~Khveshchenko, 
Robust two-qubit quantum registers,
\href{https://doi.org/10.1103/PhysRevLett.94.040506}
{Phys. Rev. Lett. \textbf{94}, 040506 (2005)}.


\bibitem{Petrosyan2002}
D.~Petrosyan and G.~Kurizki,
Scalable solid-state quantum processor using subradiant two-atom states,
\href{https://doi.org/10.1103/PhysRevLett.89.207902}
{Phys. Rev. Lett. \textbf{89}, 207902 (2002)}.


\bibitem{Liao2010}
J.-Q.~Liao, J.-F.~Huang, L.-M.~Kuang, and C.~P.~Sun,
Coherent excitation-energy transfer and quantum entanglement in a dimer,
\href{https://doi.org/10.1103/PhysRevA.82.052109}
{Phys. Rev. A \textbf{82}, 052109 (2010)}.


\bibitem{McCutcheon2011}
D.~P.~S.~McCutcheon and A.~Nazir,
Coherent and incoherent dynamics in excitonic energy transfer: Correlated fluctuations and off-resonance effects,
\href{https://doi.org/10.1103/PhysRevB.83.165101}
{Phys. Rev. B \textbf{83}, 165101 (2011)}.


\bibitem{Wang2020}
Y.-P.~Wang, and C.-M.~Hu,
Dissipative couplings in cavity magnonics,
\href{https://doi.org/10.1063/1.5144202}
{Journal of Applied Physics \textbf{127}, 130901 (2020)}.

\bibitem{DowningLuis}
C.~A.~Downing, T.~J.~Sturges, G.~Weick, M.~Stobi\'{n}ska, and L.~Mart\'{i}n-Moreno,
Topological phases of polaritons in a cavity waveguide,
\href{https://doi.org/10.1103/PhysRevLett.123.217401}
{Phys. Rev. Lett. \textbf{123}, 217401 (2019)}.


\bibitem{Haldane1988}
F.~D.~M.~Haldane,
Model for a quantum Hall effect without Landau levels: condensed-matter realization of the `parity anomaly',
\href{https://doi.org/10.1103/PhysRevLett.61.2015}
{Phys. Rev. Lett. \textbf{61}, 2015 (1988)}.


\bibitem{Poyatos1996}
J.~F.~Poyatos, J.~I.~Cirac, and P.~Zoller,
Quantum reservoir engineering with laser cooled trapped ions,
\href{https://doi.org/10.1103/PhysRevLett.77.4728}
{Phys. Rev. Lett. \textbf{77}, 4728 (1996)}.


\bibitem{delValle2010}
E.~del~Valle,
Strong and weak coupling of two coupled qubits,
\href{https://doi.org/10.1103/PhysRevA.81.053811}
{Phys. Rev. A. \textbf{81}, 053811 (2010)}.

\bibitem{Ficek2002}
Z.~Ficek, and R.~Tana\'{s},
Entangled states and collective nonclassical effects in two-atom systems,
\href{https://doi.org/10.1016/S0370-1573(02)00368-X}
{Phys. Rep. \textbf{372}, 369 (2002)}.


\bibitem{Downing2017}
C.~A.~Downing, E.~Mariani, and G.~Weick,
Radiative frequency shifts in nanoplasmonic dimers,
\href{https://doi.org/10.1103/PhysRevB.96.155421}
{Phys. Rev. B \textbf{96}, 155421 (2017)}.

\bibitem{Lembessis2013}
V.~E.~Lembessis, A.~Al~Rsheed, O.~M.~Aldossary, and Z.~Ficek,
Two-atom system as a nano-antenna for mode switching and light routing,
\href{https://doi.org/10.1103/PhysRevA.88.053814}
{Phys. Rev. A \textbf{88}, 053814 (2013)}.

\bibitem{Gardiner2014}
C.~Gardiner and P.~Zoller,
\textit{The Quantum World of Ultra-cold Atoms and Light, Book I: Foundations of Quantum Optics} (Imperial College Press, London, 2014).


\bibitem{TroianiLaussy2007}
E.~del~Valle, F.~Laussy, F.~Troiani, and C.~Tejedor,
The steady state of two quantum dots in a cavity,
\href{https://doi.org/10.1016/j.spmi.2007.07.001}
{Superlattice. Microst. \textbf{43}, 465 (2007)}.



\bibitem{Dung2002}
H.~T.~Dung, L.~Knoll, and D.-G.~Welsch,
Resonant dipole-dipole interaction in the presence of dispersing and absorbing surroundings,
\href{https://doi.org/10.1103/PhysRevA.66.063810}
{Phys. Rev. A \textbf{66}, 063810 (2002)}.


 

\bibitem{Radmore1982}
P.~M.~Radmore and P.~L.~Knight,
Population trapping and dispersion in a three-level system,
\href{https://doi.org/10.1088/0022-3700/15/4/009}
{J. Phys. B: At. Mol. Phys. \textbf{15}, 561 (1982)}.

\bibitem{Dalton1982}
B.~J.~Dalton and P.~L.~Knight,
The effects of laser field fluctuations on coherent population trapping,
\href{https://doi.org/10.1088/0022-3700/15/21/019}
{J. Phys. B: At. Mol. Phys. \textbf{15}, 3997 (1982)}.

\bibitem{Swain1982}
S.~Swain,
Conditions for population trapping in a three-level system,
\href{https://doi.org/10.1088/0022-3700/15/19/010}
{J. Phys. B: At. Mol. Phys. \textbf{15}, 3405 (1982)}.


\bibitem{Akram2000}
U.~Akram, Z.~Ficek, and S.~Swain,
Decoherence and coherent population transfer between two coupled systems,
\href{https://doi.org/10.1103/PhysRevA.62.013413}
{Phys. Rev. A \textbf{62}, 013413 (2000)}.



\bibitem{Dorfman2016}
K.~E.~Dorfman, F.~Schlawin, and S.~Mukamel,
Nonlinear optical signals and spectroscopy with quantum light,
\href{https://doi.org/10.1103/RevModPhys.88.045008}
{Rev. Mod. Phys. \textbf{88}, 045008 (2016)}.



\bibitem{Aspect1988}
A.~Aspect, E.~Arimondo, R.~Kaiser, N.~Vansteenkiste, and C.~Cohen-Tannoudji,
Laser cooling below the one-photon recoil energy by velocity-selective coherent population trapping,
\href{https://doi.org/10.1103/PhysRevLett.61.826}
{Phys. Rev. Lett. \textbf{61}, 826 (1988)}.

\bibitem{Santori2006}
C.~Santori, P.~Tamarat, P.~Neumann, J.~Wrachtrup, D.~Fattal, R.~G.~Beausoleil, J.~Rabeau, P.~Olivero, A.~D.~Greentree, S.~Prawer, F.~Jelezko, and P.~Hemmer,
Coherent population trapping of single spins in diamond under optical excitation,
\href{https://doi.org/10.1103/PhysRevLett.97.247401}
{Phys. Rev. Lett. \textbf{97}, 247401 (2006)}.

\bibitem{Agarwal2006}
G.~S.~Agarwal, and K.~T.~Kapale,
Subwavelength atom localization via coherent population trapping,
\href{https://doi.org/10.1088/0953-4075/39/17/002}
{J. Phys. B: At. Mol. Opt. Phys. \textbf{39}, 3437 (2006)}.

\bibitem{Amico2019}
I.~D'Amico, D.~G.~Angelakis, F.~Bussieres, H.~Caglayan, C.~Couteau, T.~Durt, B.~Kolaric, P.~Maletinsky, W.~Pfeiffer, P.~Rabl, A.~Xuereb, and M.~Agio,
Nanoscale quantum optics,
\href{https://doi.org/10.1393/ncr/i2019-10158-0}
{Riv. Nuovo Cimento \textbf{42}, 153 (2019)}.

\bibitem{Aidelsburger2013}
M.~Aidelsburger, M.~Atala, M.~Lohse, J.~T.~Barreiro, B.~Paredes, and I.~Bloch,
Realization of the Hofstadter Hamiltonian with ultracold atoms in optical lattices,
\href{https://doi.org/10.1103/PhysRevLett.111.185301}
{Phys. Rev. Lett. \textbf{111}, 185301 (2013)}.

\bibitem{Celi2014}
A.~Celi, P.~Massignan, J.~Ruseckas, N.~Goldman, I.~B.~Spielman, G.~Juzeliunas, and M.~Lewenstein,
Synthetic gauge fields in synthetic dimensions,
\href{https://doi.org/10.1103/PhysRevLett.112.043001}
{Phys. Rev. Lett. \textbf{112}, 043001 (2014)}.


\bibitem{Fang2012b}
K.~Fang, Z.~Yu, and S.~Fan,
Realizing effective magnetic field for photons by controlling the phase of dynamic modulation,
\href{https://doi.org/10.1038/nphoton.2012.236}
{Nat. Photonics \textbf{6}, 782 (2012)}.

\bibitem{Fang2013}
K.~Fang, and S.~Fan,
Controlling the flow of light using the inhomogeneous effective gauge field that emerges from dynamic modulation,
\href{https://doi.org/10.1103/PhysRevLett.111.203901}
{Phys. Rev. Lett. \textbf{111}, 203901 (2013)}.

\bibitem{Tzuang2014}
L.~D.~Tzuang, K.~Fang, P.~Nussenzveig, S.~Fan, and M.~Lipson,
Non-reciprocal phase shift induced by an effective magnetic flux for light,
\href{https://doi.org/10.1038/nphoton.2014.177}
{Nat. Photonics \textbf{8}, 701 (2014)}.

\bibitem{Mirza2017}
I.~M.~Mirza, J.~G.~Hoskins, and J.~C.~Schotland,
Chirality, band structure, and localization in waveguide quantum electrodynamics,
\href{https://doi.org/10.1103/PhysRevA.96.053804}
{Phys. Rev. A \textbf{96}, 053804 (2017)}.  

\bibitem{Jen2019}
H.~H.~Jen,
Selective transport of atomic excitations in a driven chiral-coupled atomic chain,
\href{https://doi.org/10.1088/1361-6455/ab04c1}
{J. Phys. B: At. Mol. Opt. Phys. \textbf{52}, 065502 (2019)}.

\bibitem{Mirza2018}
I.~M.~Mirza, J.~G.~Hoskins, and J.~C.~Schotland,
Dimer chains in waveguide quantum electrodynamics,
\href{https://doi.org/10.1016/j.optcom.2020.125427}
{Opt. Commun. \textbf{463}, 125427 (2020)}.

\bibitem{Buonaiuto2019}
G.~Buonaiuto, R.~Jones, B.~Olmos, and I.~Lesanovsky,
Dynamical creation and detection of entangled many-body states in a chiral atom chain,
\href{https://doi.org/10.1088/1367-2630/ab4f50}
{New J. Phys. \textbf{21}, 113021 (2019)}.

\bibitem{Lin2020}
H.~H.~Jen, M.-S.~Chang, G.-D.~Lin, and Y.-C.~Chen,
Subradiance dynamics in a singly excited chirally coupled atomic chain,
\href{https://doi.org/10.1103/PhysRevA.101.023830}
{Phys. Rev. A \textbf{101}, 023830 (2020)}.

\bibitem{Jenny2020}
H.~H.~Jen,
Steady-state phase diagram of a weakly driven chiral-coupled atomic chain,
\href{https://doi.org/10.1103/PhysRevResearch.2.013097}
{Phys. Rev. Research \textbf{2}, 013097 (2020)}.

\bibitem{Jen2019b}
H.~H.~Jen,
Quantum-coherence-enhanced subradiance in a chiral-coupled atomic chain,
\href{https://arxiv.org/abs/1903.05352}
{arXiv:1903.05352}.

\bibitem{Schrapp2012}
M.~Schrapp, E.~del Valle, J.~J.~Finley, and F.~P.~Laussy,
Quantum dynamics of damped and driven anharmonic oscillators,
\href{https://doi.org/10.1002/pssc.201100195}
{Phys. Status Solidi c \textbf{9}, 1296 (2012)}.


\bibitem{ValleBook2010}
E.~del~Valle,
\textit{Microcavity Quantum Electrodynamics} (VDM Verlag, Saarbr\"{u}cken, 2010).

\bibitem{Huang2020}
L.~Huang, L.~Xu, M.~Woolley, and A.~E.~Miroshnichenko,
Trends in quantum nanophotonics,
\href{https://doi.org/10.1002/qute.201900126}
{Adv. Quantum Technol. \textbf{42}, 153 (2020)}.

\bibitem{ValleLaussy2009}
E.~del Valle, F.~P.~Laussy, and C.~Tejedor,
Quantum regression formula and luminescence spectra of two coupled modes under incoherent continuous pumping,
\href{https://doi.org/10.1103/PhysRevA.84.043816}
{AIP Conference Proceedings \textbf{1147}, 238 (2009)}.

\bibitem{ValleLaussy2011}
E. del Valle and F. P. Laussy,
Regimes of strong light-matter coupling under incoherent excitation,
\href{https://doi.org/10.1103/PhysRevA.84.043816}
{Phys. Rev. A \textbf{84}, 043816 (2011)}.

\bibitem{Kavokin2007}
A.~V.~Kavokin, J.~J.~Baumberg, G.~Malpuech, and F.~P.~Laussy,
\textit{Microcavities} (2nd edition, Oxford University Press, 2017).

\bibitem{Laussy2008}
F.~P.~Laussy, E.~del~Valle, and C.~Tejedor,
Strong coupling of quantum dots in microcavities,
\href{https://doi.org/10.1103/PhysRevLett.101.083601}
{Phys. Rev. Lett. \textbf{101}, 083601 (2008)}.

\bibitem{Laussy2009}
F.~P.~Laussy, E.~del~Valle, and C.~Tejedor,
Luminescence spectra of quantum dots in microcavities. I. Bosons,
\href{https://doi.org/10.1103/PhysRevB.79.235325}
{Phys. Rev. B \textbf{79}, 235325 (2009)}.

\bibitem{Valle2009}
E.~del~Valle, F.~P.~Laussy, and C.~Tejedor,
Luminescence spectra of quantum dots in microcavities. II. Fermions,
\href{https://doi.org/10.1103/PhysRevB.79.235326}
{Phys. Rev. B \textbf{79}, 235326 (2009)}.

\bibitem{Villas2011}
F.~P.~Laussy, A.~Laucht, E.~del Valle, J.~J.~Finley, and J.~M.~Villas-Boas,
Luminescence spectra of quantum dots in microcavities. III. Multiple quantum dots,
\href{https://doi.org/10.1103/PhysRevB.84.195313}
{Phys. Rev. B \textbf{84}, 195313 (2011)}.



\bibitem{ValleSuper2011}
E.~del~Valle and F.~P.~Laussy,
Effective cavity pumping from weakly coupled quantum dots,
\href{https://doi.org/10.1016/j.spmi.2010.05.005}
{Superlatt. Microstruct. \textbf{49}, 241 (2011)}.



\bibitem{Valle208}
E.~del Valle, F.~P.~Laussy, F.~M.~Souza, and I.~A.~Shelykh,
Optical spectra of a quantum dot in a microcavity in the nonlinear regime,
\href{https://doi.org/10.1103/PhysRevB.78.085304}
{Phys. Rev. B \textbf{78}, 085304 (2008)}.

\bibitem{Degenfeld2012}
P.~Degenfeld-Schonburg, E.~del Valle, and M.~J.~Hartmann,
Signatures of single-site addressability in resonance fluorescence spectra,
\href{https://doi.org/10.1103/PhysRevA.85.013842}
{Phys. Rev. A \textbf{85}, 013842 (2012)}.





\bibitem{Valle211}
E.~del Valle and A.~Kavokin,
Terahertz lasing in a polariton system: quantum theory,
\href{https://doi.org/10.1103/PhysRevB.83.193303}
{Phys. Rev. B \textbf{83}, 193303 (2011)}.






\end{thebibliography}
\end{document}